\documentclass[a4paper,11pt]{article}
\usepackage{pos}
\usepackage{mciteplus}
\makeatletter
\let\orig@mcitethebibliography\mcitethebibliography
\renewcommand{\mcitethebibliography}[1]{%
  \orig@mcitethebibliography{#1}%
  \setlength{\itemsep}{3pt}%
  \setlength{\parsep}{0pt}%
  \setlength{\topsep}{0pt}%
  \setlength{\parskip}{0pt}%
}
\makeatother
\title{Hadron Structure from lattice QCD in the context of the Electron-Ion Collider}

\author{Constantia Alexandrou}

\affiliation{Department of Physics, University of Cyprus, PO Box 20537, 1678 Nicosia, Cyprus and\\
  The Cyprus Institute, 20 Konstantinou Kavafi Str., 
2121 Aglantzia, Nicosia, Cyprus}

\emailAdd{alexand@ucy.ac.cy}

\abstract{Hadron structure calculations using lattice Quantum Chromodynamics (QCD) have advanced significantly in recent years. Results for charges, form factors, and lower Mellin moments can be obtained to high precision, generalized parton distributions can now be computed either directly or reconstructed from  moments, and transverse-momentum–dependent distributions can be accessed through direct lattice calculations. Together, these quantities provide detailed and complementary insights into the internal structure of hadrons.
These theoretical developments are highly relevant to the experimental program of the Electron–Ion Collider (EIC) and of other facilities. We review the most pertinent lattice QCD results for hadron structure that inform the EIC scientific agenda, with particular emphasis on the pion, kaon, and nucleon.}

\FullConference{The 42nd International Symposium on Lattice Field Theory (LATTICE2025)\\
2-8 November 2025\\
Tata Institute of Fundamental Research, Mumbai, India\\}

\begin{document}
\maketitle

\section{Introduction}\vspace*{-0.5cm}
Lattice Quantum Chromodynamics (LQCD)~\cite{Wilson:1974sk} 
is nowadays considered the primary non-perturbative framework for computing the properties of hadrons and their interactions.  
Its current level of precision and scope has been achieved after more than four decades of theoretical and algorithmic developments following the pioneering simulation of the SU(2) pure Yang–Mills
theory
in 1980~\cite{Creutz:1980zw}.
In this overview, we focus on some of the essential inputs that LQCD can provide in connection to the physics program of the
Electron--Ion Collider (EIC) being constructed at Brookhaven National Lab (BNL), as articulated, for instance, in the EIC reports
~\cite{AbdulKhalek:2021gbh, Aschenauer:2019kzf}.

A central pillar of the EIC science case is the three-dimensional  (3D) imaging
of hadrons in terms of quark and gluon degrees of freedom.
In this context, LQCD has made substantial progress in two directions:
i) Percent-level precision is routinely achieved for a range of quantities, such as the nucleon
 axial and tensor charges, electromagnetic and axial
form factors, and second moments of parton distribution functions
(PDFs), matching the
accuracy targets required to fully exploit EIC measurements; ii) The  
 Bjorken-$x$ dependence of PDFs and generalized parton distributions
(GPDs) can now be extracted within LQCD using e.g. the Large-Momentum Effective Theory (LaMET)~\cite{Ji:2013dva,Ji:2020ect} and related
approaches~\cite{Ma:2017pxb,Radyushkin:2017cyf, Chambers:2017dov, Liu:2016djw, Detmold:2005gg}. Calculations employing hadron
boosts $P_z\sim 1.5$--$2$~GeV, nonperturbative renormalization, and 
perturbative matching up to NNLO  yield results consistent with phenomenological
extractions in controlled kinematic regions~\cite{Cichy:2018mum}.
Ongoing efforts target flavor separation, gluon distributions,  and higher-twist effects, all of which are identified as
key deliverables in the EIC theory program~\cite{Accardi:2012qut}.

Transverse-momentum--dependent distributions (TMDs) constitute another
major component of the EIC physics agenda. Exploratory lattice  QCD studies
using staple-shaped Wilson lines have begun to address process
dependence, soft factors, and rapidity evolution
\cite{Musch:2010ka, Ji:2018hvs, Ebert:2018gzl}, laying the groundwork for quantitative
connections to semi-inclusive deep-inelastic scattering and Drell--Yan
measurements at the EIC. While these calculations remain 
 challenging, they are progressing in parallel with
advances in factorization and phenomenology, consistent with the
coordinated theory effort envisioned for the EIC~\cite{Scimemi:2019cmh}. In this overview, we will not cover progress in the determination of TMDs within LQCD, see Ref.\cite{Boussarie:2023izj} for a recent review.

Achieving the precision and systematic control demanded by the EIC
program requires iterative refinement of lattice spacing, volume,
momentum reach, and operator renormalization, closely coupled to
phenomenological needs.
\begin{wrapfigure}[8]{L}{0.7\linewidth}\vspace*{-0.5cm}
\includegraphics[width=\linewidth]{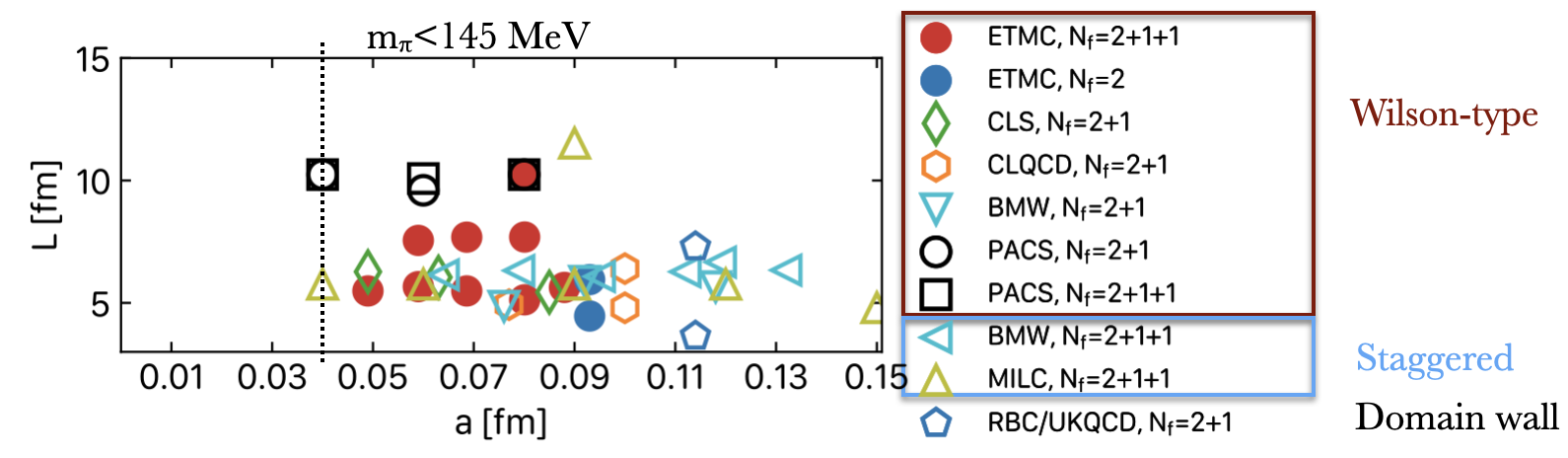}\vspace{-0.2cm}
    \caption{Gauge ensembles used for hadron structure simulated for $m_\pi<145$~ MeV.}\label{fig:sims}\vspace*{-0.7cm}
\end{wrapfigure}

\noindent
Near-term goals in LQCD for EIC physics include extending form factors to larger momentum transfers, computing higher Mellin moments,   reducing discretization 
uncertainties, and reaching higher
boosts ($P_z\gtrsim 3$~GeV) in the computations of GPDs, as well as  incorporating QED effects for quantities  where percent precision can now be reached.
Algorithmic advances---such as multilevel methods, improved momentum
smearing, and machine-learning--assisted  reconstruction
techniques~\cite{Shanahan:2022ifi}---are expected to significantly enhance
statistical precision and kinematic reach. 

\noindent
QCD  can provide critical input on  pion, kaon  and proton structure that can aid the design of detectors at the EIC but also  at other future experiments, such as AMBER at CERN. LQCD results can also help  the interpretation of results from EIC and current facilities, such as  Jefferson Lab (JLab) and CERN. Continued development is
essential for transforming  high-quality experimental data
into a quantitative, three-dimensional understanding of hadron
structure and the dynamics of QCD.

State-of-the-art calculations are routinely performed with
$N_f=2+1$ and $N_f=2+1+1$ dynamical fermions at the physical pion mass,
lattice spacings as small as $a\simeq 0.05$~fm, and spatial volumes
$L\gtrsim 6$~fm, enabling controlled continuum  and
infinite-volume extrapolations. In Fig.~\ref{fig:sims}, we show gauge ensembles simulated with pion mass $m_\pi<145$ MeV using Wilson-improved, staggered and domain wall fermions. Simulations at smaller lattice spacings become challenging due to the large autocorrelation times. \vspace*{-0.3cm}

\section{Mellin moments}\vspace*{-0.3cm}
While light-cone matrix elements cannot be computed using a Euclidean lattice formulation, the operator product expansion expresses such matrix elements as a tower of matrix elements of  local operators that are  connected to Mellin moments computable in LQCD. To leading order, the relevant  operators  are \vspace*{-0.3cm}
 \begin{equation} {\cal O}_{V[A]}^{\mu_1\mu_2\ldots\mu_{n}}      = \bar{\psi}  \gamma^{\{\mu_1}[\gamma_5]i\overleftrightarrow{D}^{\mu_2}\ldots i\overleftrightarrow{D}^{\mu_{n}\}} \psi\,\,\, {\rm and}\,\,\,
   {\cal O}^{\rho\mu_1\mu_2\ldots\mu_{n}}_T = \bar{\psi}  i\sigma^{\rho\{\mu_1}i\overleftrightarrow{D}^{\mu_2}\ldots i\overleftrightarrow{D}^{\mu_{n}\}} \psi\vspace*{-0.3cm}
\end{equation} 
 made traceless, where    $\{\}$ denotes symmetrization and $\overleftrightarrow{D}=1/2(\overrightarrow{D}-\overleftarrow{D})$~\cite{Hagler:2009ni}. They are related through moments in the momentum fraction $x$, with the $n^{\rm th}$ moment given by $\int_{-1}^{1}dxx^{n-1}f(x)$. Such computations were first performed in the early 90s.
 Precise results  exist mostly for the two lowest moments of the  nucleon and  pion, the structure of which will be studied in great detail at the EIC. The structure of the kaon,  another targeted particle at the EIC, has received less attention and it is a case where LQCD can provide precise results on its structure.\vspace*{-0.3cm}

\subsection{Pion and Kaon structure} 
\begin{figure}[h!]
   \includegraphics[width=\linewidth]{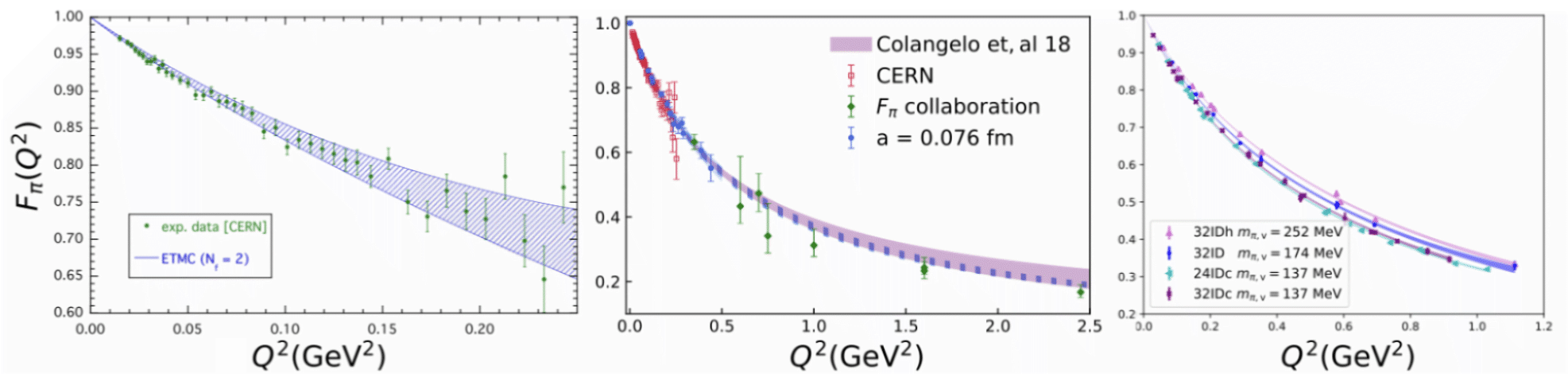}
    \caption{Pion vector form factor from: i) Left panel: ETMC results extrapolated to the chiral  limit for low $Q^2$ compared to data from CERN~\cite{ETM:2017wqc}; ii) Middle panel: BNL/JLab  results compared to experimental data and phenomenology~\cite{Gao:2021xsm}; iii) Right panel: $\chi$QCD for $m_\pi=252, 174$ and  137~MeV for intermediate  $Q^2$~\cite{Wang:2020nbf}. }\label{fig:Fpi}\vspace*{-0.6cm}
\end{figure}

For spin-0 particles, the lowest moment ($n=1$)
yields the    vector, $g_V$, and tensor, $g_T$, charges and the corresponding form factors, $F_V(Q^2)$ and $F_T(Q^2)$.  
One can also consider the matrix elements of the scalar operator, which yields the scalar charge $g_S$  and $F_S(Q^2)$. In Fig.~\ref{fig:Fpi}, we show results from three groups focusing on different Euclidean momentum transfer squared ($Q^2$) ranges: i) The Extended Twisted Mass Collaboration (ETMC), which  used $N_f=2$   twisted mass fermion (TMF) ensembles with $m_\pi=140, 240, 340$~ MeV and $a=0.09$~fm, for low $Q^2$. Volume corrected results are extrapolated to the chiral  limit reproducing   experimental results from CERN.  ii) The BNL/JLab  group, which  analyzed  one ensemble of Wilson-clover valence quarks and HISQ sea quarks with $m_{\pi}=140$~MeV and $a=0.076$ in the Breit frame, allowing them to reach  $Q^2$ as large as 2.5~GeV$^2$. Their values are in agreement with experimental and phenomenological results. iii) The  $\chi$QCD collaboration that performed the calculation using  seven ensembles of
domain-wall  and  valence overlap fermions spanning pion  masses $m_\pi \in (137$--$340)$ MeV in the intermediate $Q^2$ range, investigating the pion mass dependence.

\begin{wrapfigure}[12]{L}{0.35\linewidth}\vspace*{-0.5cm}
    \includegraphics[width=\linewidth]{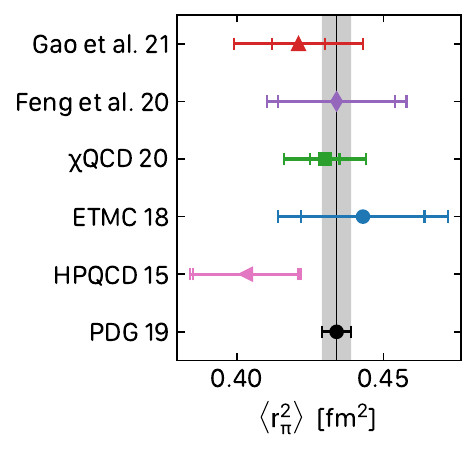}\vspace{-0.5cm}
    \caption{LQCD results on the pion charge radius compared to the PDG value (gray band).}
    \label{fig:rpi}
\end{wrapfigure}
The pion charge radius can be extracted from the slope of the form factor as $Q^2\rightarrow 0$. In Fig.~\ref{fig:rpi}, we show results for the root mean square charge radius (r.m.s.) $\langle r_\pi^2\rangle$ from various recent LQCD determinations that show agreement with  the PDG value, albeit having larger errors. Although LQCD may not reach  the large $Q^2$-values of $\sim 30$~GeV$^2$ targeted by JLab and EIC, extending the $Q^2$-range, as well as increasing the accuracy for the r.m.s radius  will still be important. LQCD results on the pion scalar and tensor form factors and on all three kaon form factors are scarce and this is an area where LQCD can provide valuable results.

\begin{figure}[h!]
\begin{minipage}{0.33\linewidth}
\hspace*{-0.8cm}\includegraphics[width=1.35\linewidth]{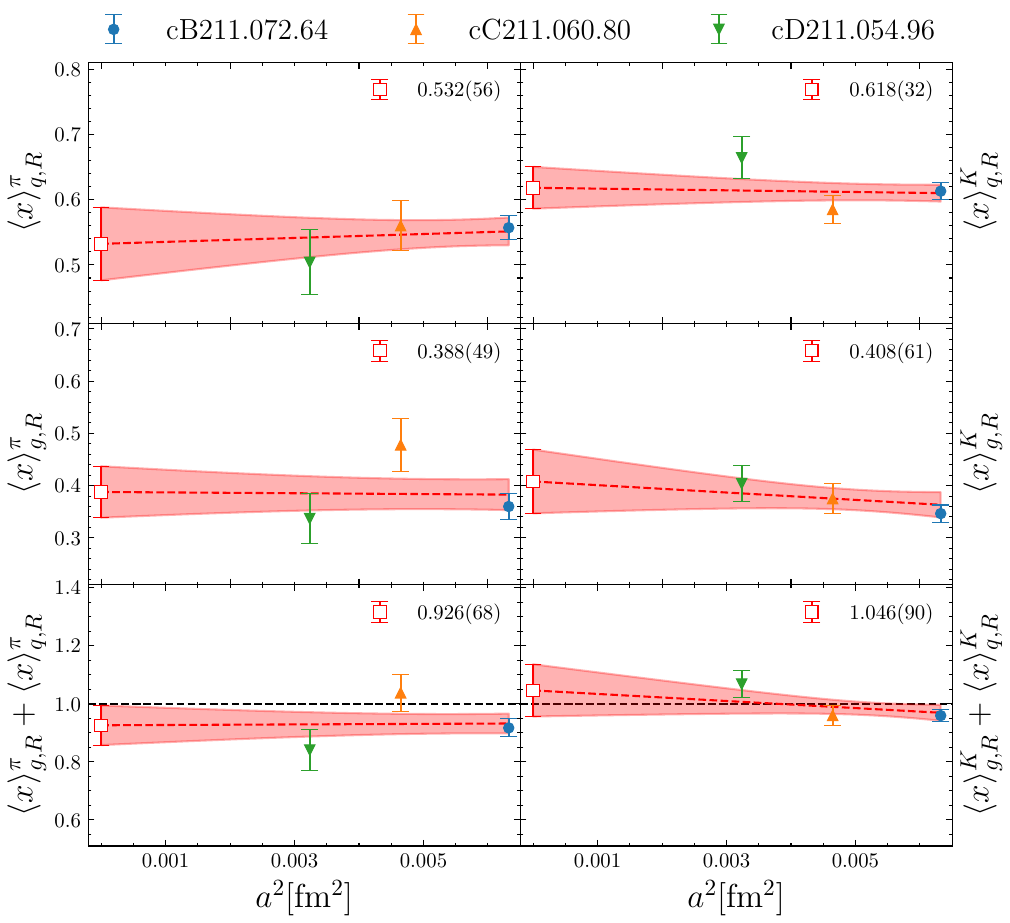}
\end{minipage}\hfill
\begin{minipage}{0.3\linewidth}
\hspace*{0.8cm}\includegraphics[width=0.9\linewidth]{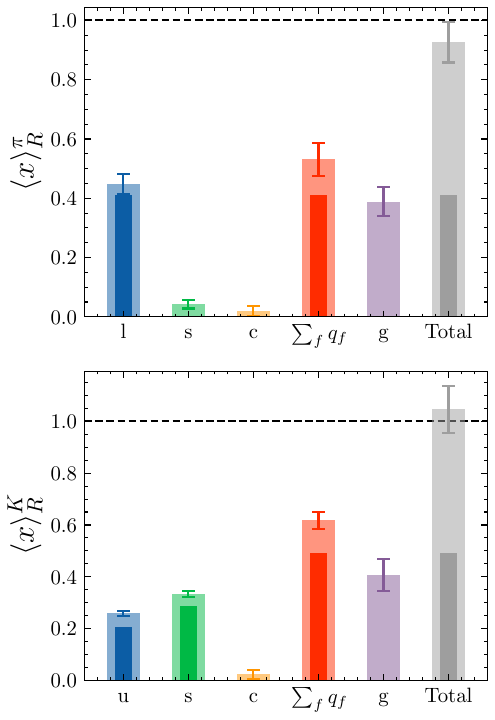}
\end{minipage}\hfill
\begin{minipage}{0.33\linewidth}
\hspace*{0.2cm}\includegraphics[width=1.1\linewidth]{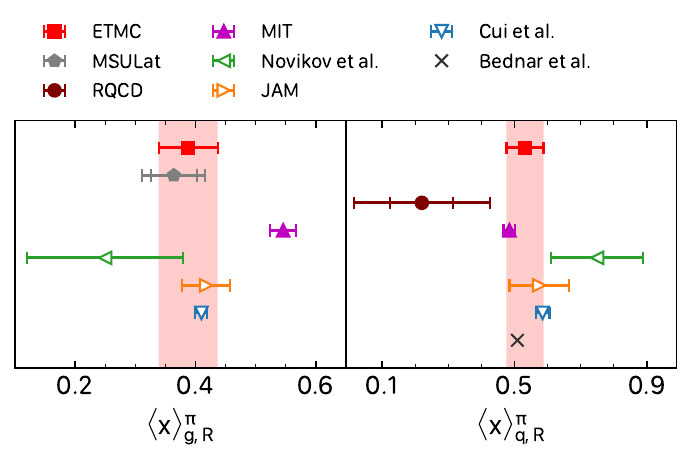}\\
\hspace*{0.2cm}\includegraphics[width=1.1\linewidth]{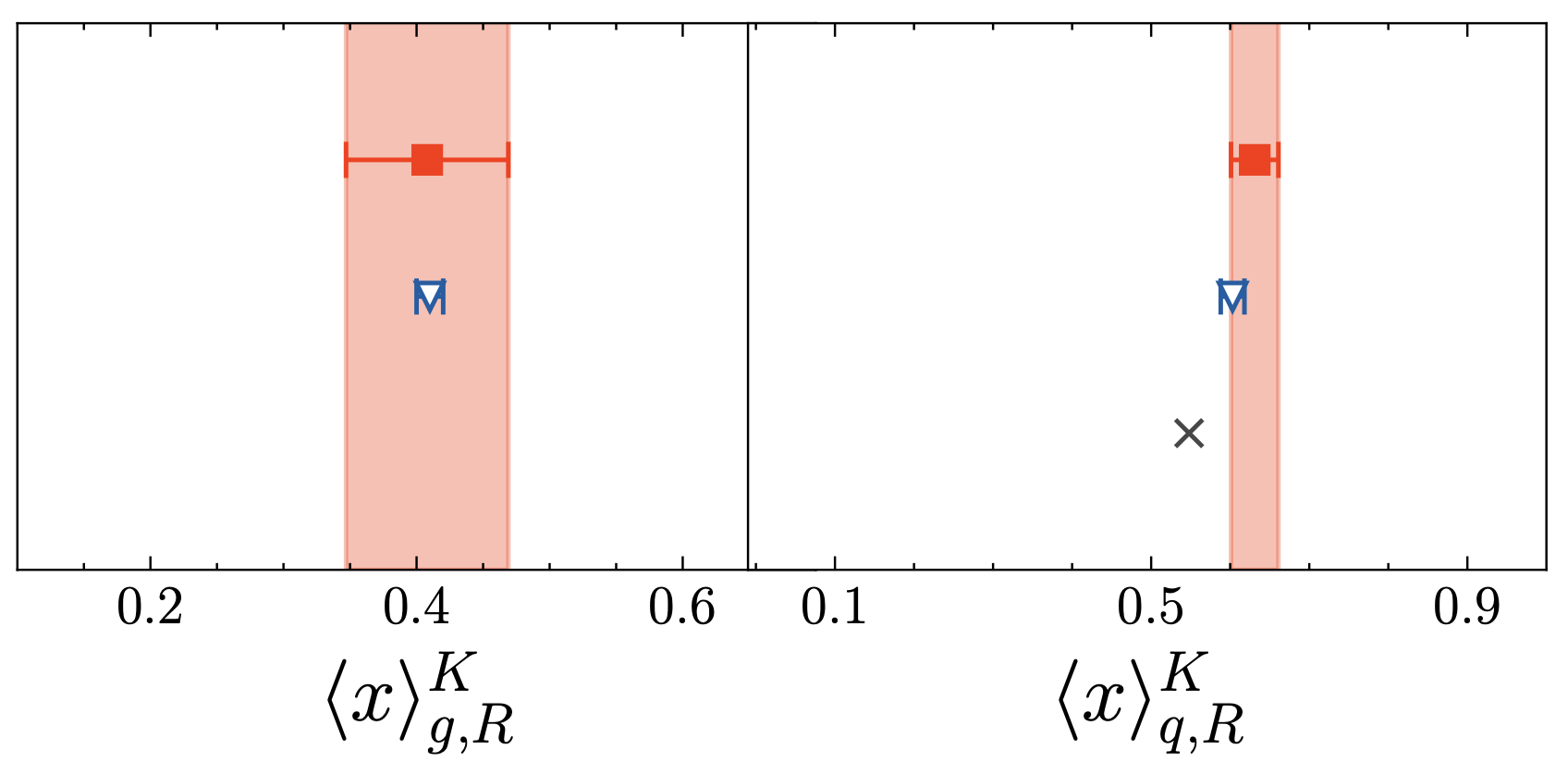}
\end{minipage}\vspace*{-0.3cm}
    \caption{Left: Continuum extrapolation of the momentum fraction of the pion and kaon using ETMC ensembles; Middle: Contributions of quarks and gluons to the pion and kaon momentum faction; Right: Comparison of the pion and kaon momentum fractions with other recent LQCD and phenomenology results. Figures are  from Ref.~\cite{ExtendedTwistedMass:2024kjf}. The pion comparison figure  includes the recent MIT group values~\cite{Hackett:2023nkr}.}\label{fig:xpion-kaon}\vspace*{-0.2cm}
\end{figure}

\begin{figure}[h!]
\begin{minipage}{0.49\linewidth}
\hspace*{-0.6cm}\includegraphics[width=1.1\linewidth,height=0.7\linewidth]{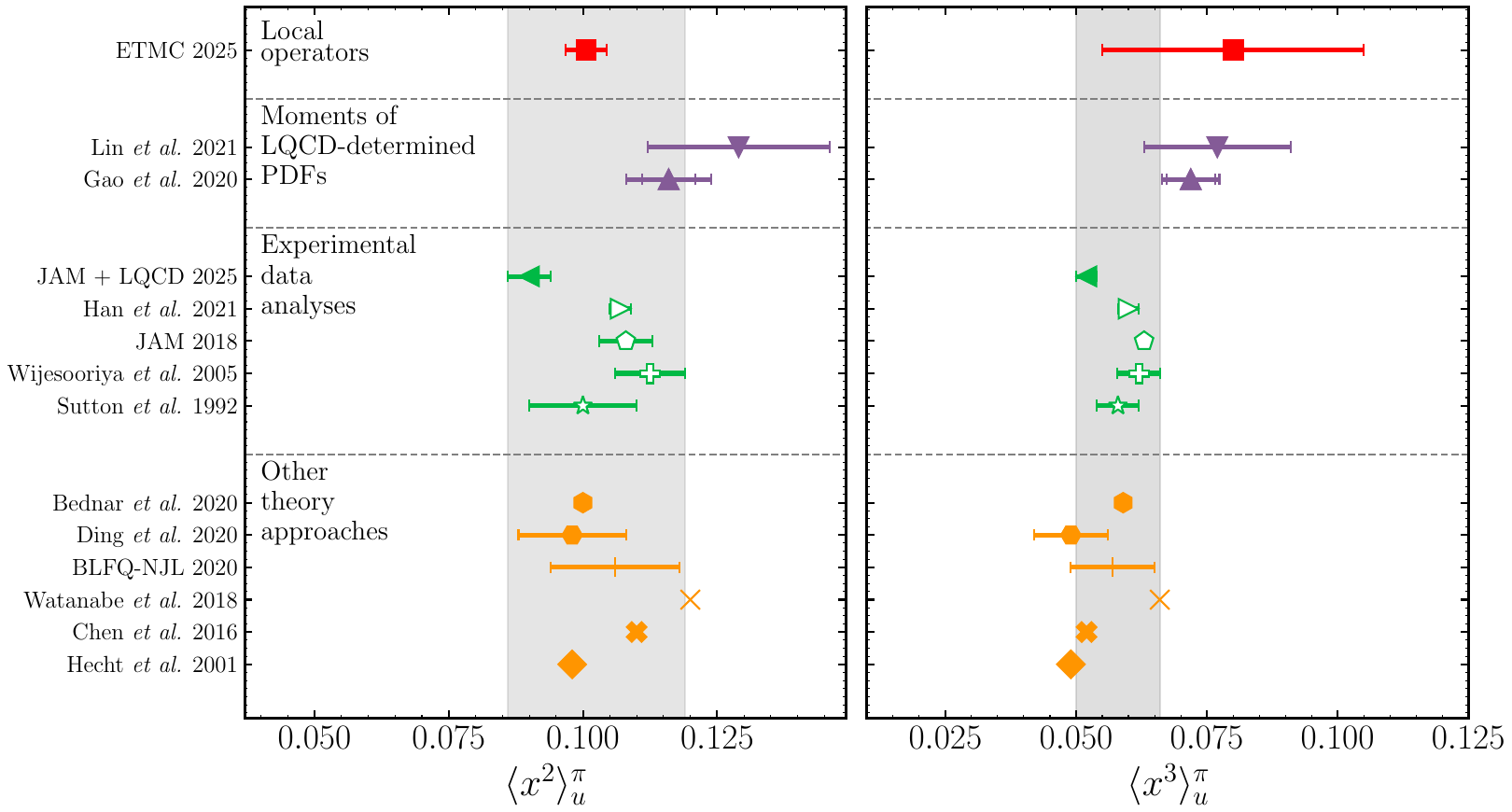}\\\vspace*{-0.7cm}
\caption{Results on the third and fourth unpolarized moments for the pion (left), and the u-~(right top) and s-~(right bottom) quark in the kaon. LQCD data are compared to  analyses of experimental data  and other theoretical determinations. The gray band denotes the spread of  experimental analyses. All results are in the $\overline{\text{MS}}$ scheme at 2~GeV.}\label{fig:x23}
\end{minipage}\hfill
\begin{minipage}{0.49\linewidth}\vspace*{-0.3cm}
\hspace*{-0.3cm}\includegraphics[width=1.05\linewidth]{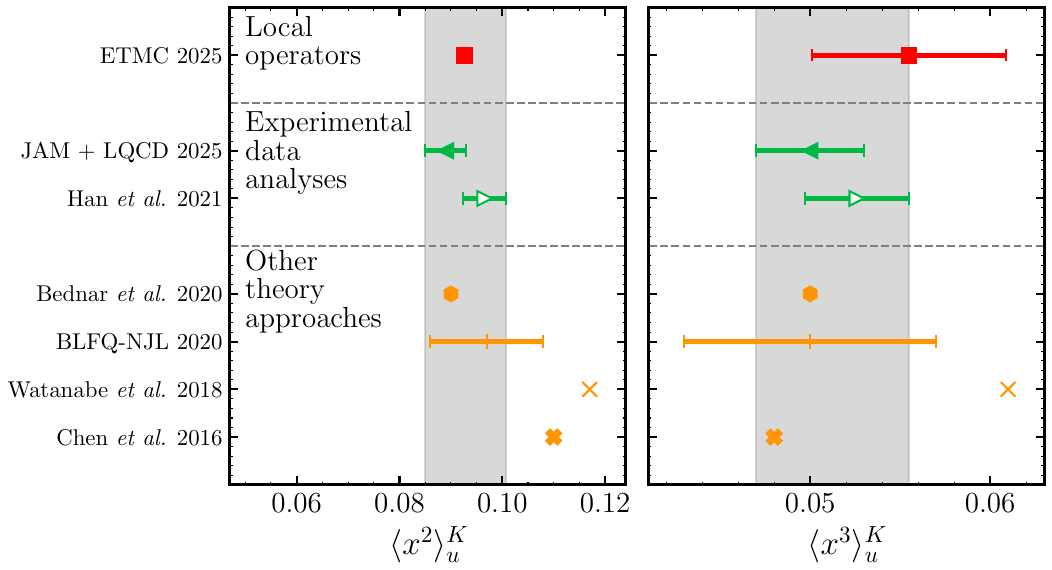}\\
 \hspace*{-0.3cm} \includegraphics[width=1.05\linewidth]{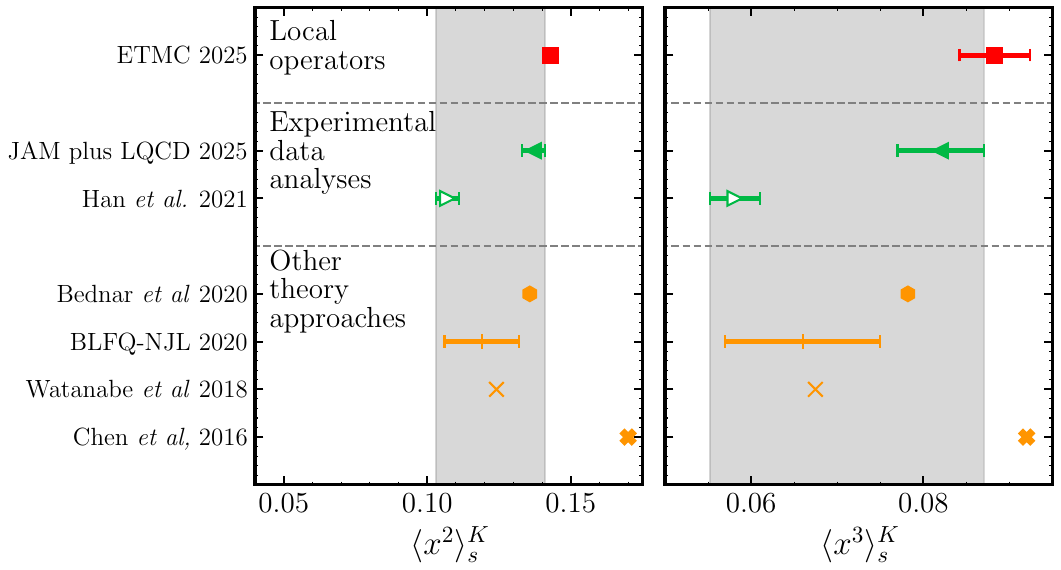}
\end{minipage}\vspace*{-0.5cm}
\end{figure}

\noindent
Matrix elements of first derivative operators ($n=2$) or of the energy and momentum tensor for quarks and gluons give access to the    unpolarized and tensor generalized form factors (GFFs), $A^{q,g}_{20}(Q^2),\, C^{q,g}_{20}(Q^2), \,B^{q,g}_{T20}(Q^2)$. Most of LQCD computations focus on the momentum fraction, $A^{q,g}_{20}(0)=\langle x\rangle_{q,g}$.  Recently,  ETMC used three $N_f=2+1+1$ TMF ensembles  simulated at approximately physical $m_\pi$  and three different lattice spacings to  compute the quark and gluon momentum fractions for the pion and kaon including all disconnected contributions. In Fig.~\ref{fig:xpion-kaon}, we show the continuum extrapolations and the contributions of the various quark flavors and gluons to the momentum fractions, as well as a comparison of recent LQCD and phenomenological data on these momentum fractions. All results are in the $\overline{\rm MS}$ scheme at 2~GeV. In general, there is agreement among the various results with some tension for the gluon $\langle x_g\rangle $ in the pion that maybe due to some systematics that need further study.
 The third and fourth  unpolarized moments of the pion and kaon  were  also recently computed by ETMC~\cite{Alexandrou:2021mmi}. These moments were also extracted from  direct computations of PDFs~\cite{Lin:2020ssv,Gao:2020ito}, as discussed in the next section.    In Fig.~\ref{fig:x23}, we show  recent LQCD results on the third and fourth Mellin moments compared to phenomenology and other theoretical determinations.  It is interesting to examine the ratio of moments for the u- and s-quarks in the kaon, where we include the disconnected contributions in $\langle x \rangle$ but not in the higher moments for which they are  expected to be small. The ratios read\vspace*{-0.4cm}
 
 \begin{equation}
   \frac{\langle x \rangle^{K}_{u}}{\langle x \rangle^{K}_{s}} = 0.810(11),\,\,
    \frac{\langle x^2 \rangle^{K}_{u}}{\langle x^2 \rangle^K_{s}} = 0.647(8),\,\,
    \frac{\langle x^3 \rangle^{K}_{u}}{\langle x^3 \rangle^K_{s}} = 0.632(67).
 \end{equation}\vspace*{-0.cm}

\begin{figure}[h!]
\begin{minipage}{0.45\linewidth}
\includegraphics[width=1.05\linewidth,height=0.5\linewidth]{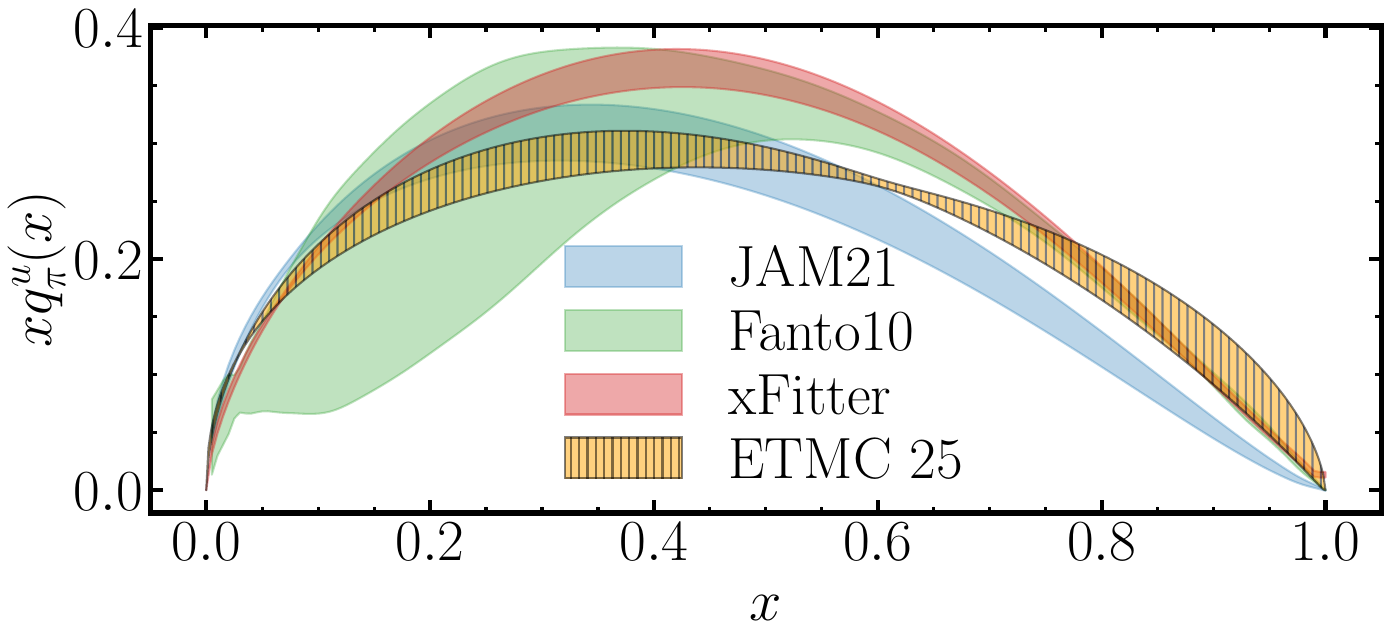}\\
\includegraphics[width=1.05\linewidth,height=0.5\linewidth]{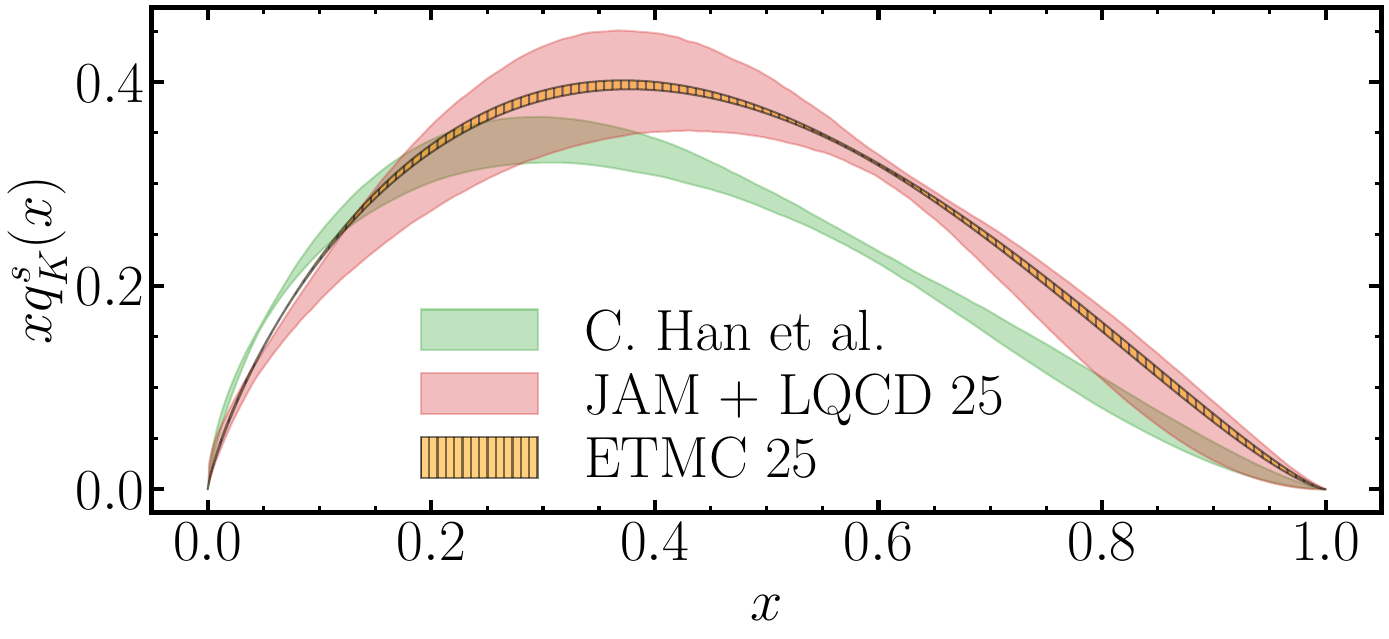}
\end{minipage}\hfill
\begin{minipage}{0.45\linewidth}
\hspace*{-0.7cm}\includegraphics[width=1.05\linewidth,height=0.5\linewidth]{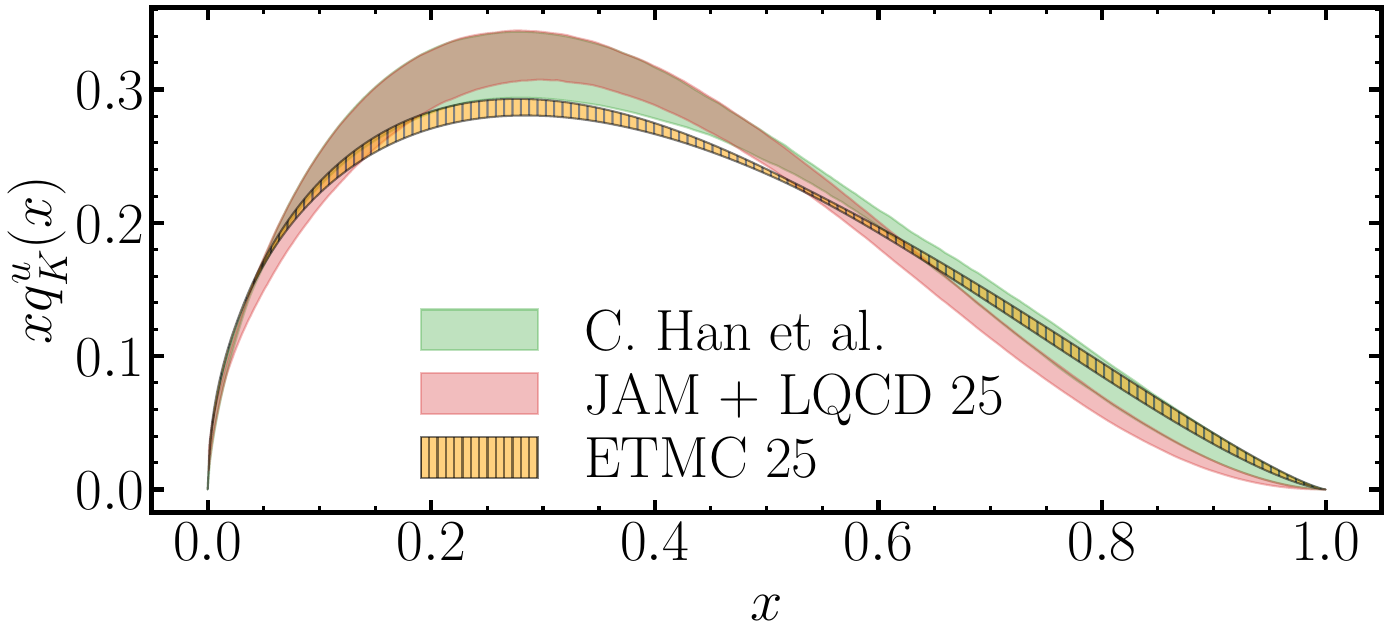}\\
\hspace*{-0.7cm}\includegraphics[width=1.05\linewidth,height=0.5\linewidth]{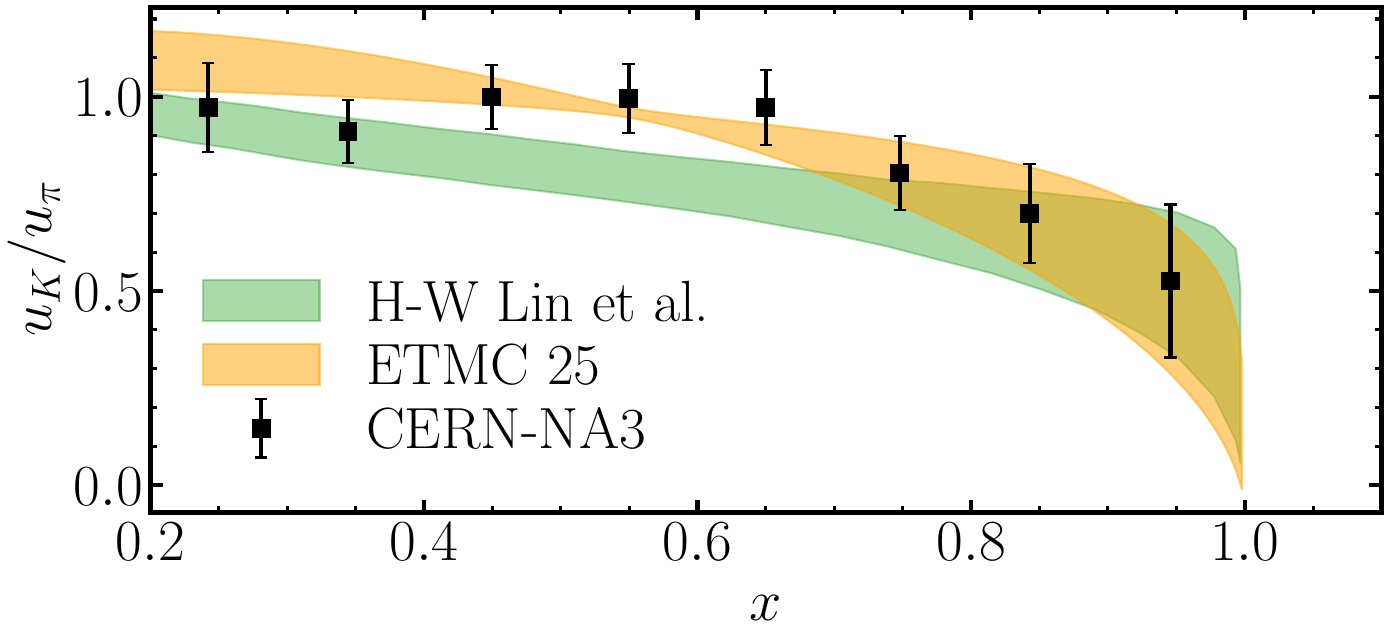}
\end{minipage}\vspace*{-0.3cm}
    \caption{Results on the unpolarized valence PDF for  $xq^\pi_u(x)$ (left top),  $xq^K_u(x)$ (right top) and $xq^K_s(x)$ (left bottom) constructed using  ETMC data (hatched orange band) compared to phenomenological results: for  $xq^\pi_u(x)$  from  JAM (blue band)~\cite{Barry:2021osv}, FANTO (green band)~\cite{Kotz:2025lio}, and xFitter (red band)~\cite{Novikov:2020snp}; for   $xq^K_u(x)$ and $xq^K_s$ from Ref.~\cite{Han:2020vjp} (green band) and JAM with LQCD input  (red band)~\cite{Barry:2025wjx}. The right bottom panel shows  ETMC results on $q_u^K(x)/q_u^\pi(x)$ (orange band) compared with the CERN-NA3 experimental data~\cite{Saclay-CERN-CollegedeFrance-EcolePoly-Orsay:1980fhh} and the direct PDF determination (green band)~\cite{Lin:2020ssv}. Results are in $\overline{\text{MS}}$ at 2~$\text{GeV}$ for $xq^\pi_u(x)$, at 5~$\text{GeV}$ for $xq^K_u(x)$ and $xq^K_s(x)$ and at 5.2~$\text{GeV}$ for $q_u^K(x)/q_u^\pi(x)$.}\label{fig:PDFs}\vspace*{-0.3cm}
\end{figure}

 \noindent
 The ratios, computed using one $N_f=2+1+1$ TMF ensemble with $m_\pi=140$~MeV and $a=0.08$~fm, point to the  strange quark PDF having its support at larger values of $x$ than the up quark PDF  in the kaon. They also indicate that
SU(3) symmetry breaking is more pronounced for higher moments. This conclusion holds  also if we use only disconnected for $\langle x \rangle$, the value then being 0.715(5).  Having moments up to the fourth, one can reconstruct the PDFs using the standard parametrization 
$q=Nx^{\alpha}(1-x)^\beta$, where $N$ is determined by normalizing the first moment of the distribution. We show in Fig.~\ref{fig:PDFs} the resulting distributions  of the pion and kaon using the ETMC Mellin moments only for the connected contributions compared to recent phenomenological determinations. We note that for the kaon the only phenomenological analysis~\cite{Han:2020vjp} is based on old data from the NA3 and NA10 experiments at CERN (NA3 and NA10), Fermilab (E615) and HERA (ZEUS and
H1),  while a recent analysis by  the JAM collaboration~\cite{Barry:2025wjx} used LQCD  from Refs.~\cite{ExtendedTwistedMass:2024kjf,Alexandrou:2021mmi}. The  ratio of $q_u^K(x)/q_u^\pi(x)$ is  compared with the CERN-NA3 experimental data~\cite{Saclay-CERN-CollegedeFrance-EcolePoly-Orsay:1980fhh} and from the direct determination of the PDFs using the quasi-PDF approach and  extrapolated to the physical pion mass and the continuum~\cite{Lin:2020ssv}. The development of  these complementary approaches to extract PDFs from LQCD provides a valuable path for cross-checks on the Mellin moments.  Quark and gluon  PDFs for the pion and kaon will be investigated  at the EIC but also at the AMBER experiment at CERN and precise LQCD determinations can provide crucial inputs.
 
While computing  Mellin moments up to the fourth is  feasible, it requires increased statistics  as the order of the moment increases. Furthermore, one can not go beyond the fourth moment due to operator mixing. 
 New ideas may enable computation of higher order moments using local operators, e.g. using 
Wilson flow~\cite{Shindler:2023xpd, Francis:2025pgf}  or the heavy-quark operator product expansion (HOPE) method~\cite{Detmold:2021uru, Detmold:2025lyb}.
A proof of concept  is presented in this conference for the pion Mellin moments  for  both of these approaches. Although the computations were done  for gauge ensembles of heavy pion mass, the results are promising, see Refs.~\cite{Francis:2025rya, Chang2026}. 

\subsection{Nucleon structure}
\begin{figure}[h!]
\begin{minipage}{0.33\linewidth}
\hspace*{-0.6cm}\includegraphics[width=1.05\linewidth]{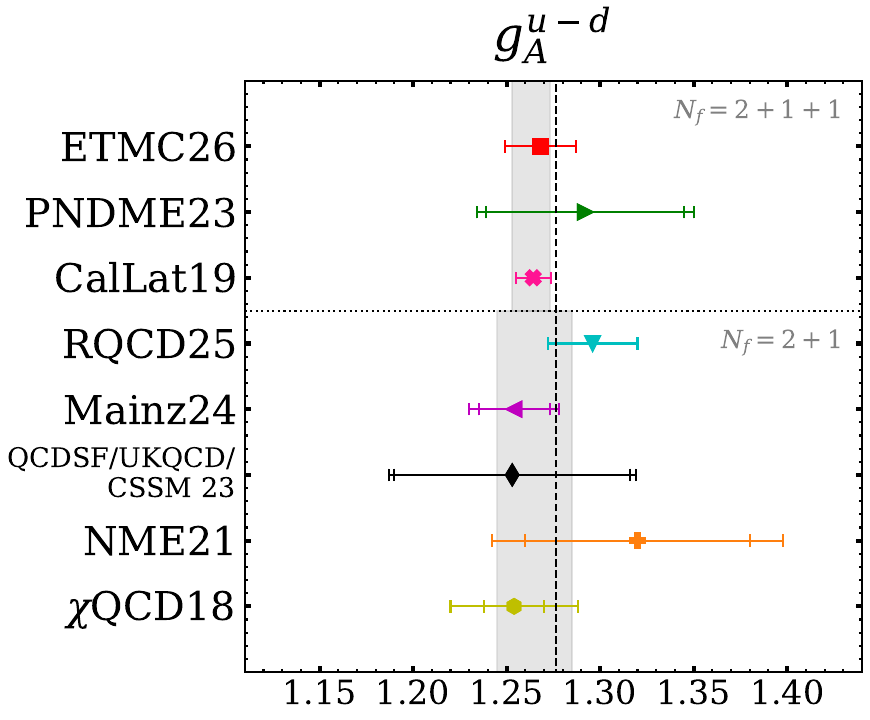}
\end{minipage}\hfill
\begin{minipage}{0.33\linewidth}
\hspace*{-0.3cm}\includegraphics[width=1.05\linewidth]{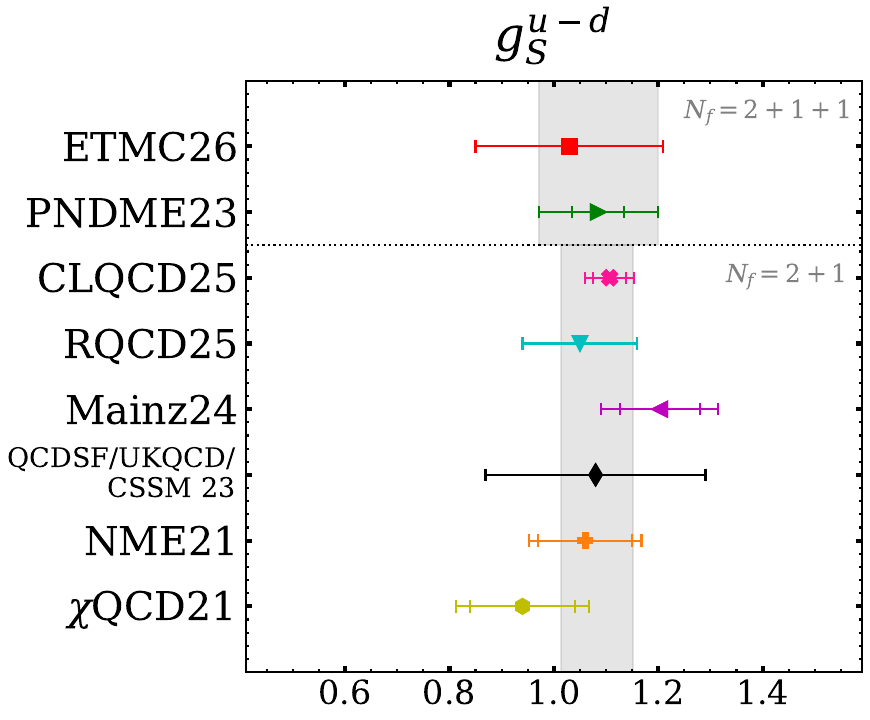}
\end{minipage}\hfill
\begin{minipage}{0.33\linewidth}
\hspace*{0.cm}\includegraphics[width=1.05\linewidth]{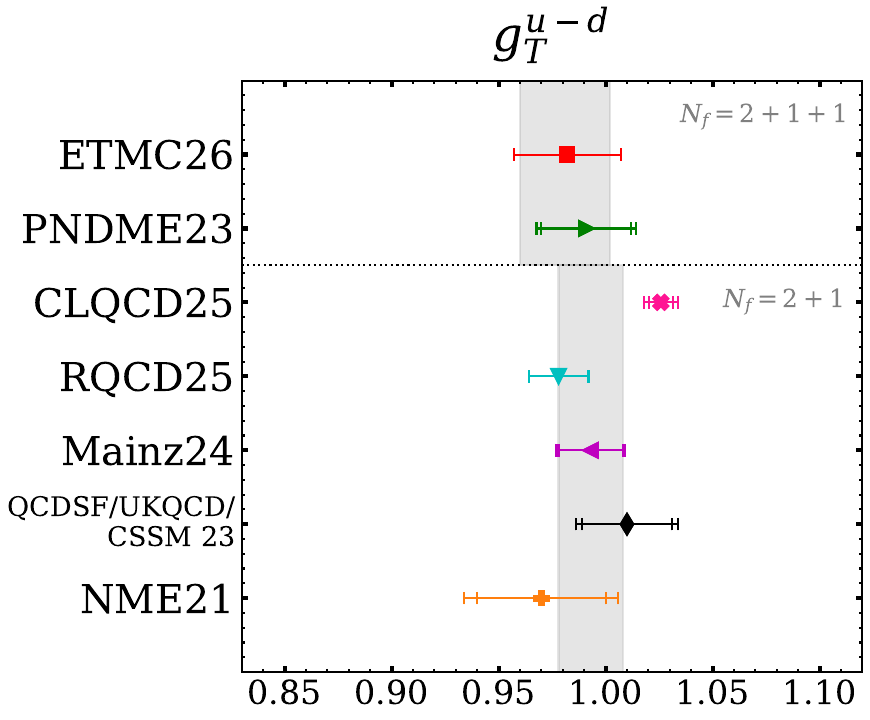}
\end{minipage}\vspace*{-0.3cm}
    \caption{Recent result on the nucleon isovector charges. The gray bands show the  FLAG 2024 average of  then published data, see Ref.~\cite{FlavourLatticeAveragingGroupFLAG:2024oxs} and references within. The dotted line is the experimental value of $g_A^{u-d}$.}\label{fig:N isocharges}
\end{figure}

\begin{figure}[h!]
\begin{minipage}{0.44\linewidth}\vspace*{-0.5cm}
\includegraphics[width=\linewidth]{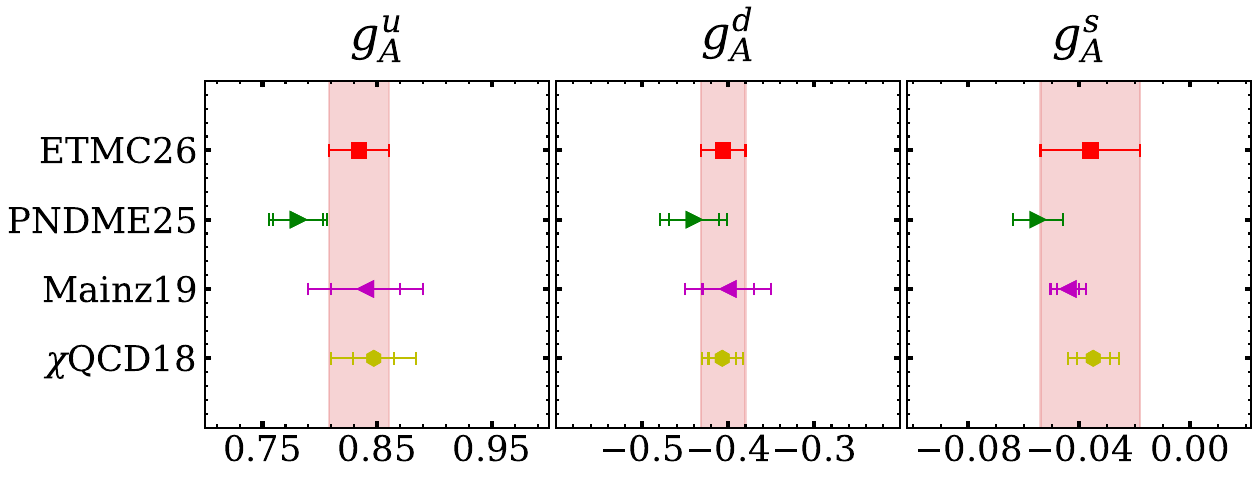}\\
\includegraphics[width=\linewidth]{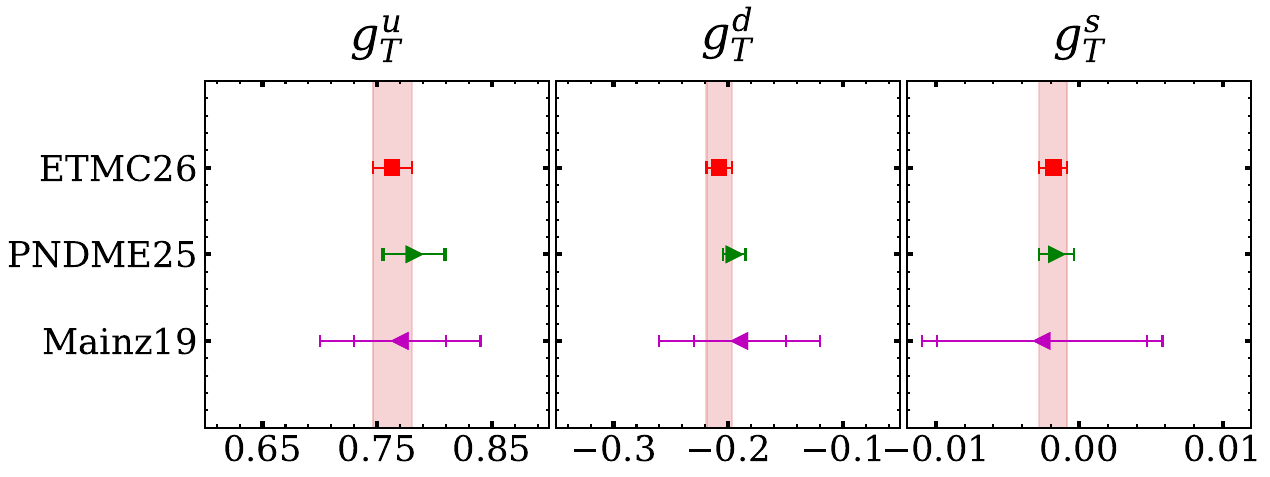}
\end{minipage}\hfill
\begin{minipage}{0.55\linewidth}\
\includegraphics[width=\linewidth]{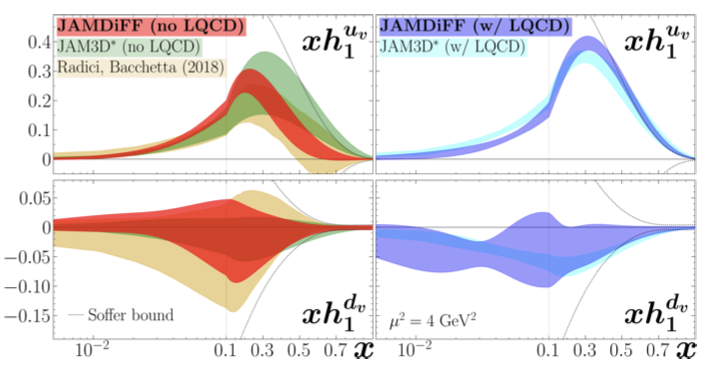}\
\end{minipage}
    \caption{Results on the axial and tensor charges for the u-, d- and s-quarks (left panel) from ETMC~\cite{Alexandrou:2024ozj}, PNDME~\cite{Park:2020axe}, Mainz~\cite{Djukanovic:2019gvi} and $\chi$QCD~\cite{Liang:2018pis}. The right panel shows results on the u- and d-quark transversity without and with LQCD input on $g^{u,d}_T$ (figure taken from Ref.~\cite{Cocuzza:2023oam}).}\label{fig:N flavor-charges}\vspace*{-0.3cm}
\end{figure}

Nucleon matrix elements for $n=1$ and $n=2$ yield the nucleon charges and second Mellin moments of the unpolarized, helicity and transversity PDFs. Allowing for momentum transfer, one obtains, for $n=1$ the electromagnetic, axial and tensor form factors, while for $n=2$ the corresponding moments of  GPDs or GFFs. The nucleon isovector axial $g^{u-d}_A$ and tensor $g_T^{u-d}$ charges are well-studied by many collaborations. Recent results are obtained by:  i) RQCD, using 47 CLS ensembles at six values of the lattice spacing, $a\in (0.038-0.098)$~fm, $m_{\pi}\in (480-130)$~MeV and multiple lattice sizes~\cite{Wang:2025nsd}; ii) ETMC using 4 ensembles at four  $a\in (0.08-0.05)$~fm and  $m_\pi\approx 140$~MeV; and iii)
CLQCD using 16 Clover ensembles at four $a\in(0.105-0.052)$~fm, $m_\pi~(340-134)$~MeV and  multiple lattice sizes. In Fig.~\ref{fig:N isocharges},  we compare them   with other  recent results and the FLAG2024 average~\cite{FlavourLatticeAveragingGroupFLAG:2024oxs}. As can be seen,  there is an overall agreement. RQCD has also computed these charges for other octet baryons~\cite{Bali:2023sdi} the values of which are less well-known. On the other hand, results on these charges for each quark flavor are limited since they require the computation of disconnected contributions. In Fig.~\ref{fig:N flavor-charges}, we show a comparison on the axial and tensor charges for each flavor. With the exception of ETMC where the continuum limit was taken  at approximately the physical value of $m_\pi$, all other computations used a combined chiral and continuum extrapolation. The LQCD values for the tensor charges  are more precise than those from phenomenology and when used as input, they improve the extraction of the transversity  PDFs, as shown in Fig.~\ref{fig:N flavor-charges}. This is an example of the crucial input that LQCD can provide for EIC but also for   the SoLID experiment at JLab that focuses on parity violation.\vspace*{-0.3cm}
\begin{figure}[h!]
\begin{minipage}{0.6\linewidth}
\hspace*{-0.5cm}\includegraphics[width=1.1\linewidth]{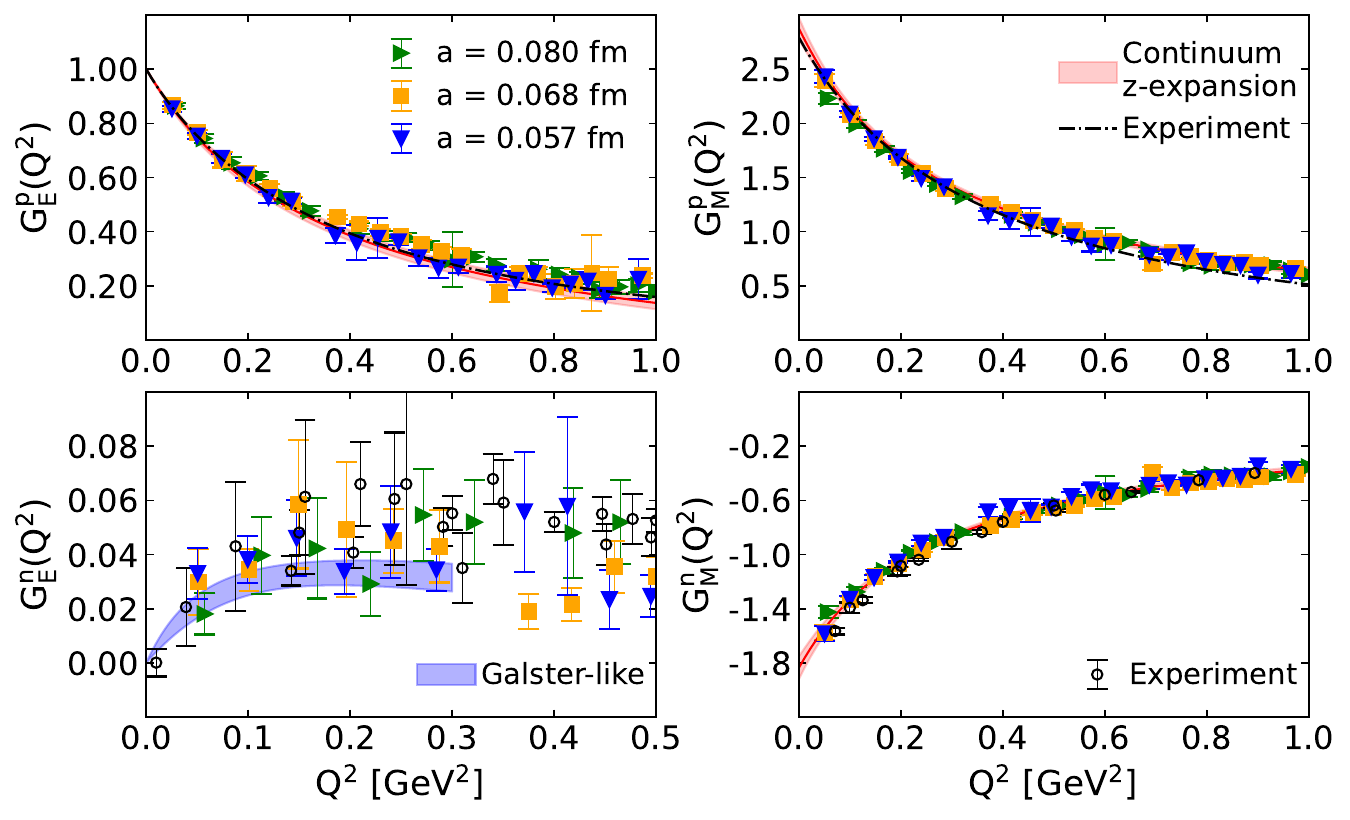}\\
\end{minipage}\hfill
 \begin{minipage}{0.39\linewidth}\vspace*{-0.5cm}
\includegraphics[width=\linewidth, height=\linewidth]{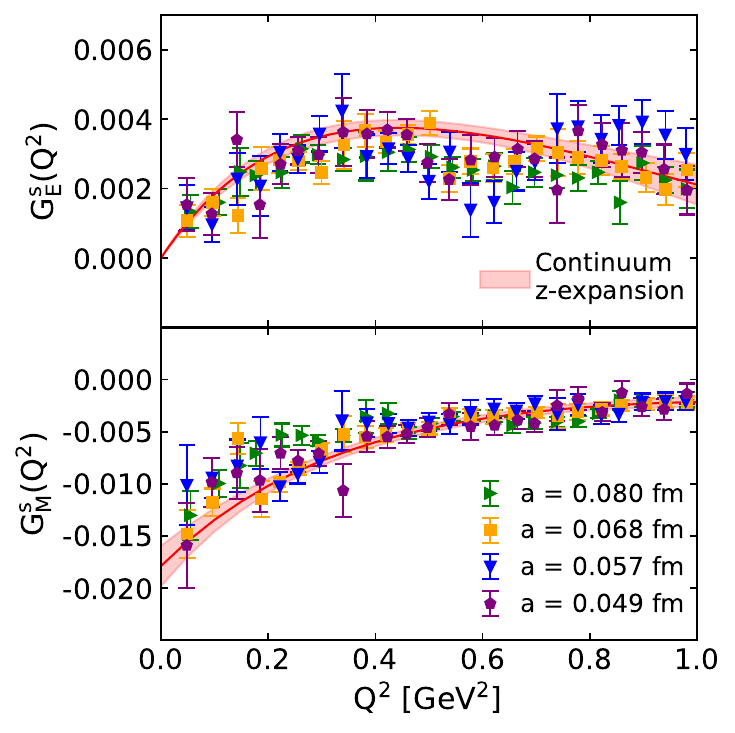}\\\vspace*{-0.3cm}
\end{minipage}\vspace*{-0.7cm}
\caption{ETMC results on the EM form factors. Left: proton (top) and neutron (bottom) electric; Middle: proton (top) and neutron (bottom) magnetic~\cite{Alexandrou:2025vto}; Right: strange electric (top) and magnetic  (bottom)~\cite{Prasad:2025pos,Alexandrou:2026xcl,Alexandrou:2026ait}. Red bands show fits to the z-expansion while the blue one is a fit to the Galster form. Bottom:  LQCD results on the proton and neutron radii and magnetic moments (four plots on the left) and proton  strange radii and magnetic moments (three plots on the right)~\cite{Prasad:2025pos}.  }\label{fig:EM}\vspace*{-0.7cm}
\end{figure}
\begin{figure}[h!]
\begin{minipage}{0.6\linewidth}
\hspace*{-0.3cm}\includegraphics[width=1.05\linewidth]{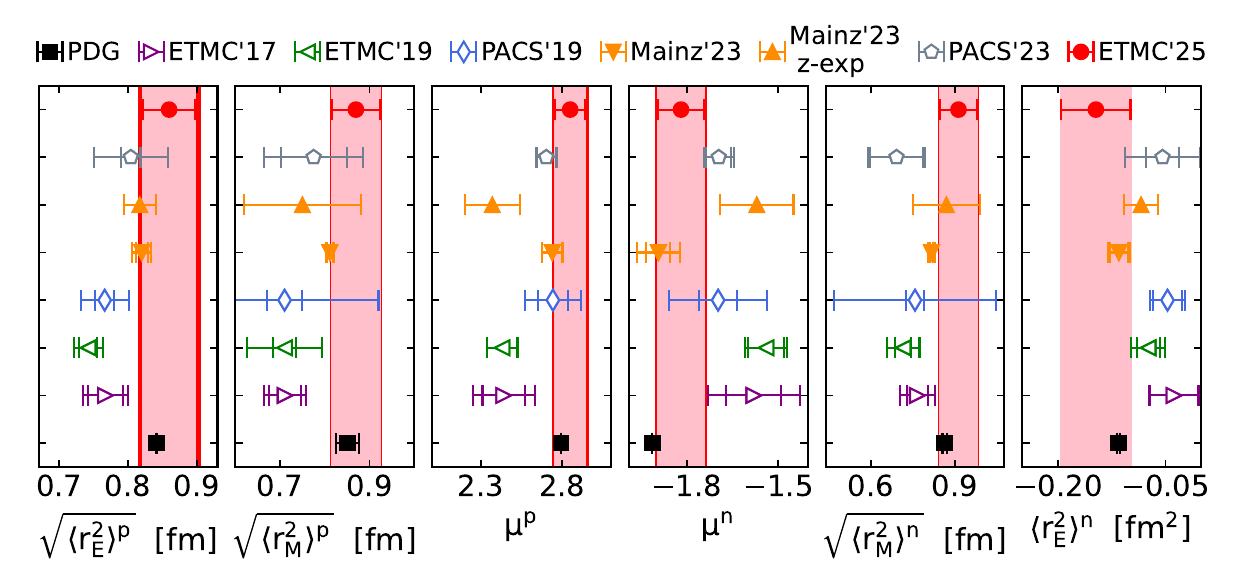}
\end{minipage}\hfill
 \begin{minipage}{0.39\linewidth}\vspace*{-0.5cm}
\includegraphics[width=1.05\linewidth]{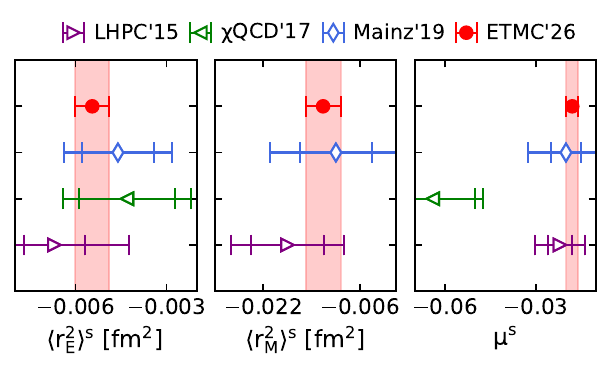}
\end{minipage}\vspace*{-0.3cm}
\caption{LQCD results on EM radii and magnetic moments for the proton and neutron (6 panels on the left) and strange radii and magnetic moment (3 panels on the right)~\cite{Prasad:2025pos}.  }\label{fig:EM radii}\vspace*{-0.7cm}
\end{figure}

Nucleon electromagnetic (EM) form factors have been computed over many years. However, it is only recently that LQCD results have included disconnected contributions and took into account lattice systematics. Recent examples are results by the  Mainz group, which obtained unprecedented accuracy after modeling the chiral and continuum extrapolations and accounting for finite size effects using CLS ensembles with one having $m_\pi=130$~MeV~\cite{Djukanovic:2023beb, Djukanovic:2023jag} and from ETMC, which   used three TMF ensembles with $m_\pi\sim140$~MeV and took the continuum limit without any chiral extrapolation~\cite{Alexandrou:2025vto}. The
 ETMC results on the proton and neutron electric and magnetic form factors are shown  in Fig.~\ref{fig:EM} and they are in agreement with experimental results. In the case of the neutron electric form factor, LQCD results are more precise than experimental ones. The strange electromagnetic form factors, which are not well measured, can also be determined  precisely from LQCD. We show recent results from ETMC using four TMF ensembles with $m_\pi\approx 140$~MeV extrapolated to the continuum limit. In Fig.~\ref{fig:EM radii}, we  show a comparison of  recent LQCD results on the radii and magnetic moments where recent results by Mainz, ETMC and PACS are in agreement with the experimental values.  We also include a comparison among  LQCD determinations of the nucleon strange radii and magnetic moment, which  provide valuable input for  experiments  on parity violation, such as  Q-weak, G0 and  HAPPEX  at JLab.

Having reproduced the EM proton form factors that provide a benchmark for the LQCD methodology, one  applies a similar analysis  to determine the axial form factors. These   are important for weak interactions, parity
violation and neutrino scattering experiments, such as  NO$\nu$A, MINER$\nu$A and MicroBooNE at Fermilab, T2K  at KEK  and  the upcoming DUNE experiment. Recent results are produced by RQCD~\cite{RQCD:2019jai}, NME~\cite{Park:2021ypf}, the Mainz group~\cite{Djukanovic:2022wru}, ETMC~\cite{Alexandrou:2023qbg} and PNDME~\cite{Jang:2023zts}, all of which are in agreement and already have impacted the analysis of MINER$\nu$A data.
\begin{figure}[h!]
\includegraphics[width=\linewidth]{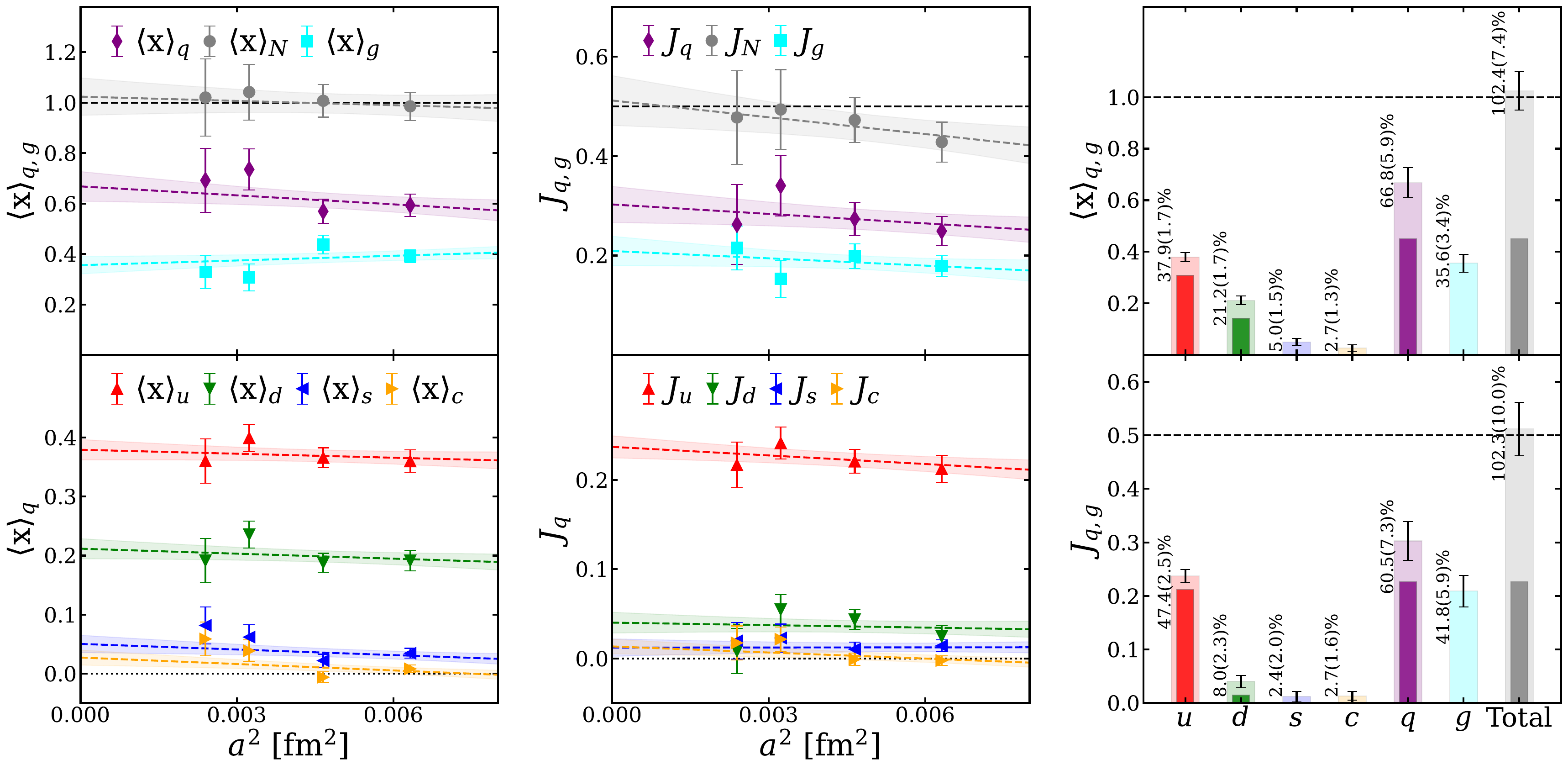}\vspace*{-0.2cm}
%
\caption{Preliminary continuum extrapolated results at physical pion mass  by ETMC in the $\overline{\rm MS}$ at 2 GeV. Left two panels show the continuum extrapolation of the momentum fraction and angular momentum for quarks and gluons; Right two panels show the contribution of quarks and gluons to the momentum fraction and spin of the nucleon.   }\vspace*{-0.6cm}\label{fig:spin sum}
\end{figure}

 \begin{figure}[h!]
 \begin{minipage}{0.45\linewidth}
\includegraphics[width=\linewidth]{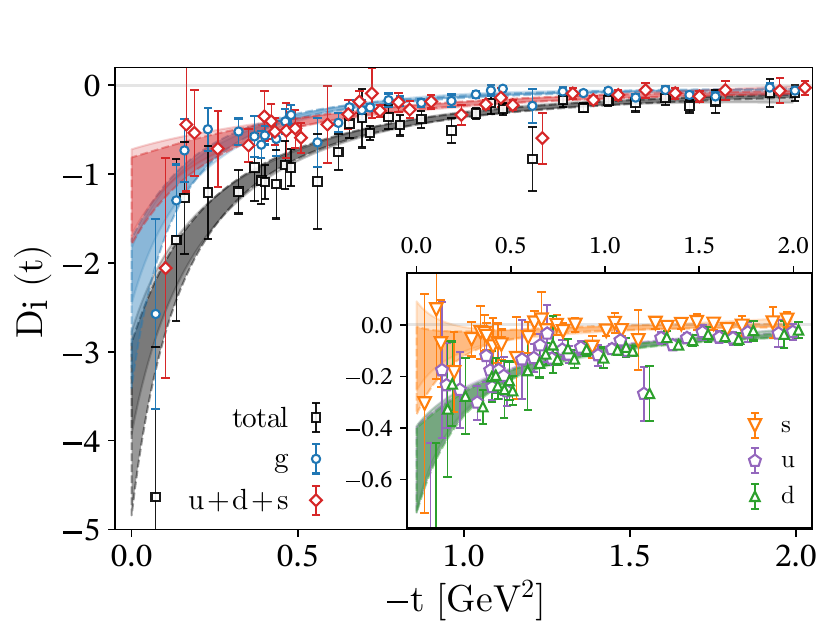}\\\vspace*{0.5cm}
\includegraphics[width=\linewidth]{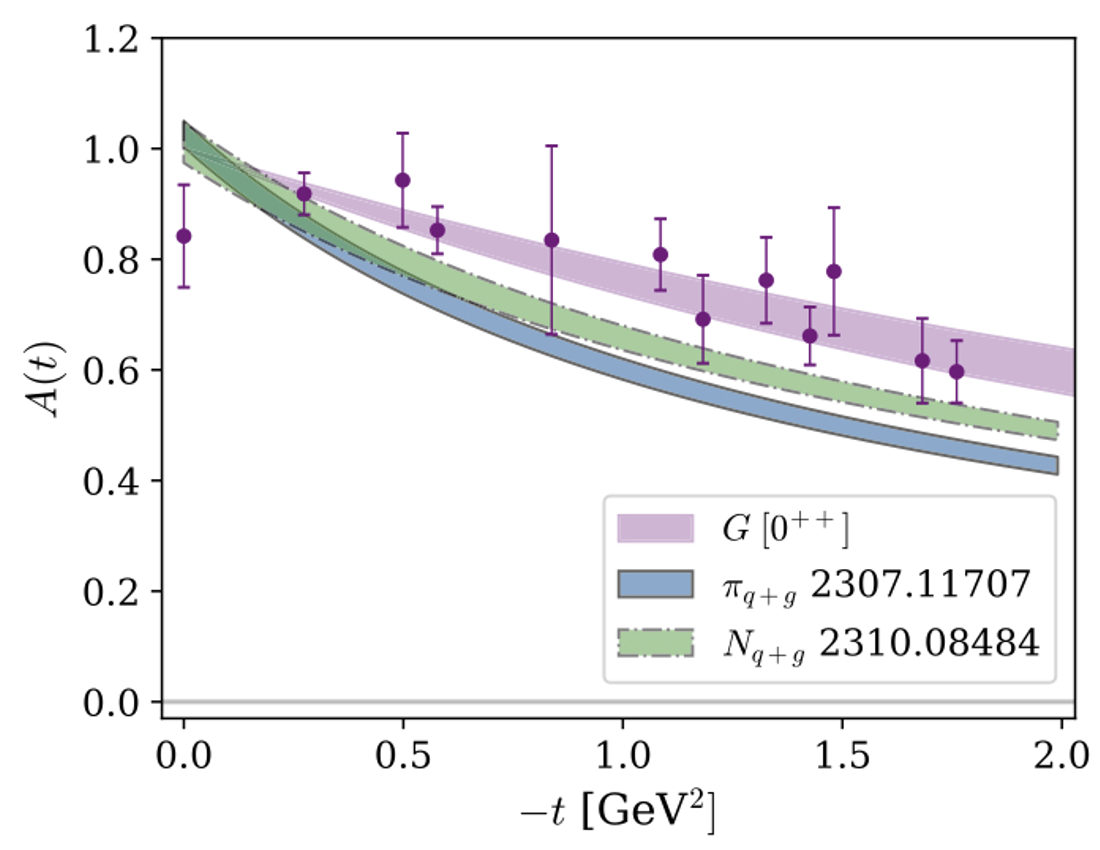}
\end{minipage}\hfill
\begin{minipage}{0.45\linewidth}
\includegraphics[width=\linewidth,height=0.7\linewidth]{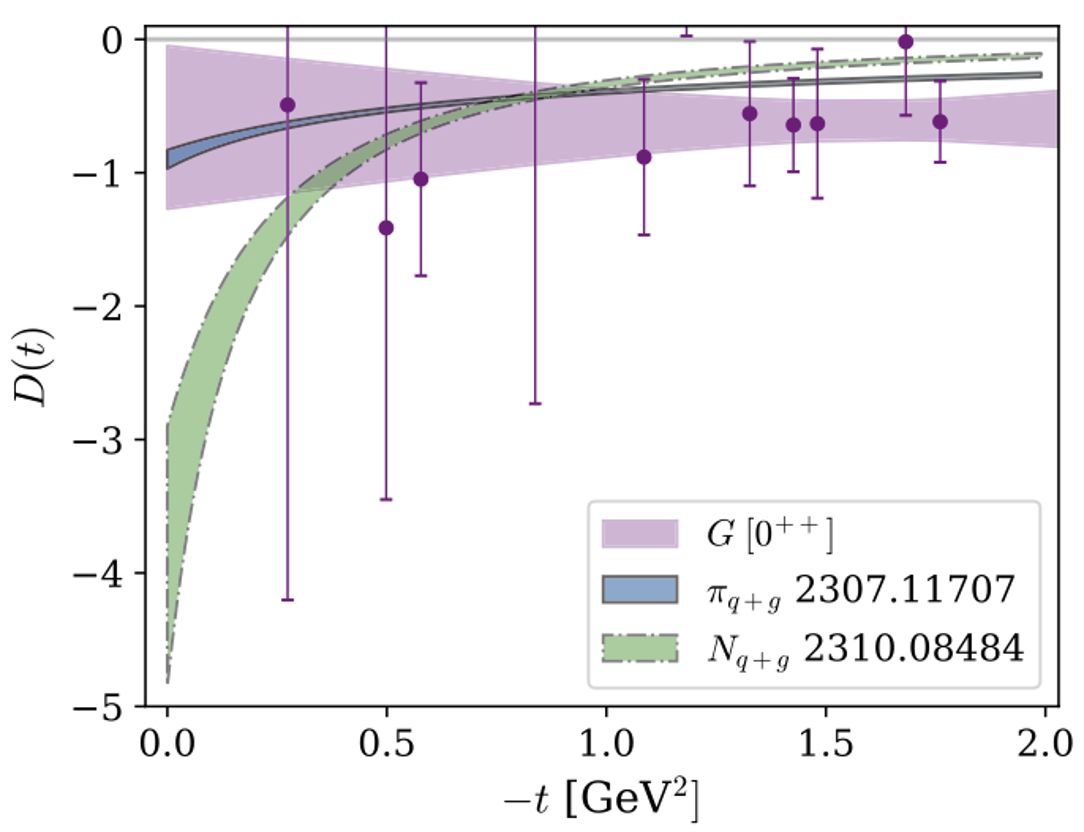}\\
\includegraphics[width=\linewidth]{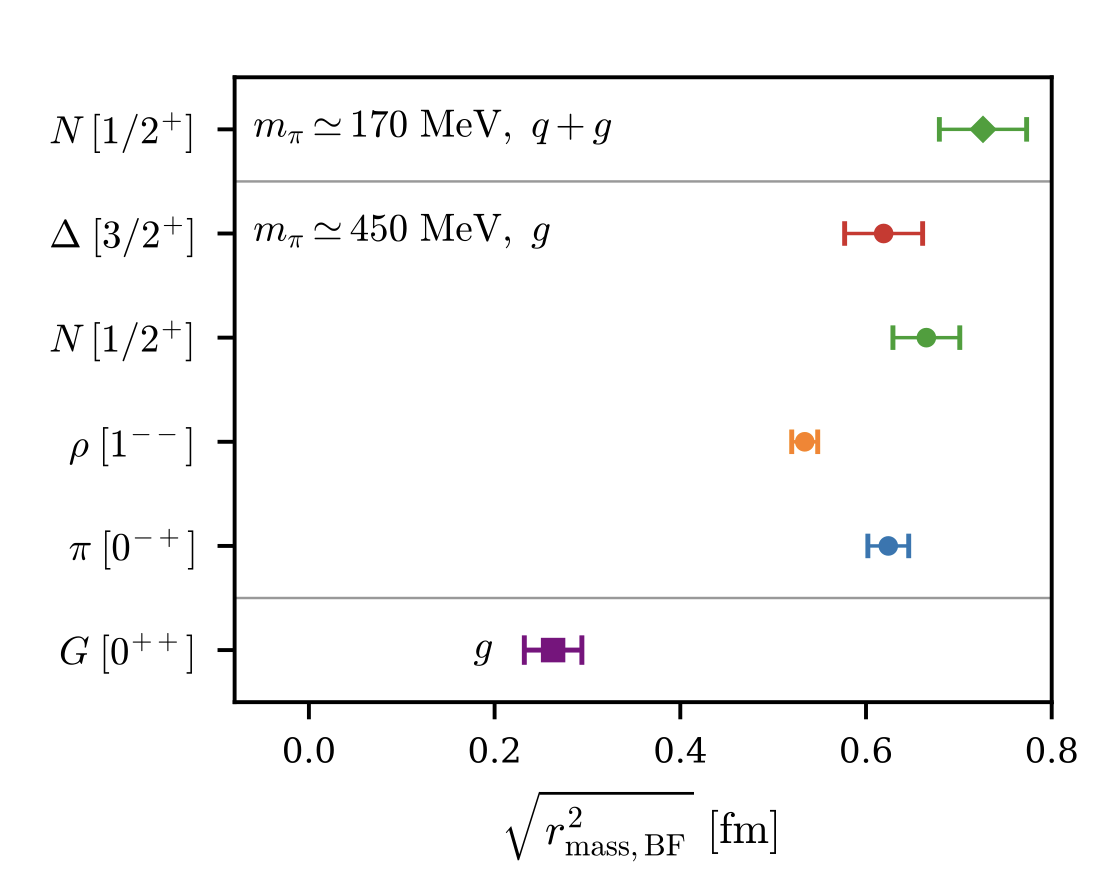}
\end{minipage}\hfill
\vspace*{-0.7cm}
\caption{Left: The nucleon $D(t)$-term    vs $-t=Q^2$~ (left top)~\cite{Hackett:2023rif}; GFF $A_{20}(t)$ (right top) and $D(t)$-term (left bottom)  and  r.m.s mass radius (right bottom) for a scalar glueball  compared to the other hadrons~\cite{Abbott:2025irb}. }\label{fig:MIT GFFs}\vspace*{-0.5cm}
\end{figure}

Turning to the second nucleon moments, the isovector second Mellin moments  have been computed by a number of collaborations over the years.  A very recent analysis by RQCD~\cite{Collins2025} used  47 CLS ensembles and performed a chiral and continuum extrapolation taking into account finite volume effects. Their values for the  unpolarized and transversity moments are in agreement with the FLAG2024 average~\cite{FlavourLatticeAveragingGroupFLAG:2024oxs}. Their value for the helicity moment, $\langle x\rangle_{\Delta u-\Delta d}$, is larger pointing to the need for further investigation. 
For the isoscalar, strange and charm Mellin moments there are only a few computations. The ETMC collaboration performed a flavor decomposition of the unpolarized moments using  four TMF ensembles with  approximately physical pion mass and $a \in(0.08-0.049)$~fm  producing results for the quark and gluon  momentum fractions in the continuum limit, as shown in Fig.~\ref{fig:spin sum}. Allowing for momentum transfer, gives access to the GFFs $A_{20}(Q^2), B_{20}(Q^2)$ and $C_{20}(Q^2)$. $B_{20}(0)$ enters the determination of the nucleon spin: $J_N = \frac{1}{2}\sum_q [A^{q,g}_{20}(0) + B^{q,g}_{20}(0)]=\frac{1}{2}\sum_q[\Delta\Sigma_q+ L_q]+J_g$. Preliminary results on the continuum extrapolation and flavor decomposition of the nucleon total spin are shown 
in Fig.~\ref{fig:spin sum}~\cite{Alexandrou2026}. Gluons contribute about 45\% to the total momentum and spin sums with   the up-quark being the other main contributor. The MIT group recently analyzed all three unpolarized GFFs and confirmed the momentum and spin sums using an ensemble of clover-improved fermions with $m_\pi=170$~MeV and $a=0.09$~fm~\cite{Hackett:2023rif}. They find consistent results with ETMC results with about equal  contributions of quarks and gluons to the momentum fraction and spin. A notable outcome is the computation of the GFF $C_{20}(Q^2)$ or $D$-term  related to the pressure and shear forces in the nucleon that are of experimental interest~\cite{Meziani:2025dwu}. Their results on the $D$-term are shown in Fig.~\ref{fig:MIT GFFs}. 
The MIT group also computed the GFFs $A_{20}(Q^2)$ and $D$-term using an $SU(3)$ ensemble for a scalar glueball and extracted the r.m.s. radius of the energy density in the Breit frame. They showed that it is smaller than the corresponding radius of the nucleon providing a distinct signature for the scalar glueball~\cite{Abbott:2025irb}. We show their results in Fig.~\ref{fig:MIT GFFs}.

A central goal of the EIC program
is to better study  multi-parton correlations by investigating higher twist PDFs and TMDs. A prominent example is the so-called nucleon third moment $d_2$ of the twist-3
contribution to the helicity structure function $g_2$ and a key quantity for studying quark-gluon correlations.  In the impact parameter zero limit, $b_\perp\rightarrow 0$, $g_2$ is connected to the Sivers function, one of the important distributions to be measured at JLab and EIC. The only LQCD computations using matrix elements of local operators are by QCDSF 20 years ago~\cite{Gockeler:2005vw}, updated recently in Ref.~\cite{Crawford:2024wzx} using three ensembles of $m_\pi\approx 400$~MeV, and   by RQCD~\cite{Burger:2021knd}.   RQCD analyzed CLS gauge ensembles with six different lattice spacings $a\in (0.039-0.098)$~fm and $220 <m_\pi <420$ MeV to perform chiral and continuum extrapolations. Their analysis is also one of the very few examples where the third moment of the nucleon helicity PDF is  computed. For the third moment of the helicity twist-2 structure function $g_1$, $2\int_0^1 dx x^2 g_1(x,Q^2)$, in the $\overline{\rm MS}$ at 2~GeV, they find 
$\langle x^2\rangle_{\Delta p} =0.035(3)(8),\,\,\langle x^2\rangle_{\Delta n}=0.0034(17)(41)$ for the proton and neutron, respectively. Their result for the twist-3 third moment 
$d_2(\mu)=4\int_0^1 dx x^2\left[g_1(x,Q^2)+\frac{3}{2}g_2(x,Q^2)\right]$ is consistent  with  experimental analyses.
In this conference, we have seen further progress in the computation of  the third nucleon moment for the unpolarized, helicity and transversity twist-2 PDFs. They were extracted from the analysis of two ensembles with  $m_\pi$ close to its physical value and $a=0.116$ and $a=0.093$~fm. All three moments were found to be of comparable magnitude~\cite{Taggi2026}.\vspace*{-0.4cm}

 \section{Direct determination of pion, kaon and nucleon parton distributions}\vspace*{-0.3cm}

\begin{wrapfigure}[10]{L}{0.4\linewidth}\vspace*{-0.5cm}
    \includegraphics[width=\linewidth]{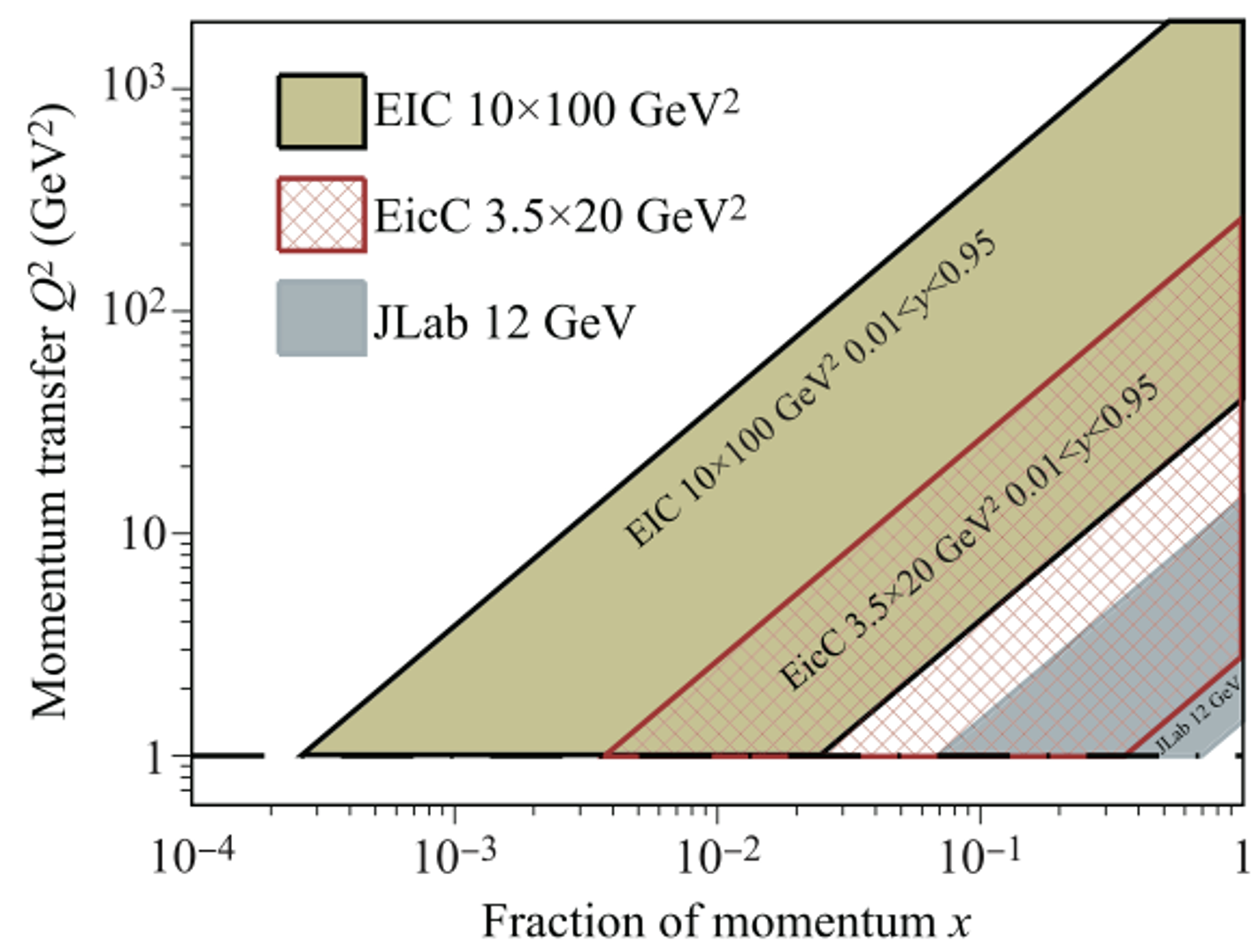}\vspace{-0.3cm}
    \caption{Low-$x$ region covered by EIC.}
    \label{fig:x region}
\end{wrapfigure}
\noindent
In the last ten years, there is a lot of progress in the direct computation of PDFs and GPDs but also TMDs, all of which are quantities targeted by the EIC scientific program. 
PDFs will be accessible in the very low-$x$ region, as shown in Fig.~\ref{fig:x region}, enabling a more accurate extraction  of Mellin moments.  Higher twist PDFs and moments probing multi-parton correlations, e.g. nucleon twist-3 PDFs,  such as the scalar $e(x)$, transversity $g_T(x)$ and the $d_2$ term, will be measured. In what follows we summarize recent progress in the LQCD computation of PDFs and GPDs. 

\noindent
Most of the results are produced using  either   the quasi-distribution approach within the framework of the large momentum effective theory (LaMET)~\cite{Ji:2013dva,Ji:2020ect} or the  pseudo-distribution approach~\cite{Radyushkin:2017cyf,Radyushkin:2019mye}  based on the short-distance factorization (SDF). For reviews, see e.g. Refs~\cite{Cichy:2019, Zhao:2019,Radyushkin:2020, Ji:2021, Constantinou:2021EPJA, Constantinou:2021PPNP, Cichy:2022PoS}. Quasi- and pseudo-distributions start from   the same matrix element and provide complementary information allowing cross-checks of results~\cite{Ji:2022ezo}.  To extract them one computes space-like matrix elements of nonlocal operators, $M_\Gamma(z,P_3) =\langle P_3\vert\,\overline{\psi}(0)\, \Gamma W(0,z)\, \psi(z)|\,P_3\rangle$, with boosted hadron states $|P_3\rangle$, where $W$ is a straight Wilson line from $0$ to $z$ and $P_3$ is the boost along the $z$-axis. In the quasi-distribution approach one renormalizes and  takes the Fourier transform to obtain the quasi-distribution $ \tilde{F}_\Gamma(x,P_3,\mu) =2P_3
  \int_{-\infty}^{\infty}\frac{dz}{4\pi}e^{-ixP_3z}\, M(z,P_3) |_{\mu}$. The light-cone PDF is then constructed within  LaMET via \vspace*{-0.3cm} %
  \begin{equation}
     \tilde{F}_\Gamma(x,P_3,\mu)=\int_{-1}^1
\frac{dy}{|y|} \, C\left(\frac{x}{y},\frac{\mu}{yP_3}\right)\, {F}_\Gamma(y,\mu) +{\cal{O}}\left(\frac{\Lambda_{QCD}^2}{x^2P_3^2},\frac{\Lambda_{QCD}^2}{(1-x)^2P_3^2}\right) ,\vspace*{-0.3cm}
  \end{equation}
 using a perturbatively computed matching kernel $C\left(\frac{x}{y},\frac{\mu}{yP_3}\right)$. Due to higher twist effects, this procedure is expected to yield reliable results for $x \in (x_{\rm min}-x_{\rm max}) \approx (0.2-0.8)$.  In the pseudo-distribution approach one computes the ratio $\tilde{\cal{M}}_\Gamma(\nu,z^2)=\frac{\bar{M}_\Gamma(\nu,z^2)}{\bar{M}_\Gamma(0,z^2)}$, where $\bar{M}$ is the same space-like matrix element computed in the quasi-distribution approach but written in term of the Ioffe time $\nu=zP$. One then performs a matching  in coordinate space via short distance factorization, 
 \begin{equation}
   \tilde{\cal{M}}_\Gamma(z^2,\nu)=\int_{-1}^1\,dy\,C(y){\cal{M}}_\Gamma(y\nu,\mu) +{\cal{O}}(z^2\Lambda^2_{\rm QCD}),  
 \end{equation} 
 where $C(y)$ is a perturbative kernel and the neglected higher twist terms limit the maximum value of  $\nu_{\rm max}$,  allowing the computation of moments only up to $x^{\nu_{\rm max}}$. The    Ioffe distribution ${\cal{M}}_\Gamma(y\nu,\mu) $ is related to the light-cone PDF via ${\cal{M}}_\Gamma(\nu,\mu) =\int_{-1}^1\,d\nu\,e^{ix\nu}F_\Gamma(x,\mu)$.
 The light-cone PDFs can be successfully extracted within the quasi- and pseudo-distribution approaches with each method having its own systematics but also yielding complementary information. These approaches were compared for the case of the pion~\cite{Gao:2020ito} and for the pion and kaon in Ref.~\cite{Miller:2025wgr} in each case using the same LQCD correlators. 

\subsection{Pion, kaon and nucleon PDFs}

\begin{wrapfigure}[15]{L}{0.7\linewidth}\vspace*{-0.6cm}
\begin{minipage}{0.5\linewidth}
\includegraphics[width=\linewidth]{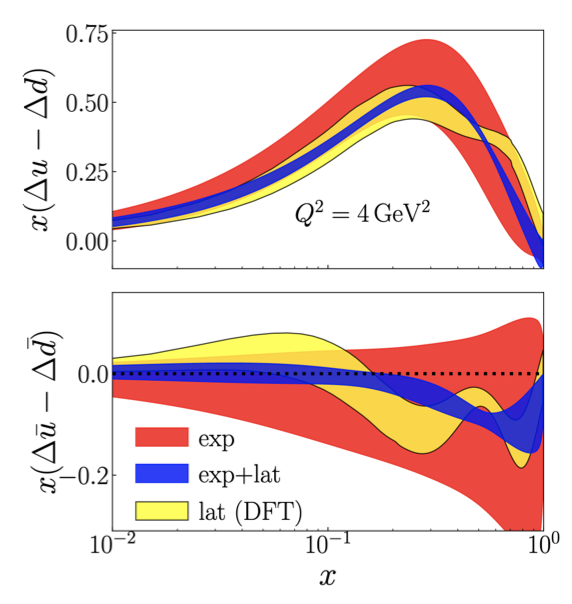}
\end{minipage}
\begin{minipage}{0.5\linewidth}
\includegraphics[width=\linewidth]{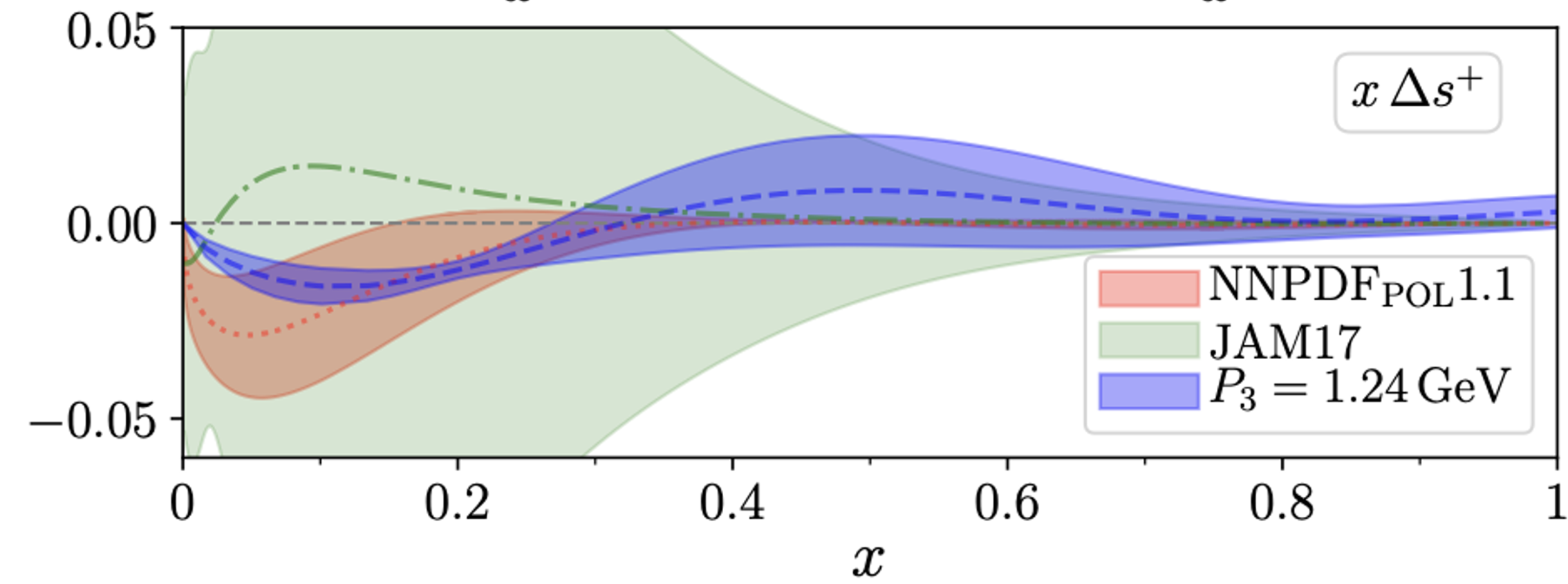}\\
\includegraphics[width=\linewidth]{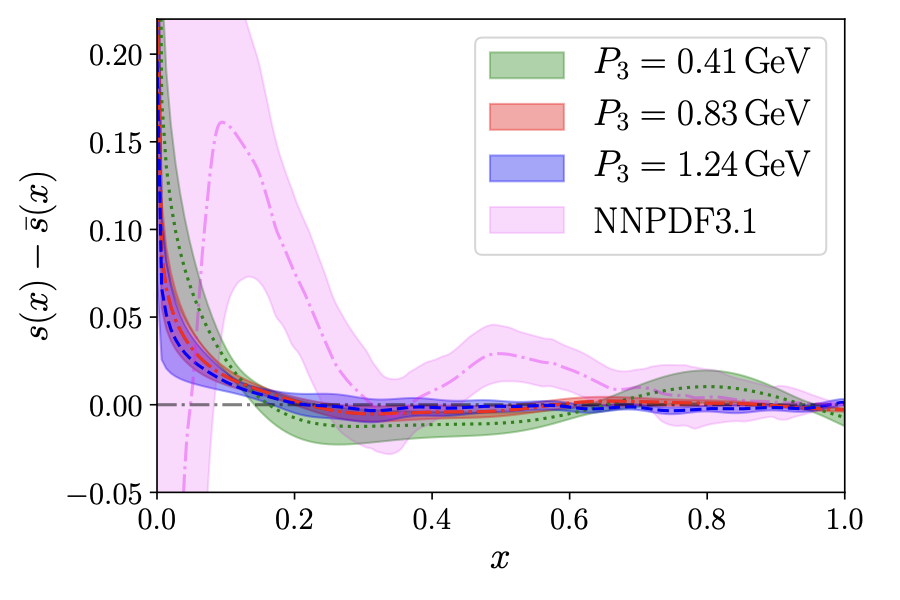}
\end{minipage}
\vspace*{-0.5cm}
\caption{Left: Results from an analysis of experimental data by JAM with (blue band) and without (red band) using LQCD input. Figure from Ref.\cite{Bringewatt:2020ixn}. Right top: Nucleon strange helicity PDF (blue band) compared to  results from JAM (green band) and NNPDF (red band)~\cite{Alexandrou:2020uyt}. Right bottom: Strange-antistrange asymmetry for the unpolarized PDF~\cite{Alexandrou:2021oih}.}\vspace*{-0.5cm}\label{fig:helicity}
\end{wrapfigure}

\noindent
Early exploratory studies  were done within the LaMET approach in Refs.~\cite{Zhang:2018nsy, Izubuchi:2019lyk} for the unpolarized distributions of the pion, of the pion and kaon in Ref.~\cite{Lin:2020ssv} and of the nucleon~\cite{Alexandrou:2015rja,Alexandrou:2018pbm,Lin:2018pvv} but also  within the pseudo-distribution approach~for the the pion unpolarized PDF~\cite{Joo:2019bzr}. Other approaches were also developed. For example, employing matrix elements of two local,
spacelike-separated and gauge-invariant currents --also referred to as "lattice cross sections (LCSs)"~\cite{Ma:2014jla}-- the pion PDF was computed~\cite{Sufian:2019bol, Sufian:2020vzb}. These early computations both established  the feasibility of these alternative approaches and demonstrated the impact on phenomenological analyses. 
This can be seen in   Fig.~\ref{fig:helicity}, where we show    results by the JAM collaboration, which illustrate that, including LQCD input, reduced uncertainties in the isovector distributions. The nucleon strange quark unpolarized~\cite{Zhang:2020dkn}, helicity~\cite{Alexandrou:2020uyt} and transversity PDFs~\cite{Alexandrou:2021oih} were also computed for heavier than physical pion mass ensembles, again showcasing the better precision that can be achieved as comped to phenomenological analyses, as shown in Fig.~\ref{fig:helicity}.

During the last five years, improvements were implemented in renormalization and matching by various groups. In addition, lattice systematics were more carefully investigated.  

\begin{figure}[h!]\vspace*{-0.6cm}
 \begin{minipage}{0.45\linewidth}
\includegraphics[width=\linewidth]{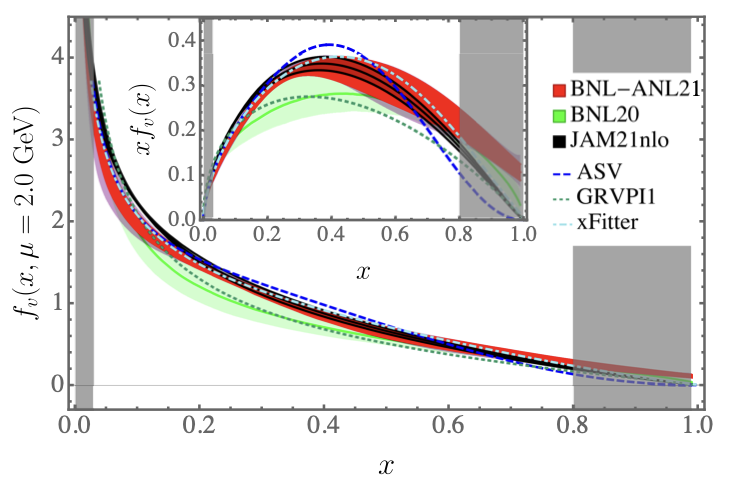}\\
\includegraphics[width=0.96\linewidth, height=0.6\linewidth]{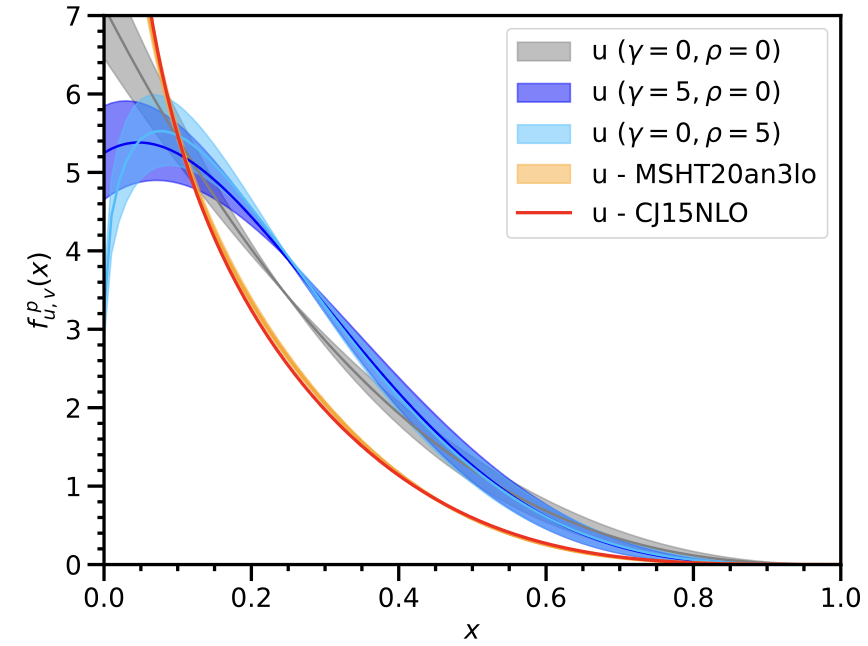}
\end{minipage}\hfill
\begin{minipage}{0.45\linewidth}
\includegraphics[width=\linewidth,height=0.6\linewidth]{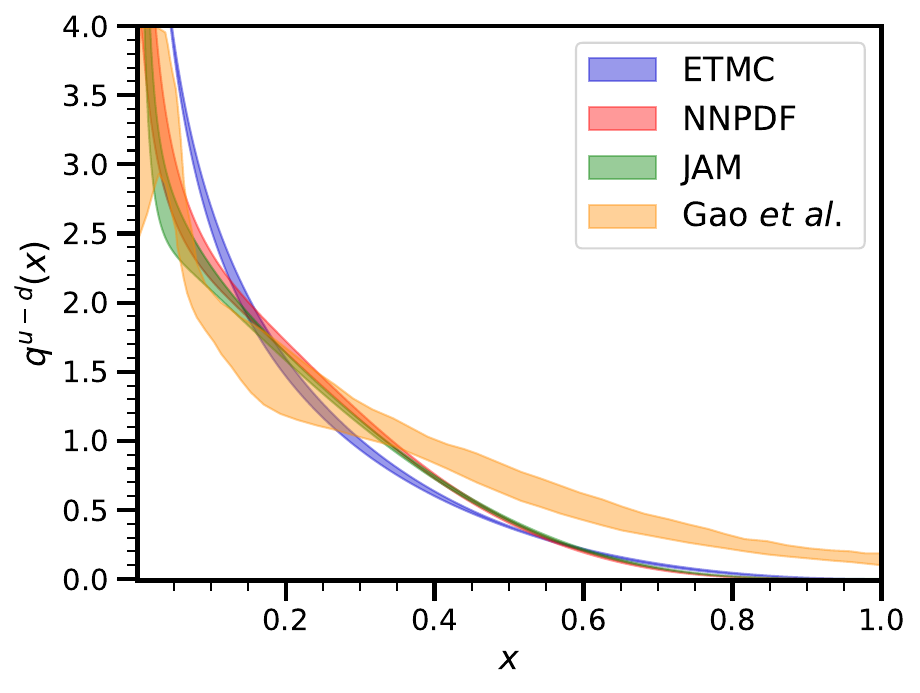}\\
\includegraphics[width=\linewidth,height=0.66\linewidth]{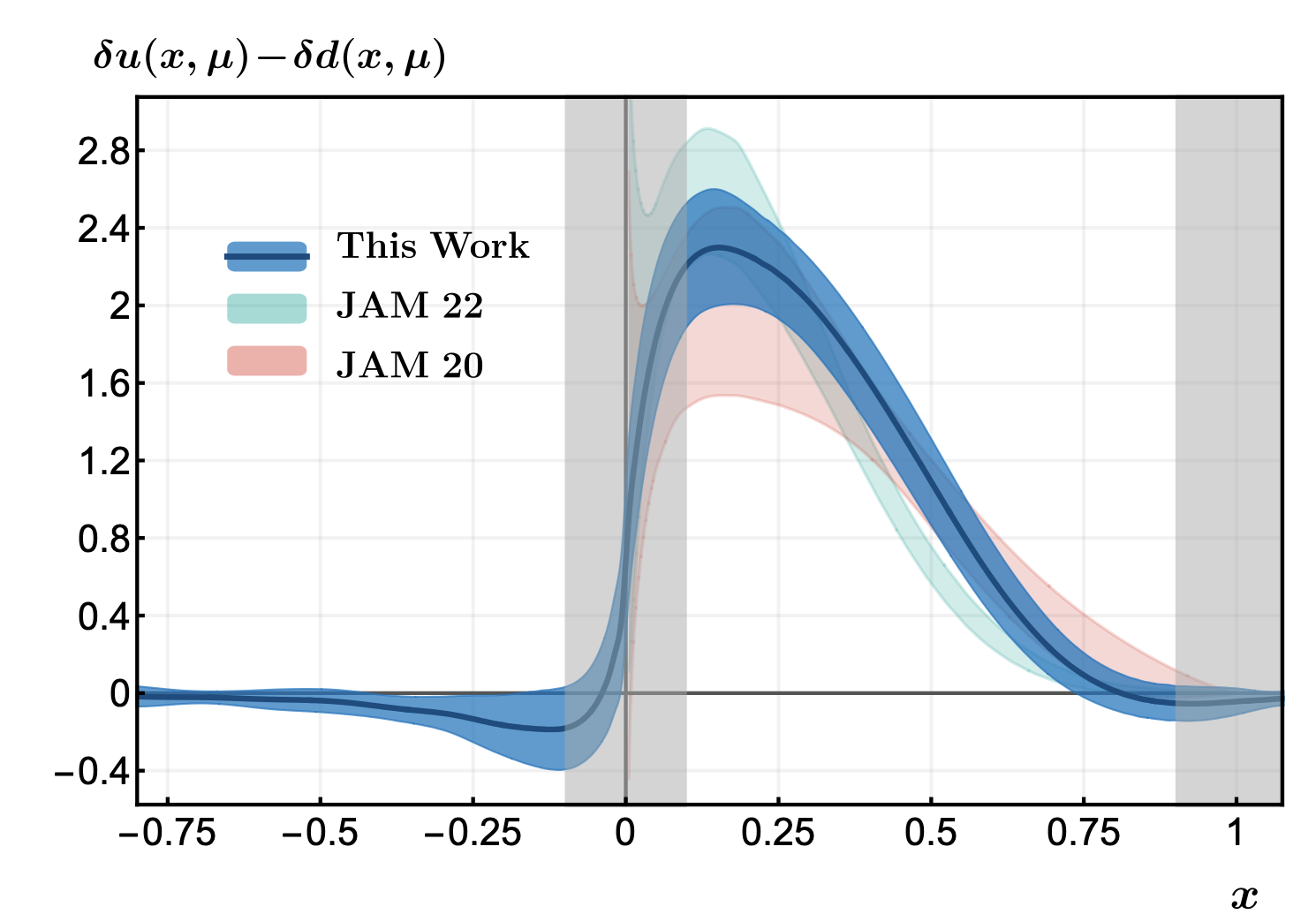}
\end{minipage}\hfill
\vspace*{-0.3cm}
\caption{Left top: Valence pion unpolarized PDF  using NNLO matching and the hybrid renormalization scheme (red band) compared  to JAM (black band)~\cite{Gao:2021dbh}. The green band shows previous results without the improvements. The gray horizontal  bands show the $x$-values where the PDF should not be trusted;  Right top: Nucleon isovector unpolarized PDF from Ref.~\cite{Gao:2022uhg} (green band)   and ETMC (orange band); Left bottom: Isovector nucleon PDF using the auxillary heavy quark method~\cite{Zimmermann:2024zde};  Right bottom: Nucleon isovector transvesity PDF at the physical point (blue band) compared to JAM analyses~\cite{LatticeParton:2022xsd}.}\label{fig:pi-nucleon PDFs}\vspace*{-0.7cm}
\end{figure}

\noindent
An early example of an improved computation is the evaluation of the valence pion unpolarized PDF computed within LaMET using one ensemble of $N_f=2+1$ HISQ sea  and Wilson–Clover
valence quarks with $m_\pi=300$~MeV and $a=0.06, 0.04$~fm~\cite{Gao:2021dbh}. They employed  a NNLO matching kernel and a hybrid renormalization scheme \cite{Ji:2020brr} their results for  the pion valence unpolarized PDF are shown in Fig.~\ref{fig:pi-nucleon PDFs}. In a followup work, they included a third ensemble with $a=0.076$~fm and $m_\pi=140$~MeV and, by combining LaMET and SDF, they determined  the Mellin moments, $\langle x^2\rangle, \langle x^4\rangle, \langle x^6\rangle$ in the continuum limit~\cite{Gao:2022iex}. They obtained values   in agreement with phenomenological determinations. They applied a similar approach  to compute the isovector nucleon unpolarized PDF using the $140$~MeV ensemble extracting results on the Mellin moments up to $\langle x^4\rangle$~\cite{Gao:2022uhg}. Their nucleon PDF is  compared in Fig.~\ref{fig:pi-nucleon PDFs} to the one by ETMC reconstructed with Mellin moments up to $\langle x^3 \rangle$  using one TMF  ensemble with $m_\pi\simeq140$~MeV.  There are differences, especially at larger $x$-values that indicate  lattice  systematics, which need further study.  

In Fig.~\ref{fig:pi-nucleon PDFs}, we also show a proof of concept  for computing the nucleon isovector PDF within  the auxillary heavy-quark approach~\cite{Zimmermann:2024zde}. A recent example, which includes a more complete  investigation of lattice systematics was done using  CLS ensembles with four $a\in (0.098-0.049)$~fm, pion masses ranging from 220~MeV to 350 MeV and momentum boosts from 1.6 GeV to 2.8 GeV by the Lattice Parton Collaboration (LPC)~\cite{LatticeParton:2022xsd}. They computed the nucleon isovector transversity with renormalization done using a hybrid scheme separating the short and  long distances followed by taking the large momentum limit. Their results, after  performing the 
continuum and chiral extrapolations, are shown in Fig.~\ref{fig:pi-nucleon PDFs} and they are in agreement with phenomenological analyses.

\begin{wrapfigure}[19]{L}{0.7\linewidth}\vspace*{-0.3cm}
  \includegraphics[width=\linewidth]{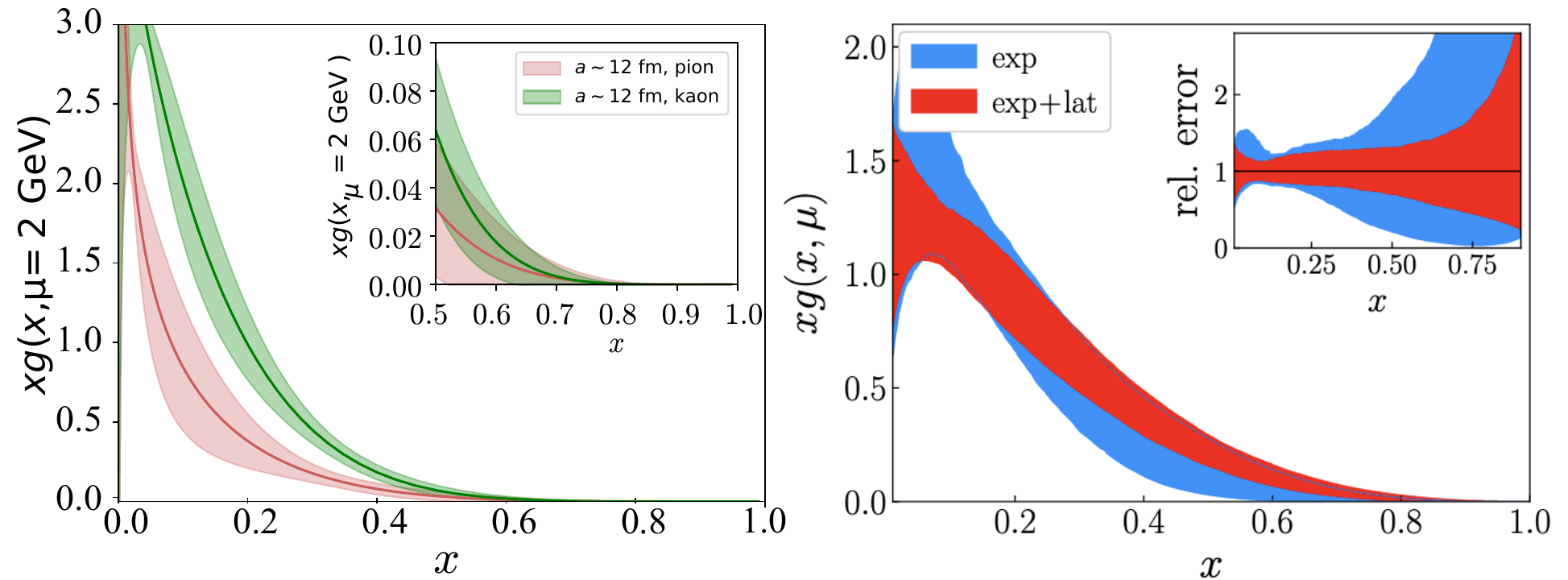}
  \caption{Left:  Results for the  gluon unpolarized PDFs for the pion (red band) and kaon (green band) from MSULAT~\cite{NieMiera:2025inn}. Right: JAM analysis for the unpolarized gluon PFD for the pion  with (red) and without (blue) LQCD data. Figure from Ref.~\cite{JeffersonLabAngularMomentumJAM:2022aix}.}\label{fig:helicity meson}
\includegraphics[width=\linewidth]{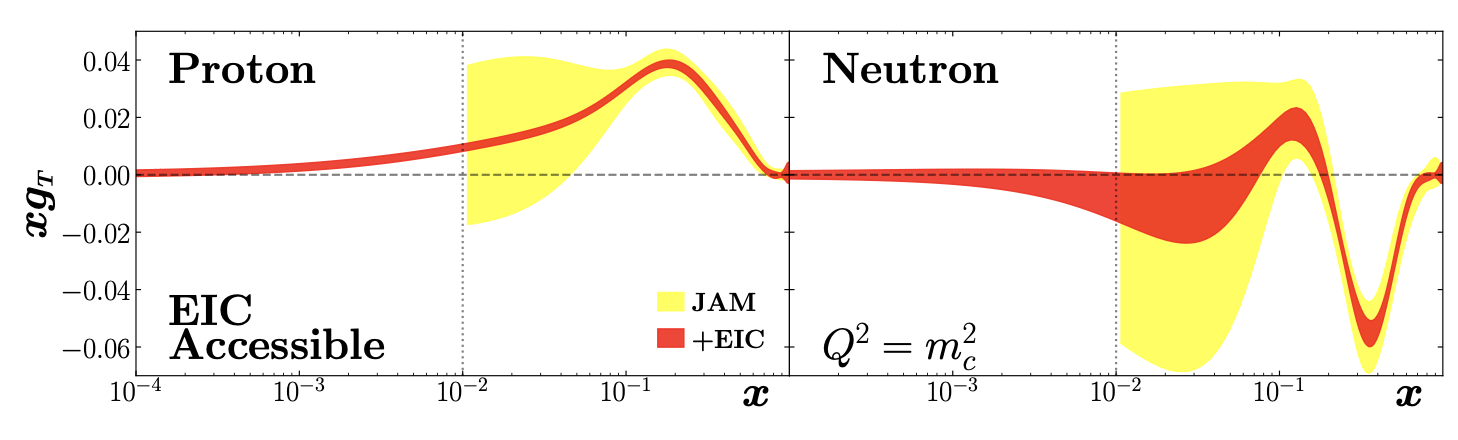}\vspace*{-0.3cm}
\caption{Impact of EIC data on the determination of the twist-3 PDF $g_T(x)$. Figure from Ref.~\cite{AbdulKhalek:2021gbh}.}\label{fig:twist-3}
\vspace*{-0.5cm}

\end{wrapfigure}

Computations of gluon PDFs  mostly used the pseudo-distribution approach, which simplifies  the renormalization. 
The gluon  unpolarized PDFs for the pion and kaon have been computed recently using one HISQ  ensemble  of $m_\pi=310$~MeV and $a=0.12$~fm~\cite{Good:2023ecp,NieMiera:2025inn} within the pseudo-distribution approach. We show their results in Fig.~\ref{fig:helicity meson} as well as  the impact they already are having on phenomenological analyses of gluon PDFs.

\begin{figure}[h!]
\includegraphics[width=\linewidth]{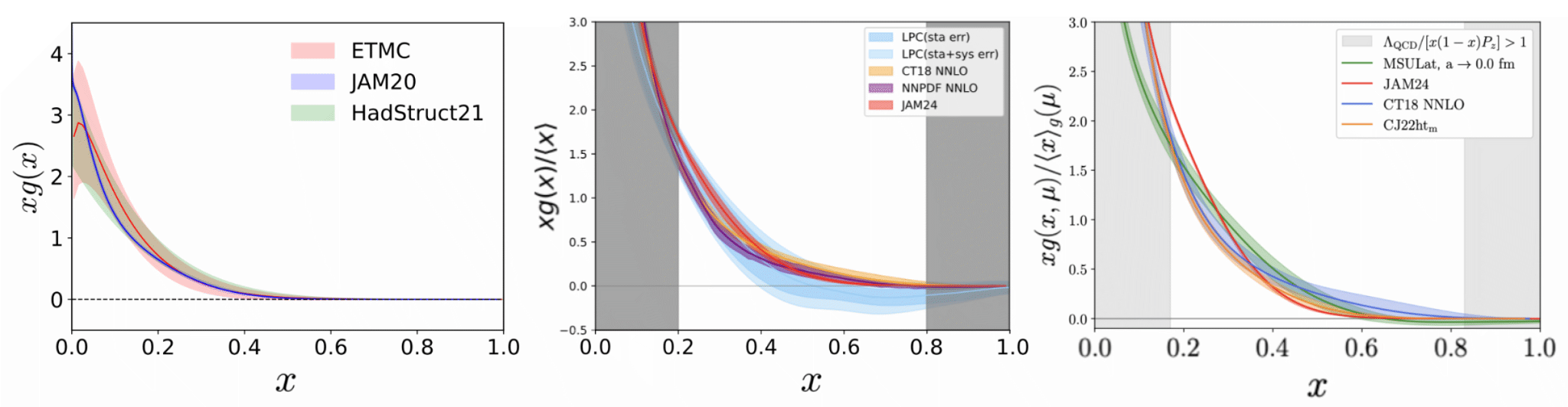}\vspace*{-0.3cm}
\caption{Nucleon gluon unpolarized PDF. Left: ETMC (red)~\cite{Delmar:2023agv} compared to HadStruc collabotaion (green)~\cite{HadStruc:2021wmh} and JAM (blue). Middle: LPC (blue) compared to CT18 (yellow), NNPDF (purple) and JAM (red)~\cite{ChenChen:2025amm}. Right: MSULAT (green) compared to JAM (red), CT18 (blue) and CJ22 (yellow)~\cite{NieMiera:2025mwj}.}\label{fig:N-gluon}\vspace*{-0.3cm}
\end{figure}

The gluon unpolarized  PDF has the complication that it   mixes with the quark singlet PDF.
However, it was shown for the nucleon  in Ref.~\cite{Delmar:2023agv}, using a TMF ensemble with $m_\pi=260$~MeV, that this mixing  is negligible compared to the statistical errors, which justifies PDF computations neglecting it. In Fig.~\ref{fig:N-gluon}, we show results on the gluon unpolarized nucleon PDF from four collaborations, two using the pseudo-distribution approach~\cite{HadStruc:2021wmh,Delmar:2023agv} and two recent ones using the quasi-distribution approach~\cite{ChenChen:2025amm}. 

 
\noindent
The pseudo-distribution approach was used by the HadStruc and ETM collaborations. The first  analyzed one HISQ ensemble with $m_\pi\sim 360$~MeV and $a\sim 0.09$~fm  and the second one TMF ensemble with $m_\pi=260$~MeV and $a\sim 0.09$~fm. Their results are compatible. The quasi-distribution approach was employed by LPC  and  the MSULAT collaboration, the first using  three CLS ensembles with   $a=0.105, 0.0897, 0.0775$~fm and  $m_\pi\sim310$ MeV, and the second employing  three HISQ ensembles with $a=0.15,~0.12,~0.09$~fm and $m_\pi\sim 300$~MeV. In both cases, the renormalization is done using a hybrid scheme. Their final results are presented  in the  continuum limit with LPC after taking the infinite momentum limit and the MSULAT collaboration at momentum boost of 2.12~GeV. The extracted PDFs are compatible.

\noindent
There are only two computations of the nucleon gluon helicity PDF $\Delta g$ by two groups~\cite{HadStruc:2022yaw, Khan:2022vot,Chowdhury:2024ymm} within the pseudo-distribution approach. LQCD input for the gluon helicity  can have a direct impact  on the sign
of $\Delta g$ at intermediate parton momentum fraction $x$, where positive and negative values are allowed by the data  in the absence of parton positivity constraints. However,  present LQCD data
cannot discriminate between positive and negative $\Delta g$ solutions~\cite{Karpie:2023nyg}. This calls for more precise data so
systematic uncertainties can be quantified.

\noindent
Important quantities for the EIC physics program include higher twist PDFs. The three twist-3 PDFs, scalar  $e(x)$,  helicity $g_2(x)$ or $g_T(x)$ and $h_2(x)$ or $h_L(x)$ have no density interpretation but, for example, the three Mellin  moments of $e(x)$ and $g_2(x)$ can yield information on the transverse force and encode information on quark-gluon correlations. Promising results on the $g_2(x)$ and $h_2(x)$ were obtained  using one TMF ensemble  with $m_\pi=260$~MeV by ETMC  using the quasi-distribution approach, 
see Ref.~\cite{Bhattacharya:2024} for a review. Due to their importance for the EIC program, further LQCD studies are called for. In Fig.~\ref{fig:twist-3}, we show the expected impact of EIC on the  extraction of the $g_T(x)$ distribution. An LQCD determination with controlled systematics would provide crucial input.\vspace*{-0.3cm}

\subsection{Generalized parton distributions}
GPDs can be computed within both the quasi- and pseudo-distribution approaches~\cite{Ji:2015qla, Xiong:2015nua, Liu:2019urm,Radyushkin:2017cyf,Radyushkin:2020} by allowing different boosted initial and final hadron states. In the original formulation, the spatial correlators are evaluated in the Breit-frame. Quasi-distributions are then computed by taking the Fourier transform of the renormalized matrix elements and performing a perturbative matching \vspace*{-0.3cm}
\begin{equation}
\small
  \tilde{F}_\Gamma(z,\tilde{\xi},Q^2,P_3,\mu^0,\mu^0_3)=\int_{-1}^1\, \frac{dy}{y} C_\Gamma\left(\frac{x}{y},\frac{\mu}{yP_3}, \frac{\mu^0_3}{yP_3},\frac{(\mu^0)^2}{(\mu^0_3)^2} \right)\, F_\Gamma(y,Q^2,\xi,\mu)+{\cal {O}}\left(\frac{m^2}{P_3^2},\frac{Q^2}{P_3^2},\frac{\Lambda^2_{\rm QCD}}{x^2P_3^2},\frac{\Lambda^2_{\rm QCD}}{(1-x)^2P_3^2}\right),  
\end{equation}
 where $\tilde{\xi}=-\frac{Q_3}{2P_3}$ is the quasi-skewness and $C_\Gamma$ the matching kernel. 
 
 \begin{figure}[h!]\vspace*{-0.4cm}
   \center{\includegraphics[width=0.9\linewidth]{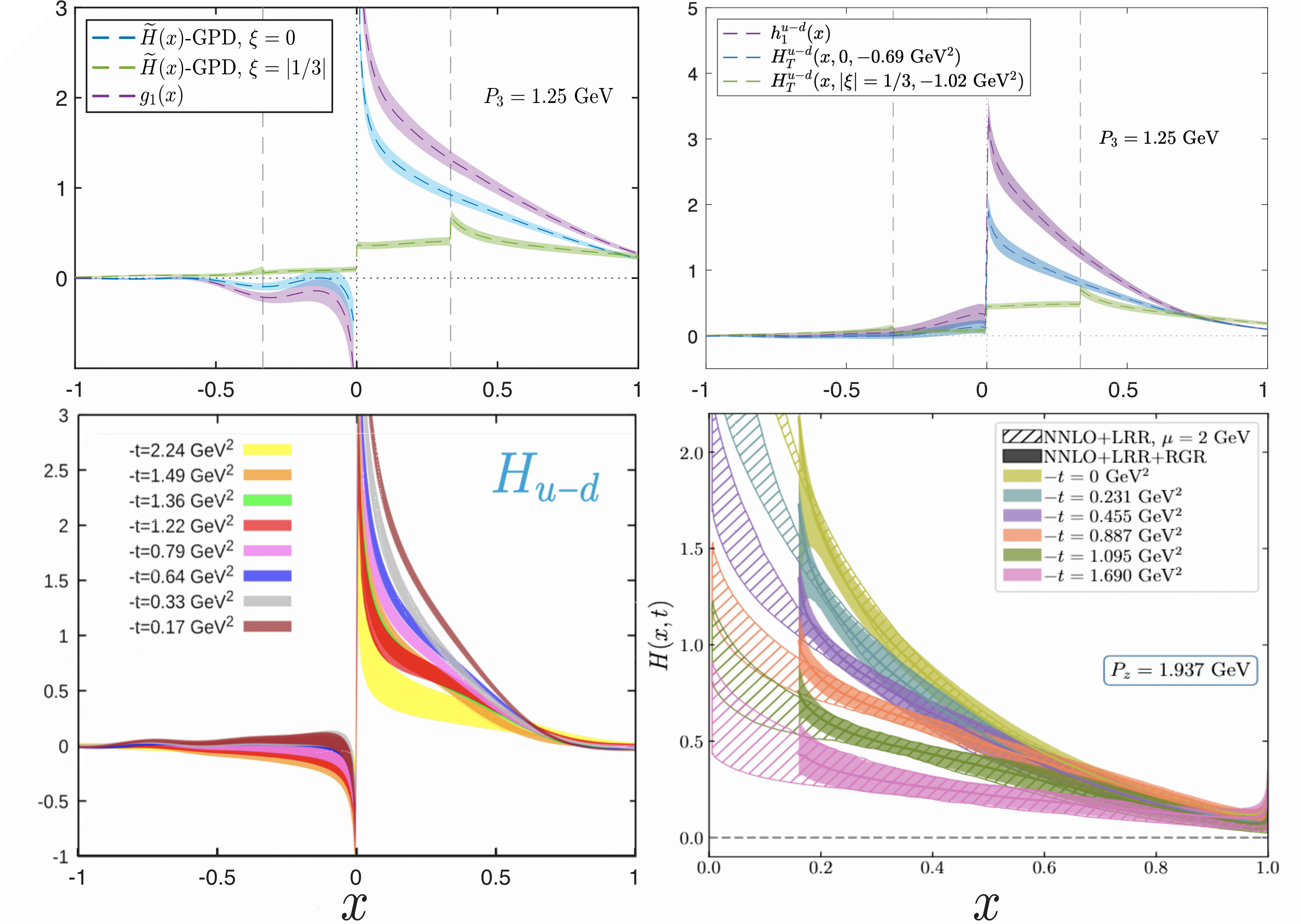}}
\caption{Results by ETMC for the  nucleon isovector helicity (left top)~\cite{Alexandrou:2020zbe} and   transversity (right top)~\cite{Alexandrou:2021bbo} for  $Q^2=-t{=}0,0.69,1.02$~GeV$^2$ in the Breit-frame;  Nucleon isovector unpolarized GPD in the asymmetric frame at various $-t=Q^2$-values and $\xi{=}0$ (Left bottom) courtesy of K. Cichy, EINN 2025); Pion isovector unpolarized GPD in the asymmetric frame for several values of $-t$~\cite{Ding:2024saz} (right bottom).   }\label{fig:GPDs}\vspace*{-0.7cm}
\end{figure}

To extract GPDs in the pseudo-distribution,  one follows the similar steps as for PDFs. As in the case of  PDFs, quasi- and pseudo-GPDs use the same lattice correlators as their starting point. First results  were obtained for the isovector pion  unpolarized~\cite{Chen:2019lcm} and nucleon isovector unpolarized, helicity and transversity~\cite{Alexandrou:2020zbe,Alexandrou:2021bbo} GPDs,  in the quasi-distribution approach. In Fig.~\ref{fig:GPDs}, we show early ETMC results on the nucleon helicity $\tilde{H}^{u-d}$  and transversity ${H}_T^{u-d}$ for zero and non-zero skewness.

\begin{wrapfigure}[16]{L}{0.5\linewidth}\vspace*{-0.3cm}
\hspace*{-0.5cm}\includegraphics[width=1.1\linewidth]{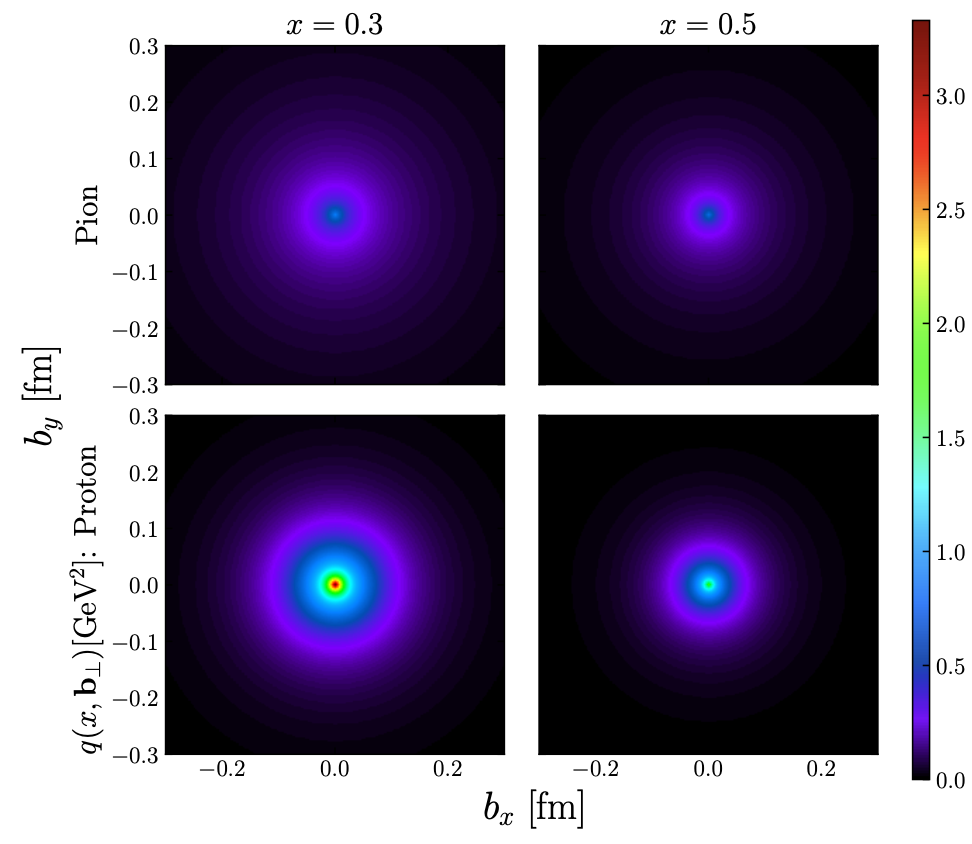}\vspace*{-0.5cm}
\caption{Distribution in impact parameter for the pion (top) and nucleon (bottom) for $x=0.3$ (left) and $x=0.5$ (right)~\cite{Bhattacharya:2022aob}.}\label{fig:imaging}
\end{wrapfigure}

\noindent
The computation was done  using one $N_f=2+1+1$ TMF ensemble with $m_\pi=260$~MeV and $a\sim 0.09$~fm, demonstrating the feasibility of the method. However, computing matrix elements in the Breit frame is very expensive since for each value of the momentum  transfer one needs another three-point function computation.  A more efficient way to compute these matrix elements is to use the lab frame and expand the matrix element in terms of gauge invariant functions that then can be related to the  GPDs defined in the Breit frame. This formalism has been developed and tested for the unpolarized, helicity and transversity GPDs in Refs.~\cite{Bhattacharya:2022aob, Bhattacharya:2023jsc, Bhattacharya:2025yba} and allows to compute the GPDs at multiple momentum transfers and skewness at once. Results using this co-called  asymmetric frame are shown for the unpolarized $H^{u-d}$~\cite{Chu:2025kew}  in Fig.~\ref{fig:GPDs} for multiple values of the momentum transfer computed for one TMF ensemble with $m_\pi=260$~MeV and $a\sim 0.09$~fm. Corresponding results for  the pion unpolarized GPD  obtained using one HISQ ensemble and clover-improve valence fermions with $m_\pi=300$~MeV and $a=0.04$~fm ~\cite{Ding:2024saz} are also shown. Both computations were done in the quasi-distributions approach  and for zero skewness. A clear feature that emerges from these results is that GPDs flatten as the momentum transfer increases. Since  pseudo-distributions  use the same lattice correlators, the asymmetric frame can  also be employed to extract pseudo-GPDs~\cite{Bhattacharya:2024qpp}.

\begin{wrapfigure}[31]{L}{0.6\linewidth}\vspace*{-0.5cm}
\hspace*{-0.3cm}\includegraphics[width=1.05\linewidth]{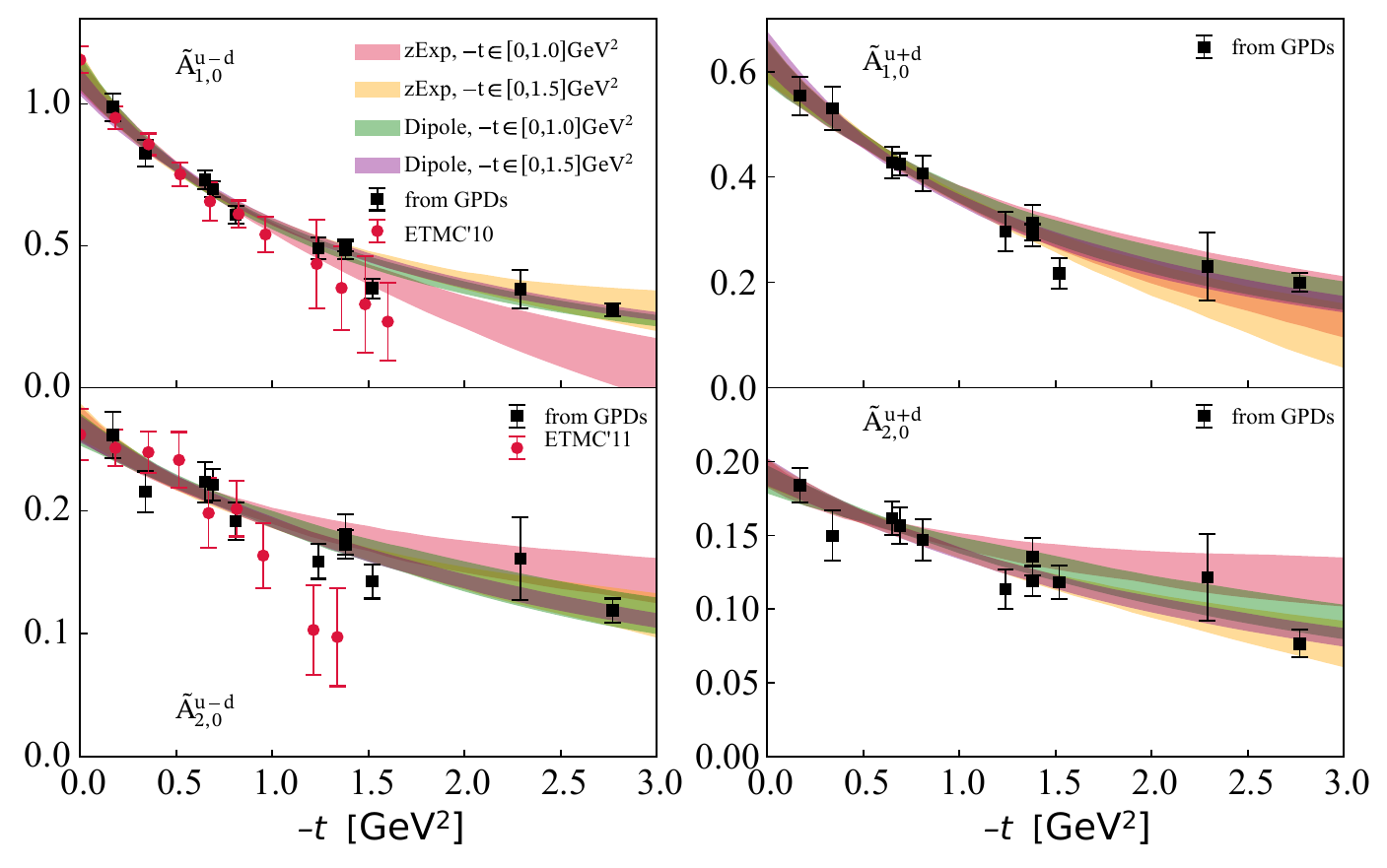}\\
 \begin{minipage}{0.49\linewidth}\vspace*{-0.cm}
\hspace*{-0.2cm}\includegraphics[width=1.05\linewidth]{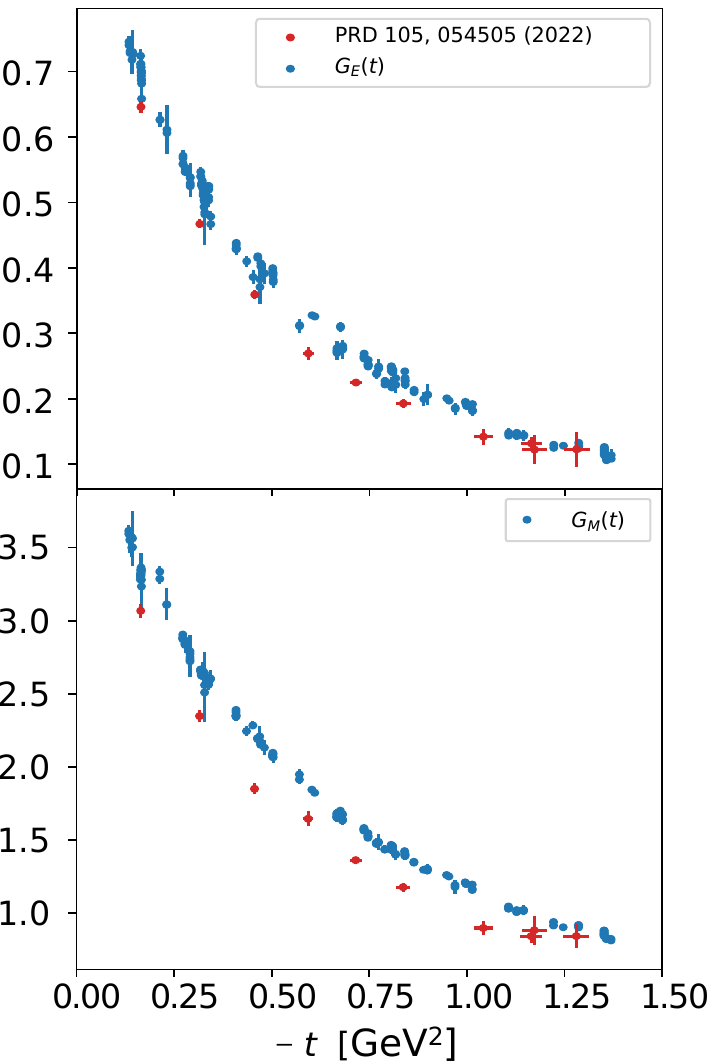}
\end{minipage}\hfill
\begin{minipage}{0.49\linewidth}\vspace*{-0.cm}
\hspace*{-0.2cm}\includegraphics[width=1.03\linewidth]{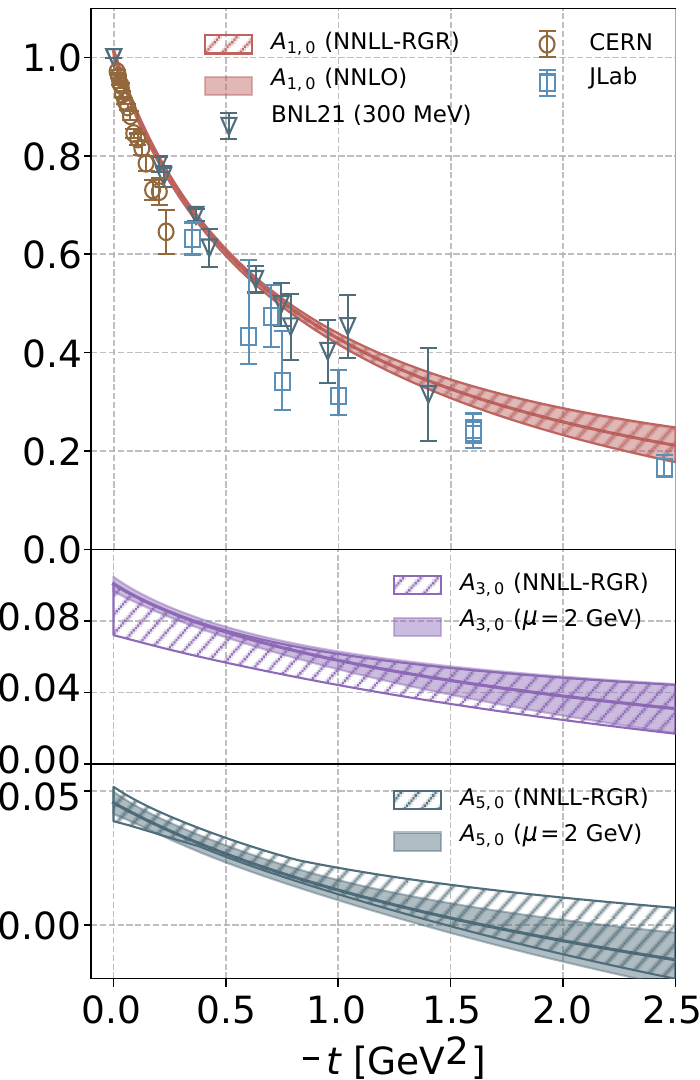}
\end{minipage}\vspace*{-0.1cm}
\caption{Top 4 panels:   Nucleon axial GFFs for i) isovector  $\tilde{A}^{u-d}_{10}$  and $\tilde{A}^{u-d}_{20}$ (left); ii) isoscalar   $\tilde{A}^{u+d}_{10}$  and $A^{u+d}_{20}$ (right), shown with black points compared to those computed from local operators (red points)~\cite{Bhattacharya:2024wtg}. Bottom 4 panels:  i) Electric  and magnetic  form factors (left) extracted from LQCD-determined GPDs (blue) and from the matrix element of the electromagnetic current (red)~\cite{HadStruc:2024rix}; ii) Pion GFFs (right) $A_{10}(t)$, $A_{30}(t)$  and $A_{50}(t)$~\cite{Gao:2025inf}.}\label{fig:GFFs-from-GPDs}\vspace*{-0.3cm}
\end{wrapfigure}

\noindent
In Fig.~\ref{fig:imaging}, LQCD data on the unpolarized isovector pion~\cite{Ding:2024saz} and nucleon~\cite{Bhattacharya:2022aob}  GPDs at zero skewness were used to  construct the distributions in impact parameter space, \vspace*{-0.15cm}
\begin{align} q(x,b_\perp)=\int_{-\infty}^\infty\,\frac{d^2\Delta_\perp}{(2\pi)^2}\,&e^{{-i\Delta_\perp}.b_\perp}\nonumber\\ & \hspace*{-0.5cm}\times H(x,\xi=0,t),\nonumber
\end{align}
for different values of $x$, providing detailed information of the 3-dimensional  structure of these hadrons.

Determining the GPDs  at multiple momentum transfers allows one to compute their moments and  extract GFFs within the SDF procedure as done for computing moments  of   LQCD-determined PDFs. 
In Fig.~\ref{fig:GFFs-from-GPDs}, we show recent results on the extraction of  GFFs from LQCD-determined GPDs at zero and non-zero skewness from three groups  that demonstrate the feasibility of the approach. All  computations were done for one gauge ensemble with heavier than physical pion mass. ETMC computed the  nucleon isovector and isoscalar helicity GPDs  for zero skewness   within the quasi-distribution approach~\cite{Bhattacharya:2024wtg}. Using SDF, the renormalized axial GPDs are expanded in terms of the  GFFs $\tilde{A}_{n0}$, \vspace*{-0.2cm}
  \begin{equation} {\tilde{H}^R}(z_3,P_3, \Delta)=\sum_{n=1}^\infty {C^{\overline{\rm MS}}_n(\mu^2 z^2)}\frac{(-iz_3P_3)^{n-1}}{(n-1)!}\nonumber\tilde{A}_{n,0}(t)+{\cal O}(z^2\Lambda_{QCD}^2).\vspace*{-0.2cm}
  \end{equation} 
Results are  in agreement with those extracted from  local operators using the same ensemble. 

The two other groups computed GFFs at non-zero skewness, which presents another recent advancement.  One group computed the  nucleon isovector unpolarized  GPDs~\cite{HadStruc:2024rix} and the other the unpolarized pion GPDs~\cite{Gao:2025inf} for zero and non-zero skewness  within the pseudo- and quasi-distribution approach, respectively. 
In Fig.~\ref{fig:GFFs-from-GPDs}, we  show results for the nucleon isovector electric and magnetic form factors compared to a direct computation of the matrix element of the electromagnetic current. Given that lattice artifacts are not fully accounted for, the agreement among these two sets of results is satisfactory. 
The other group computed the pion unpolarized  GFFs  up to the fourth moment (n=5). 
As can be seen, their results for the pion form factor is in  good agreement with the direct determination from matrix elements of the electromagnetic current computed using the same ensemble.  We also include their results on the GFFs $A_{30}$ and $A_{50}$, which are small but non-zero. These two cases demonstrate the wealth of information that LQCD can provide.    Since,  currently, experiments only give
indirect access to GPDs, LQCD is in an ideal position to  provide systematically improvable information  on these GPDs and, thus, on the 3D  structure of hadrons. \vspace*{-0.2cm}

\section{Conclusions}\vspace*{-0.1cm}
The Electron–Ion Collider  will be a uniquely versatile facility, delivering an unprecedented wealth of precision data on the internal structure of the pion, kaon, and proton. By extending measurements of electromagnetic and transition form factors to large momentum transfer, the EIC will probe hadron structure at short distances, directly accessing the dynamics of quarks and gluons in previously unexplored regimes.
A major breakthrough will come from exploring the low-$x$ region of PDFs and GPDs, significantly improving our knowledge of sea quark and gluon distributions. These measurements are central to addressing fundamental questions in QCD, such as  the spin structure of the proton, the origin of its mass, and its mechanical properties. Through exclusive and semi-inclusive processes, the EIC will enable true three-dimensional imaging of the pion, kaon, and proton, providing spatial and momentum tomography of  these hadrons.

Furthermore, precision measurements of helicity and transversity transverse-momentum-dependent PDFs (TMDs)—including the Sivers and Boer–Mulders functions—as well as higher-twist PDFs, will illuminate multi-parton correlations and the dynamical interplay between spin and orbital motion. This program will deepen our understanding of QCD beyond the leading-twist approximation and open a window into correlated quark–gluon dynamics.

In parallel, lattice QCD is poised to play a transformative role in the EIC scientific program. State-of-the-art lattice computations now provide increasingly precise determinations of hadron charges, form factors, and Mellin moments of PDFs and GPDs, with systematic uncertainties under quantitative control. As percent-level precision is reached in several of these key observables, the inclusion of isospin-breaking effects becomes mandatory, ushering in a new era of combined QCD+QED calculations. 

Remarkable progress has also been achieved in the direct computation of PDFs, GPDs, and TMDs—including twist-3 distributions—supported by improved matching frameworks and non-perturbative renormalization procedures. These developments significantly strengthen the synergy between lattice QCD  and experimental programs at JLab, EIC and CERN. 

Together,  EIC and lattice QCD define a powerful, mutually reinforcing program: high-precision experimental measurements and {\it ab initio} theoretical calculations that can converge to deliver a quantitative, multidimensional understanding of hadron structure rooted directly in QCD. The realization of this ambitious lattice QCD program critically depends on continued investments in high performance computing  and in the availability of large computational resources.


\medskip
\noindent
{\bf Acknowledgments.}
I would like to thank all members of ETMC for their valuable contributions. Special thanks to my close collaborators S. Bacchio, G. Koutsou, Y. Li and G. Spanoudes, and PhD students L. Chacon, C. Iona, P. Jana, C. Kummer and B. Prasad   for providing me with  results for this writeup. I selected examples of results on Mellin moments and PDFs/GPDs that I thought are of high relevance to EIC physics  from the highlights sent to me by  A. Avkhadiev, S. Collins, Ch. Monahan,  W. Morris, D. Pefkou, A. Shindler, W. Wang, Z. Yong, and Ch. Zimmerman, all of whom I would like to thank for sharing their work some of which still unpublished. I acknowledge partial funding from the AQTIVATE European Joint Doctorate, grant agreement No. 101072344, the MSCA co-funded project ENGAGE grant agreement No. 101034267 and the project IMAGE-N  (EXCELLENCE/0524/0459) co-financed by the European Regional Development Fund and the Republic of Cyprus through the Research and Innovation Foundation within the framework of the Cohesion Policy Programme “THALIA 2021-2027”.
\vspace*{-0.4cm}
\bibliography{refs}\vspace*{-0.7cm}

@article{FlavourLatticeAveragingGroupFLAG:2024oxs,
    author = "Aoki, Y. and others",
    collaboration = "Flavour Lattice Averaging Group (FLAG)",
    title = "{FLAG review 2024}",
    eprint = "2411.04268",
    archivePrefix = "arXiv",
    primaryClass = "hep-lat",
    reportNumber = "CERN-TH-2024-192, FERMILAB-PUB-24-0785-T",
    doi = "10.1103/nfzp-p5dn",
    journal = "Phys. Rev. D",
    volume = "113",
    number = "1",
    pages = "014508",
    year = "2026"
}

@article{Cocuzza:2023oam,
    author = "Cocuzza, C. and Metz, A. and Pitonyak, D. and Prokudin, A. and Sato, N. and Seidl, R.",
    collaboration = "JAM",
    title = "{Transversity Distributions and Tensor Charges of the Nucleon: Extraction from Dihadron Production and Their Universal Nature}",
    eprint = "2306.12998",
    archivePrefix = "arXiv",
    primaryClass = "hep-ph",
    reportNumber = "JLAB-THY-23-3859",
    doi = "10.1103/PhysRevLett.132.091901",
    journal = "Phys. Rev. Lett.",
    volume = "132",
    number = "9",
    pages = "091901",
    year = "2024"
}

@article{Han:2020vjp,
    author = "Han, Chengdong and Xie, Gang and Wang, Rong and Chen, Xurong",
    title = "{An Analysis of Parton Distribution Functions of the Pion and the Kaon with the Maximum Entropy Input}",
    eprint = "2010.14284",
    archivePrefix = "arXiv",
    primaryClass = "hep-ph",
    doi = "10.1140/epjc/s10052-021-09087-8",
    journal = "Eur. Phys. J. C",
    volume = "81",
    number = "4",
    pages = "302",
    year = "2021"
}

@article{Barry:2021osv,
    author = "Barry, P. C. and Ji, Chueng-Ryong and Sato, N. and Melnitchouk, W.",
    collaboration = "Jefferson Lab Angular Momentum (JAM)",
    title = "{Global QCD Analysis of Pion Parton Distributions with Threshold Resummation}",
    eprint = "2108.05822",
    archivePrefix = "arXiv",
    primaryClass = "hep-ph",
    reportNumber = "JLAB-THY-21-3482",
    doi = "10.1103/PhysRevLett.127.232001",
    journal = "Phys. Rev. Lett.",
    volume = "127",
    number = "23",
    pages = "232001",
    year = "2021"
}

@article{Barry:2025wjx,
    author = "Barry, P. C. and Ji, Chueng-Ryong and Melnitchouk, W. and Sato, N. and Steffens, Fernanda",
    collaboration = "JAM",
    title = "{First simultaneous global QCD analysis of kaon and pion parton distributions with lattice QCD constraints}",
    eprint = "2510.11979",
    archivePrefix = "arXiv",
    primaryClass = "hep-ph",
    reportNumber = "JLAB-THY-25-4569",
    month = "10",
    year = "2025"
}

@article{Bhattacharya:2024,
    author = "Bhattacharya, Sh.",
    title = "{Hadron structure via Generalized Parton
     Distributions}",
    journal = "PoS",
    volume = "LATTICE 2024",
    pages = "013",
    year = "2025"
}

@article{Kotz:2025lio,
    author = "Kotz, Lucas and Courtoy, Aurore and Nadolsky, Pavel and Ponce-Chavez, Maximiliano",
    title = "{Epistemic and nuclear uncertainties for the parton distributions of the pion}",
    eprint = "2505.13594",
    archivePrefix = "arXiv",
    primaryClass = "hep-ph",
    doi = "10.1103/h2vn-1wxp",
    journal = "Phys. Rev. D",
    volume = "112",
    number = "7",
    pages = "L071502",
    year = "2025"
}

@article{Novikov:2020snp,
    author = "Novikov, Ivan and others",
    title = "{Parton Distribution Functions of the Charged Pion Within The xFitter Framework}",
    eprint = "2002.02902",
    archivePrefix = "arXiv",
    primaryClass = "hep-ph",
    reportNumber = "DESY-20-013, DESY 20-013",
    doi = "10.1103/PhysRevD.102.014040",
    journal = "Phys. Rev. D",
    volume = "102",
    number = "1",
    pages = "014040",
    year = "2020"
}

@article{Saclay-CERN-CollegedeFrance-EcolePoly-Orsay:1980fhh,
    author = "Badier, J. and others",
    collaboration = "Saclay-CERN-College de France-Ecole Poly-Orsay",
    title = "{Measurement of the $K^- / \pi^-$ Structure Function Ratio Using the {Drell-Yan} Process}",
    reportNumber = "CERN-EP/80-48",
    doi = "10.1016/0370-2693(80)90530-4",
    journal = "Phys. Lett. B",
    volume = "93",
    pages = "354--356",
    year = "1980"
}

@article{Meziani:2025dwu,
    author = "Meziani, Zein-Eddine",
    title = "{Gluonic Energy-Momentum Tensor Form Factors of the Proton}",
    eprint = "2505.05671",
    archivePrefix = "arXiv",
    primaryClass = "nucl-ex",
    doi = "10.22323/1.483.0076",
    journal = "PoS",
    volume = "QCHSC24",
    pages = "076",
    year = "2025"
}

@article{JeffersonLabAngularMomentumJAM:2022aix,
    author = "Barry, P. C. and others",
    collaboration = "Jefferson Lab Angular Momentum (JAM), HadStruc",
    title = "{Complementarity of experimental and lattice QCD data on pion parton distributions}",
    eprint = "2204.00543",
    archivePrefix = "arXiv",
    primaryClass = "hep-ph",
    reportNumber = "JLAB-THY-22-3592",
    doi = "10.1103/PhysRevD.105.114051",
    journal = "Phys. Rev. D",
    volume = "105",
    number = "11",
    pages = "114051",
    year = "2022"
}

@article{Karpie:2023nyg,
    author = "Karpie, J. and Whitehill, R. M. and Melnitchouk, W. and Monahan, C. and Orginos, K. and Qiu, J. -W. and Richards, D. G. and Sato, N. and Zafeiropoulos, S.",
    collaboration = "Jefferson Lab Angular Momentum, HadStruc",
    title = "{Gluon helicity from global analysis of experimental data and lattice QCD Ioffe time distributions}",
    eprint = "2310.18179",
    archivePrefix = "arXiv",
    primaryClass = "hep-ph",
    reportNumber = "JLAB-THY-23-3950",
    doi = "10.1103/PhysRevD.109.036031",
    journal = "Phys. Rev. D",
    volume = "109",
    number = "3",
    pages = "036031",
    year = "2024"
}

@article{Bringewatt:2020ixn,
    author = "Bringewatt, J. and Sato, N. and Melnitchouk, W. and Qiu, Jian-Wei and Steffens, F. and Constantinou, M.",
    title = "{Confronting lattice parton distributions with global QCD analysis}",
    eprint = "2010.00548",
    archivePrefix = "arXiv",
    primaryClass = "hep-ph",
    reportNumber = "JLAB-THY-20-3257",
    doi = "10.1103/PhysRevD.103.016003",
    journal = "Phys. Rev. D",
    volume = "103",
    number = "1",
    pages = "016003",
    year = "2021"
}

@article{Alexandrou:2021oih,
    author = "Alexandrou, Constantia and Constantinou, Martha and Hadjiyiannakou, Kyriakos and Jansen, Karl and Manigrasso, Floriano",
    title = "{Flavor decomposition of the nucleon unpolarized, helicity, and transversity parton distribution functions from lattice QCD simulations}",
    eprint = "2106.16065",
    archivePrefix = "arXiv",
    primaryClass = "hep-lat",
    doi = "10.1103/PhysRevD.104.054503",
    journal = "Phys. Rev. D",
    volume = "104",
    number = "5",
    pages = "054503",
    year = "2021"
}

@article{Alexandrou:2020uyt,
    author = "Alexandrou, Constantia and Constantinou, Martha and Hadjiyiannakou, Kyriakos and Jansen, Karl and Manigrasso, Floriano",
    title = "{Flavor decomposition for the proton helicity parton distribution functions}",
    eprint = "2009.13061",
    archivePrefix = "arXiv",
    primaryClass = "hep-lat",
    reportNumber = "DESY-20-207",
    doi = "10.1103/PhysRevLett.126.102003",
    journal = "Phys. Rev. Lett.",
    volume = "126",
    number = "10",
    pages = "102003",
    year = "2021"
}

@article{Zhang:2020dkn,
    author = "Zhang, Rui and Lin, Huey-Wen and Yoon, Boram",
    title = "{Probing nucleon strange and charm distributions with lattice QCD}",
    eprint = "2005.01124",
    archivePrefix = "arXiv",
    primaryClass = "hep-lat",
    reportNumber = "MSUHEP-20-008",
    doi = "10.1103/PhysRevD.104.094511",
    journal = "Phys. Rev. D",
    volume = "104",
    number = "9",
    pages = "094511",
    year = "2021"
}

@article{Joo:2019bzr,
    author = "Jo{\'o}, B{\'a}lint and Karpie, Joseph and Orginos, Kostas and Radyushkin, Anatoly V. and Richards, David G. and Sufian, Raza Sabbir and Zafeiropoulos, Savvas",
    title = "{Pion valence structure from Ioffe-time parton pseudodistribution functions}",
    eprint = "1909.08517",
    archivePrefix = "arXiv",
    primaryClass = "hep-lat",
    reportNumber = "JLAB-THY-19-3038",
    doi = "10.1103/PhysRevD.100.114512",
    journal = "Phys. Rev. D",
    volume = "100",
    number = "11",
    pages = "114512",
    year = "2019"
}

@article{Cichy:2019,
  author        = {Cichy, Krzysztof and Constantinou, Martha},
  title         = {A guide to light-cone PDFs from lattice QCD},
  journal       = {Advances in High Energy Physics},
  volume        = {2019},
  pages         = {3036904},
  year          = {2019},
  eprint        = {1811.07248},
  archivePrefix = {arXiv},
  primaryClass  = {hep-lat},
  doi           = {10.1155/2019/3036904}
}

@article{Zhao:2019,
  author        = {Zhao, Yong},
  title         = {Unraveling high-energy hadron structures with lattice QCD},
  journal       = {International Journal of Modern Physics A},
  volume        = {33},
  pages         = {1830033},
  year          = {2019},
  eprint        = {1812.07192},
  archivePrefix = {arXiv},
  primaryClass  = {hep-ph},
  doi           = {10.1142/S0217751X18300338}
}

@article{Radyushkin:2020,
  author        = {Radyushkin, Anatoly V.},
  title         = {Theory and applications of parton pseudodistributions},
  journal       = {International Journal of Modern Physics A},
  volume        = {35},
  pages         = {2030002},
  year          = {2020},
  eprint        = {1912.04244},
  archivePrefix = {arXiv},
  primaryClass  = {hep-ph},
  doi           = {10.1142/S0217751X20300021}
}

@article{Ji:2021,
  author        = {Ji, Xiangdong and Liu, Yu-Sheng and Liu, Yu and Zhang, Jian-Hui and Zhao, Yong},
  title         = {Large-momentum effective theory},
  journal       = {Reviews of Modern Physics},
  volume        = {93},
  pages         = {035005},
  year          = {2021},
  eprint        = {2004.03543},
  archivePrefix = {arXiv},
  primaryClass  = {hep-ph},
  doi           = {10.1103/RevModPhys.93.035005}
}

@article{Constantinou:2021EPJA,
  author        = {Constantinou, Martha},
  title         = {The x-dependence of hadronic parton distributions},
  journal       = {European Physical Journal A},
  volume        = {57},
  pages         = {77},
  year          = {2021},
  eprint        = {2010.02445},
  archivePrefix = {arXiv},
  primaryClass  = {hep-lat},
  doi           = {10.1140/epja/s10050-021-00378-1}
}

@article{Constantinou:2021PPNP,
  author        = {Constantinou, Martha and others},
  title         = {Parton distributions and lattice QCD calculations: a community white paper},
  journal       = {Progress in Particle and Nuclear Physics},
  volume        = {121},
  pages         = {103908},
  year          = {2021},
  eprint        = {2006.08636},
  archivePrefix = {arXiv},
  primaryClass  = {hep-ph},
  doi           = {10.1016/j.ppnp.2021.103908}
}

@article{Cichy:2022PoS,
  author        = {Cichy, Krzysztof},
  title         = {Status and perspectives of quasi-PDFs},
  journal       = {Proceedings of Science},
  volume        = {LATTICE2021},
  pages         = {017},
  year          = {2022},
  eprint        = {2110.07440},
  archivePrefix = {arXiv},
  primaryClass  = {hep-lat}
}

@article{Ebert:2018gzl,
    author = "Ebert, Markus A. and Stewart, Iain W. and Zhao, Yong",
    title = "{Determining the Nonperturbative Collins-Soper Kernel From Lattice QCD}",
    eprint = "1811.00026",
    archivePrefix = "arXiv",
    primaryClass = "hep-ph",
    reportNumber = "MIT-CTP 5049",
    doi = "10.1103/PhysRevD.99.034505",
    journal = "Phys. Rev. D",
    volume = "99",
    number = "3",
    pages = "034505",
    year = "2019"
}

@article{Boussarie:2023izj,
    author = "Boussarie, Renaud and others",
    title = "{TMD Handbook}",
    eprint = "2304.03302",
    archivePrefix = "arXiv",
    primaryClass = "hep-ph",
    reportNumber = "JLAB-THY-23-3780, LA-UR-21-20798, MIT-CTP/5386",
    month = "4",
    year = "2023"
}

@article{RQCD:2019jai,
    author = {Bali, Gunnar S. and Barca, Lorenzo and Collins, Sara and Gruber, Michael and L{\"o}ffler, Marius and Sch{\"a}fer, Andreas and S{\"o}ldner, Wolfgang and Wein, Philipp and Weish{\"a}upl, Simon and Wurm, Thomas},
    collaboration = "RQCD",
    title = "{Nucleon axial structure from lattice QCD}",
    eprint = "1911.13150",
    archivePrefix = "arXiv",
    primaryClass = "hep-lat",
    doi = "10.1007/JHEP05(2020)126",
    journal = "JHEP",
    volume = "05",
    pages = "126",
    year = "2020"
}

@article{Hackett:2023rif,
    author = "Hackett, Daniel C. and Pefkou, Dimitra A. and Shanahan, Phiala E.",
    title = "{Gravitational Form Factors of the Proton from Lattice QCD}",
    eprint = "2310.08484",
    archivePrefix = "arXiv",
    primaryClass = "hep-lat",
    reportNumber = "MIT-CTP/5630, FERMILAB-PUB-23-592-T",
    doi = "10.1103/PhysRevLett.132.251904",
    journal = "Phys. Rev. Lett.",
    volume = "132",
    number = "25",
    pages = "251904",
    year = "2024"
}

@article{Crawford:2024wzx,
    author = {Crawford, J. A. and Can, K. U. and Horsley, R. and Rakow, P. E. L. and Schierholz, G. and St{\"u}ben, H. and Young, R. D. and Zanotti, J. M.},
    collaboration = "QCDSF",
    title = "{Transverse Force Distributions in the Proton from Lattice QCD}",
    eprint = "2408.03621",
    archivePrefix = "arXiv",
    primaryClass = "hep-lat",
    reportNumber = "ADP-24-12/T1251, DESY-24-120, LTH 1380",
    doi = "10.1103/PhysRevLett.134.071901",
    journal = "Phys. Rev. Lett.",
    volume = "134",
    number = "7",
    pages = "071901",
    year = "2025"
}

@article{Park:2021ypf,
    author = "Park, Sungwoo and Gupta, Rajan and Yoon, Boram and Mondal, Santanu and Bhattacharya, Tanmoy and Jang, Yong-Chull and Jo{\'o}, B{\'a}lint and Winter, Frank",
    collaboration = "Nucleon Matrix Elements (NME)",
    title = "{Precision nucleon charges and form factors using (2+1)-flavor lattice QCD}",
    eprint = "2103.05599",
    archivePrefix = "arXiv",
    primaryClass = "hep-lat",
    reportNumber = "LA-UR-21-20526, JLAB-THY-22-3583",
    doi = "10.1103/PhysRevD.105.054505",
    journal = "Phys. Rev. D",
    volume = "105",
    number = "5",
    pages = "054505",
    year = "2022"
}

@article{Djukanovic:2022wru,
    author = "Djukanovic, Dalibor and von Hippel, Georg and Koponen, Jonna and Meyer, Harvey B. and Ottnad, Konstantin and Schulz, Tobias and Wittig, Hartmut",
    title = "{Isovector axial form factor of the nucleon from lattice QCD}",
    eprint = "2207.03440",
    archivePrefix = "arXiv",
    primaryClass = "hep-lat",
    reportNumber = "MITP-22-053",
    doi = "10.1103/PhysRevD.106.074503",
    journal = "Phys. Rev. D",
    volume = "106",
    number = "7",
    pages = "074503",
    year = "2022"
}

@article{Alexandrou:2023qbg,
    author = "Alexandrou, Constantia and Bacchio, Simone and Constantinou, Martha and Finkenrath, Jacob and Frezzotti, Roberto and Kostrzewa, Bartosz and Koutsou, Giannis and Spanoudes, Gregoris and Urbach, Carsten",
    collaboration = "Extended Twisted Mass",
    title = "{Nucleon axial and pseudoscalar form factors using twisted-mass fermion ensembles at the physical point}",
    eprint = "2309.05774",
    archivePrefix = "arXiv",
    primaryClass = "hep-lat",
    doi = "10.1103/PhysRevD.109.034503",
    journal = "Phys. Rev. D",
    volume = "109",
    number = "3",
    pages = "034503",
    year = "2024"
}

@article{Jang:2023zts,
    author = "Jang, Yong-Chull and Gupta, Rajan and Bhattacharya, Tanmoy and Yoon, Boram and Lin, Huey-Wen",
    collaboration = "Precision Neutron Decay Matrix Elements (PNDME)",
    title = "{Nucleon isovector axial form factors}",
    eprint = "2305.11330",
    archivePrefix = "arXiv",
    primaryClass = "hep-lat",
    reportNumber = "Los Alamos LA-UR-23-25225",
    doi = "10.1103/PhysRevD.109.014503",
    journal = "Phys. Rev. D",
    volume = "109",
    number = "1",
    pages = "014503",
    year = "2024"
}

@article{ETM:2017wqc,
    author = "Alexandrou, C. and others",
    collaboration = "ETM",
    title = "{Pion vector form factor from lattice QCD at the physical point}",
    eprint = "1710.10401",
    archivePrefix = "arXiv",
    primaryClass = "hep-lat",
    doi = "10.1103/PhysRevD.97.014508",
    journal = "Phys. Rev. D",
    volume = "97",
    number = "1",
    pages = "014508",
    year = "2018"
}

@article{ExtendedTwistedMass:2024kjf,
    author = "Alexandrou, Constantia and others",
    collaboration = "Extended Twisted Mass",
    title = "{Quark and Gluon Momentum Fractions in the Pion and in the Kaon}",
    eprint = "2405.08529",
    archivePrefix = "arXiv",
    primaryClass = "hep-lat",
    doi = "10.1103/PhysRevLett.134.131902",
    journal = "Phys. Rev. Lett.",
    volume = "134",
    number = "13",
    pages = "131902",
    year = "2025"
}

@article{Hackett:2023nkr,
    author = "Hackett, Daniel C. and Oare, Patrick R. and Pefkou, Dimitra A. and Shanahan, Phiala E.",
    title = "{Gravitational form factors of the pion from lattice QCD}",
    eprint = "2307.11707",
    archivePrefix = "arXiv",
    primaryClass = "hep-lat",
    reportNumber = "MIT-CTP/5585",
    doi = "10.1103/PhysRevD.108.114504",
    journal = "Phys. Rev. D",
    volume = "108",
    number = "11",
    pages = "114504",
    year = "2023"
}

@article{Alexandrou:2021mmi,
    author = {Alexandrou, Constantia and Bacchio, Simone and Clo{\"e}t, Ian and Constantinou, Martha and Hadjiyiannakou, Kyriakos and Koutsou, Giannis and Lauer, Colin},
    collaboration = "ETM",
    title = "{Pion and kaon {\ensuremath{\langle}}x3{\ensuremath{\rangle}} from lattice QCD and PDF reconstruction from Mellin moments}",
    eprint = "2104.02247",
    archivePrefix = "arXiv",
    primaryClass = "hep-lat",
    doi = "10.1103/PhysRevD.104.054504",
    journal = "Phys. Rev. D",
    volume = "104",
    number = "5",
    pages = "054504",
    year = "2021"
}

@article{Shindler:2023xpd,
    author = "Shindler, Andrea",
    title = "{Moments of parton distribution functions of any order from lattice QCD}",
    eprint = "2311.18704",
    archivePrefix = "arXiv",
    primaryClass = "hep-lat",
    reportNumber = "TTK-23-31",
    doi = "10.1103/PhysRevD.110.L051503",
    journal = "Phys. Rev. D",
    volume = "110",
    number = "5",
    pages = "L051503",
    year = "2024"
}

@article{Francis:2025rya,
    author = "Francis, Anthony and others",
    title = "{Gradient flow for parton distribution functions: first application to the pion}",
    eprint = "2509.02472",
    archivePrefix = "arXiv",
    primaryClass = "hep-lat",
    month = "9",
    year = "2025"
}

@article{Francis:2025pgf,
    author = "Francis, Anthony and Fritzsch, Patrick and Karur, Rohith and Kim, Jangho and Pederiva, Giovanni and Pefkou, Dimitra A. and Rago, Antonio and Shindler, Andrea and Walker-Loud, Andr{\'e} and Zafeiropoulos, Savvas",
    title = "{Moments of parton distributions functions of the pion from lattice QCD using gradient flow}",
    eprint = "2510.26738",
    archivePrefix = "arXiv",
    primaryClass = "hep-lat",
    month = "10",
    year = "2025"
}

@article{Detmold:2021uru,
    author = "Detmold, William and Grebe, Anthony V. and Kanamori, Issaku and Lin, C. -J. David and Perry, Robert J. and Zhao, Yong",
    collaboration = "HOPE",
    title = "{Parton physics from a heavy-quark operator product expansion: Formalism and Wilson coefficients}",
    eprint = "2103.09529",
    archivePrefix = "arXiv",
    primaryClass = "hep-lat",
    reportNumber = "MIT-CTP/5294",
    doi = "10.1103/PhysRevD.104.074511",
    journal = "Phys. Rev. D",
    volume = "104",
    number = "7",
    pages = "074511",
    year = "2021"
}

@article{Detmold:2025lyb,
    author = "Detmold, William and Grebe, Anthony V. and Kanamori, Issaku and Lin, C. -J. David and Perry, Robert J. and Zhao, Yong",
    collaboration = "HOPE",
    title = "{Parton physics from a heavy-quark operator product expansion: Lattice QCD calculation of the fourth moment of the pion distribution amplitude}",
    eprint = "2509.04799",
    archivePrefix = "arXiv",
    primaryClass = "hep-lat",
    reportNumber = "MIT-CTP/5913, FERMILAB-PUB-25-0615-T",
    doi = "10.1103/xs4z-nblq",
    journal = "Phys. Rev. D",
    volume = "113",
    number = "1",
    pages = "014510",
    year = "2026"
}

@unpublished{Collins2025,
    author = "Collins, S.",
    collabotation = "RQCD",
    title = "{\it Nucleon isovactor charges from RQCD}",
    year="Private communication"
}

@article{Bali:2023sdi,
    author = {Bali, Gunnar S. and Collins, Sara and Heybrock, Simon and L{\"o}ffler, Marius and R{\"o}dl, Rudolf and S{\"o}ldner, Wolfgang and Weish{\"a}upl, Simon},
    collaboration = "RQCD",
    title = "{Octet baryon isovector charges from Nf=2+1 lattice QCD}",
    eprint = "2305.04717",
    archivePrefix = "arXiv",
    primaryClass = "hep-lat",
    doi = "10.1103/PhysRevD.108.034512",
    journal = "Phys. Rev. D",
    volume = "108",
    number = "3",
    pages = "034512",
    year = "2023"
}

@article{Wang:2025nsd,
    author = "Wang, Ji-Hao and Hu, Zhi-Cheng and Ji, Xiangdong and Jiang, Xiangyu and Su, Yushan and Sun, Peng and Yang, Yi-Bo",
    collaboration = "CLQCD",
    title = "{Precision determination of nucleon iso-vector scalar and tensor charges at the physical point}",
    eprint = "2511.02326",
    archivePrefix = "arXiv",
    primaryClass = "hep-lat",
    month = "11",
    year = "2025"
}

@article{Liang:2018pis,
    author = "Liang, Jian and Yang, Yi-Bo and Draper, Terrence and Gong, Ming and Liu, Keh-Fei",
    title = "{Quark spins and Anomalous Ward Identity}",
    eprint = "1806.08366",
    archivePrefix = "arXiv",
    primaryClass = "hep-ph",
    doi = "10.1103/PhysRevD.98.074505",
    journal = "Phys. Rev. D",
    volume = "98",
    number = "7",
    pages = "074505",
    year = "2018"
}

@article{Park:2020axe,
    author = "Park, Sungwoo and Bhattacharya, Tanmoy and Gupta, Rajan and Jang, Yong-Chull and Joo, Balint and Lin, Huey-Wen and Yoon, Boram",
    title = "{Nucleon charges and form factors using clover and HISQ ensembles}",
    eprint = "2002.02147",
    archivePrefix = "arXiv",
    primaryClass = "hep-lat",
    reportNumber = "LA-UR-19-31873",
    doi = "10.22323/1.363.0136",
    journal = "PoS",
    volume = "LATTICE2019",
    pages = "136",
    year = "2020"
}

@article{Djukanovic:2019gvi,
    author = "Djukanovic, Dalibor and Meyer, Harvey and Ottnad, Konstantin and von Hippel, Georg and Wilhelm, Jonas and Wittig, Hartmut",
    title = "{Strange nucleon form factors and isoscalar charges with $N_f=2+1$ $\mathcal{O}(a)$-improved Wilson fermions}",
    eprint = "1911.01177",
    archivePrefix = "arXiv",
    primaryClass = "hep-lat",
    reportNumber = "MITP/19-069",
    doi = "10.22323/1.363.0158",
    journal = "PoS",
    volume = "LATTICE2019",
    pages = "158",
    year = "2019"
}

@article{Alexandrou:2024ozj,
    author = "Alexandrou, C. and Bacchio, S. and Finkenrath, J. and Iona, C. and Koutsou, G. and Li, Y. and Spanoudes, G.",
    title = "{Nucleon charges and {\ensuremath{\sigma}}-terms in lattice QCD}",
    eprint = "2412.01535",
    archivePrefix = "arXiv",
    primaryClass = "hep-lat",
    doi = "10.1103/PhysRevD.111.054505",
    journal = "Phys. Rev. D",
    volume = "111",
    number = "5",
    pages = "054505",
    year = "2025"
}

@article{Djukanovic:2023beb,
    author = "Djukanovic, Dalibor and von Hippel, Georg and Meyer, Harvey B. and Ottnad, Konstantin and Salg, Miguel and Wittig, Hartmut",
    title = "{Electromagnetic form factors of the nucleon from Nf=2+1 lattice QCD}",
    eprint = "2309.06590",
    archivePrefix = "arXiv",
    primaryClass = "hep-lat",
    reportNumber = "MITP-23-044",
    doi = "10.1103/PhysRevD.109.094510",
    journal = "Phys. Rev. D",
    volume = "109",
    number = "9",
    pages = "094510",
    year = "2024"
}

@article{Djukanovic:2023jag,
    author = "Djukanovic, Dalibor and von Hippel, Georg and Meyer, Harvey B. and Ottnad, Konstantin and Salg, Miguel and Wittig, Hartmut",
    title = "{Precision Calculation of the Electromagnetic Radii of the Proton and Neutron from Lattice QCD}",
    eprint = "2309.07491",
    archivePrefix = "arXiv",
    primaryClass = "hep-lat",
    reportNumber = "MITP-23-045",
    doi = "10.1103/PhysRevLett.132.211901",
    journal = "Phys. Rev. Lett.",
    volume = "132",
    number = "21",
    pages = "211901",
    year = "2024"
}

@article{Alexandrou:2025vto,
    author = "Alexandrou, Constantia and Bacchio, Simone and Koutsou, Giannis and Prasad, Bhavna and Spanoudes, Gregoris",
    title = "{Proton and neutron electromagnetic form factors from lattice QCD in the continuum limit}",
    eprint = "2507.20910",
    archivePrefix = "arXiv",
    primaryClass = "hep-lat",
    month = "7",
    year = "2025"
}

@unpublished{Alexandrou2026,
    author = "Alexandrou, C. and others",
    title = "{Flavor decomposition of the nucleon momentum and spin sums}",
    eprint = "Forthcoming publication",
    year="Forthcoming publication"
}

@unpublished{Taggi2026,
    author = "Taggi E.",
    title = "{Higher moments of parton distribution functions from Lattice QCD at the physical point}",
    eprint = "Lattice 2025",
    year="Lattice 2025"
}

@unpublished{Chang2026,
    author = "Chang,  A. and others",
    title = "{\it Parton physics from a heavy-quark operator product expansion}",
    eprint = "PoS Lattice 2025",
    year=" PoS Lattice 2025"
}

@article{Gockeler:2005vw,
    author = "Gockeler, M. and Horsley, R. and Pleiter, D. and Rakow, Paul E. L. and Schafer, A. and Schierholz, G. and Stuben, H. and Zanotti, J. M.",
    title = "{Investigation of the second moment of the nucleon's g(1) and g(2) structure functions in two-flavor lattice QCD}",
    eprint = "hep-lat/0506017",
    archivePrefix = "arXiv",
    reportNumber = "DESY-05-076, EDINBURGH-2005-04, MPP-2005-52",
    doi = "10.1103/PhysRevD.72.054507",
    journal = "Phys. Rev. D",
    volume = "72",
    pages = "054507",
    year = "2005"
}

@article{Burger:2021knd,
    author = {B{\"u}rger, S. and Wurm, T. and L{\"o}ffler, M. and G{\"o}ckeler, M. and Bali, G. and Collins, S. and Sch{\"a}fer, A. and Sternbeck, A.},
    collaboration = "RQCD",
    title = "{Lattice results for the longitudinal spin structure and color forces on quarks in a nucleon}",
    eprint = "2111.08306",
    archivePrefix = "arXiv",
    primaryClass = "hep-lat",
    doi = "10.1103/PhysRevD.105.054504",
    journal = "Phys. Rev. D",
    volume = "105",
    number = "5",
    pages = "054504",
    year = "2022"
}

@article{Alexandrou:2018pbm,
    author = "Alexandrou, Constantia and Cichy, Krzysztof and Constantinou, Martha and Jansen, Karl and Scapellato, Aurora and Steffens, Fernanda",
    title = "{Light-Cone Parton Distribution Functions from Lattice QCD}",
    eprint = "1803.02685",
    archivePrefix = "arXiv",
    primaryClass = "hep-lat",
    doi = "10.1103/PhysRevLett.121.112001",
    journal = "Phys. Rev. Lett.",
    volume = "121",
    number = "11",
    pages = "112001",
    year = "2018"
}

@article{Lin:2018pvv,
    author = "Lin, Huey-Wen and Chen, Jiunn-Wei and Ji, Xiangdong and Jin, Luchang and Li, Ruizi and Liu, Yu-Sheng and Yang, Yi-Bo and Zhang, Jian-Hui and Zhao, Yong",
    title = "{Proton Isovector Helicity Distribution on the Lattice at Physical Pion Mass}",
    eprint = "1807.07431",
    archivePrefix = "arXiv",
    primaryClass = "hep-lat",
    reportNumber = "MSUHEP-18-013, MIT-CTP/5032",
    doi = "10.1103/PhysRevLett.121.242003",
    journal = "Phys. Rev. Lett.",
    volume = "121",
    number = "24",
    pages = "242003",
    year = "2018"
}

@article{Gao:2022iex,
    author = "Gao, Xiang and Hanlon, Andrew D. and Karthik, Nikhil and Mukherjee, Swagato and Petreczky, Peter and Scior, Philipp and Shi, Shuzhe and Syritsyn, Sergey and Zhao, Yong and Zhou, Kai",
    title = "{Continuum-extrapolated NNLO valence PDF of the pion at the physical point}",
    eprint = "2208.02297",
    archivePrefix = "arXiv",
    primaryClass = "hep-lat",
    doi = "10.1103/PhysRevD.106.114510",
    journal = "Phys. Rev. D",
    volume = "106",
    number = "11",
    pages = "114510",
    year = "2022"
}

@article{Gao:2022uhg,
    author = "Gao, Xiang and Hanlon, Andrew D. and Holligan, Jack and Karthik, Nikhil and Mukherjee, Swagato and Petreczky, Peter and Syritsyn, Sergey and Zhao, Yong",
    title = "{Unpolarized proton PDF at NNLO from lattice QCD with physical quark masses}",
    eprint = "2212.12569",
    archivePrefix = "arXiv",
    primaryClass = "hep-lat",
    doi = "10.1103/PhysRevD.107.074509",
    journal = "Phys. Rev. D",
    volume = "107",
    number = "7",
    pages = "074509",
    year = "2023"
}

@article{Gao:2020ito,
    author = "Gao, Xiang and Jin, Luchang and Kallidonis, Christos and Karthik, Nikhil and Mukherjee, Swagato and Petreczky, Peter and Shugert, Charles and Syritsyn, Sergey and Zhao, Yong",
    title = "{Valence parton distribution of the pion from lattice QCD: Approaching the continuum limit}",
    eprint = "2007.06590",
    archivePrefix = "arXiv",
    primaryClass = "hep-lat",
    doi = "10.1103/PhysRevD.102.094513",
    journal = "Phys. Rev. D",
    volume = "102",
    number = "9",
    pages = "094513",
    year = "2020"
}

@article{Ji:2020brr,
    author = {Ji, Xiangdong and Liu, Yizhuang and Sch{\"a}fer, Andreas and Wang, Wei and Yang, Yi-Bo and Zhang, Jian-Hui and Zhao, Yong},
    title = "{A Hybrid Renormalization Scheme for Quasi Light-Front Correlations in Large-Momentum Effective Theory}",
    eprint = "2008.03886",
    archivePrefix = "arXiv",
    primaryClass = "hep-ph",
    doi = "10.1016/j.nuclphysb.2021.115311",
    journal = "Nucl. Phys. B",
    volume = "964",
    pages = "115311",
    year = "2021"
}

@article{Gao:2021dbh,
    author = "Gao, Xiang and Hanlon, Andrew D. and Mukherjee, Swagato and Petreczky, Peter and Scior, Philipp and Syritsyn, Sergey and Zhao, Yong",
    title = "{Lattice QCD Determination of the Bjorken-x Dependence of Parton Distribution Functions at Next-to-Next-to-Leading Order}",
    eprint = "2112.02208",
    archivePrefix = "arXiv",
    primaryClass = "hep-lat",
    doi = "10.1103/PhysRevLett.128.142003",
    journal = "Phys. Rev. Lett.",
    volume = "128",
    number = "14",
    pages = "142003",
    year = "2022"
}

@article{Abbott:2025irb,
    author = "Abbott, Ryan and Hackett, Daniel C. and Pefkou, Dimitra A. and Romero-L{\'o}pez, Fernando and Shanahan, Phiala E.",
    title = "{Lattice Evidence That Scalar Glueballs are Small}",
    eprint = "2508.21821",
    archivePrefix = "arXiv",
    primaryClass = "hep-lat",
    reportNumber = "MIT-CTP/5907, FERMILAB-PUB-25-0621-T, INT-PUB-25-021",
    doi = "10.1103/67xg-qxhz",
    journal = "Phys. Rev. Lett.",
    volume = "136",
    number = "4",
    pages = "041901",
    year = "2026"
}

@article{Sufian:2019bol,
    author = "Sufian, Raza Sabbir and Karpie, Joseph and Egerer, Colin and Orginos, Kostas and Qiu, Jian-Wei and Richards, David G.",
    title = "{Pion Valence Quark Distribution from Matrix Element Calculated in Lattice QCD}",
    eprint = "1901.03921",
    archivePrefix = "arXiv",
    primaryClass = "hep-lat",
    reportNumber = "JLAB-THY-19-2847",
    doi = "10.1103/PhysRevD.99.074507",
    journal = "Phys. Rev. D",
    volume = "99",
    number = "7",
    pages = "074507",
    year = "2019"
}

@article{Zimmermann:2024zde,
    author = {Zimmermann, Christian and Sch{\"a}fer, Andreas},
    title = "{Valence quark PDFs of the proton from two-current correlations in lattice QCD}",
    eprint = "2405.07712",
    archivePrefix = "arXiv",
    primaryClass = "hep-lat",
    doi = "10.1103/PhysRevD.110.074503",
    journal = "Phys. Rev. D",
    volume = "110",
    number = "7",
    pages = "074503",
    year = "2024"
}

@article{LatticeParton:2022xsd,
    author = "Yao, Fei and others",
    collaboration = "Lattice Parton",
    title = "{Nucleon Transversity Distribution in the Continuum and Physical Mass Limit from Lattice QCD}",
    eprint = "2208.08008",
    archivePrefix = "arXiv",
    primaryClass = "hep-lat",
    doi = "10.1103/PhysRevLett.131.261901",
    journal = "Phys. Rev. Lett.",
    volume = "131",
    number = "26",
    pages = "261901",
    year = "2023"
}

@article{Ji:2022ezo,
    author = "Ji, Xiangdong",
    title = "{Large-Momentum Effective Theory vs. Short-Distance Operator Expansion: Contrast and Complementarity}",
    eprint = "2209.09332",
    archivePrefix = "arXiv",
    primaryClass = "hep-lat",
    doi = "10.34133/research.0695",
    journal = "Research",
    volume = "8",
    pages = "0695",
    year = "2025"
}

@article{Lin:2020ssv,
    author = "Lin, Huey-Wen and Chen, Jiunn-Wei and Fan, Zhouyou and Zhang, Jian-Hui and Zhang, Rui",
    title = "{Valence-Quark Distribution of the Kaon and Pion from Lattice QCD}",
    eprint = "2003.14128",
    archivePrefix = "arXiv",
    primaryClass = "hep-lat",
    reportNumber = "MSUHEP-20-006",
    doi = "10.1103/PhysRevD.103.014516",
    journal = "Phys. Rev. D",
    volume = "103",
    number = "1",
    pages = "014516",
    year = "2021"
}

@article{Gao:2021xsm,
    author = "Gao, Xiang and Karthik, Nikhil and Mukherjee, Swagato and Petreczky, Peter and Syritsyn, Sergey and Zhao, Yong",
    title = "{Pion form factor and charge radius from lattice QCD at the physical point}",
    eprint = "2102.06047",
    archivePrefix = "arXiv",
    primaryClass = "hep-lat",
    doi = "10.1103/PhysRevD.104.114515",
    journal = "Phys. Rev. D",
    volume = "104",
    number = "11",
    pages = "114515",
    year = "2021"
}

@article{Wang:2020nbf,
    author = "Wang, Gen and Liang, Jian and Draper, Terrence and Liu, Keh-Fei and Yang, Yi-Bo",
    collaboration = "chiQCD",
    title = "{Lattice Calculation of Pion Form Factor with Overlap Fermions}",
    eprint = "2006.05431",
    archivePrefix = "arXiv",
    primaryClass = "hep-ph",
    doi = "10.1103/PhysRevD.104.074502",
    journal = "Phys. Rev. D",
    volume = "104",
    pages = "074502",
    year = "2021"
}

@article{Alexandrou:2015rja,
    author = "Alexandrou, Constantia and Cichy, Krzysztof and Drach, Vincent and Garcia-Ramos, Elena and Hadjiyiannakou, Kyriakos and Jansen, Karl and Steffens, Fernanda and Wiese, Christian",
    title = "{Lattice calculation of parton distributions}",
    eprint = "1504.07455",
    archivePrefix = "arXiv",
    primaryClass = "hep-lat",
    reportNumber = "SFB-CPP-14-124, DESY-15-059, CP3-ORIGINS-2015-013, DIAS-2015-13",
    doi = "10.1103/PhysRevD.92.014502",
    journal = "Phys. Rev. D",
    volume = "92",
    pages = "014502",
    year = "2015"
}

@article{NieMiera:2025inn,
    author = "NieMiera, Alex and Good, William and Lin, Huey-Wen",
    title = "{Kaon gluon parton distribution and momentum fraction from 2+1+1 lattice QCD with high statistics}",
    eprint = "2506.03002",
    archivePrefix = "arXiv",
    primaryClass = "hep-lat",
    reportNumber = "MSUHEP-25-025",
    doi = "10.1103/89mq-gx33",
    journal = "Phys. Rev. D",
    volume = "112",
    number = "7",
    pages = "074504",
    year = "2025"
}

@article{Khan:2022vot,
    author = "Khan, Tanjib and Liu, Tianbo and Sufian, Raza Sabbir",
    title = "{Gluon helicity in the nucleon from lattice QCD and machine learning}",
    eprint = "2211.15587",
    archivePrefix = "arXiv",
    primaryClass = "hep-lat",
    doi = "10.1103/PhysRevD.108.074502",
    journal = "Phys. Rev. D",
    volume = "108",
    number = "7",
    pages = "074502",
    year = "2023"
}

@article{Chowdhury:2024ymm,
    author = "Chowdhury, Talal Ahmed and Izubuchi, Taku and Kamruzzaman, Methun and Karthik, Nikhil and Khan, Tanjib and Liu, Tianbo and Paul, Arpon and Schoenleber, Jakob and Sufian, Raza Sabbir",
    title = "{Polarized and unpolarized gluon PDFs: Generative machine learning applications for lattice QCD matrix elements at short distance and large momentum}",
    eprint = "2409.17234",
    archivePrefix = "arXiv",
    primaryClass = "hep-lat",
    doi = "10.1103/PhysRevD.111.074509",
    journal = "Phys. Rev. D",
    volume = "111",
    number = "7",
    pages = "074509",
    year = "2025"
}

@article{HadStruc:2021wmh,
    author = "Khan, Tanjib and others",
    collaboration = "HadStruc",
    title = "{Unpolarized gluon distribution in the nucleon from lattice quantum chromodynamics}",
    eprint = "2107.08960",
    archivePrefix = "arXiv",
    primaryClass = "hep-lat",
    reportNumber = "JLAB-THY-21-3469",
    doi = "10.1103/PhysRevD.104.094516",
    journal = "Phys. Rev. D",
    volume = "104",
    number = "9",
    pages = "094516",
    year = "2021"
}

@article{Delmar:2023agv,
    author = "Delmar, Joseph and Alexandrou, Constantia and Cichy, Krzysztof and Constantinou, Martha and Hadjiyiannakou, Kyriakos",
    title = "{Gluon PDF of the proton using twisted mass fermions}",
    eprint = "2310.01389",
    archivePrefix = "arXiv",
    primaryClass = "hep-lat",
    doi = "10.1103/PhysRevD.108.094515",
    journal = "Phys. Rev. D",
    volume = "108",
    number = "9",
    pages = "094515",
    year = "2023"
}

@article{ChenChen:2025amm,
    author = "Chen, Chen and Dong, Hongxin and Liu, Liuming and Sun, Peng and Xiong, Xiaonu and Yang, Yi-Bo and Yao, Fei and Zhang, Jian-Hui and Zeng, Chunhua and Zhong, Shiyi",
    title = "{Unpolarized gluon PDF of the nucleon from lattice QCD in the continuum limit}",
    eprint = "2510.26425",
    archivePrefix = "arXiv",
    primaryClass = "hep-lat",
    month = "10",
    year = "2025"
}

@article{NieMiera:2025mwj,
    author = "NieMiera, Alex and Good, William and Lin, Huey-Wen and Yao, Fei",
    title = "{First Self-Renormalized Gluon PDF of Nucleon from Large-Momentum Effective Theory in the Continuum Limit}",
    eprint = "2510.17758",
    archivePrefix = "arXiv",
    primaryClass = "hep-lat",
    reportNumber = "MSUHEP-25-024",
    month = "10",
    year = "2025"
}

@article{Good:2023ecp,
    author = "Good, William and Hasan, Kinza and Chevis, Allison and Lin, Huey-Wen",
    title = "{Gluon moment and parton distribution function of the pion from Nf=2+1+1 lattice QCD}",
    eprint = "2310.12034",
    archivePrefix = "arXiv",
    primaryClass = "hep-lat",
    reportNumber = "MSUHEP-23-027",
    doi = "10.1103/PhysRevD.109.114509",
    journal = "Phys. Rev. D",
    volume = "109",
    number = "11",
    pages = "114509",
    year = "2024"
}

@article{HadStruc:2022yaw,
    author = "Egerer, Colin and others",
    collaboration = "HadStruc",
    title = "{Toward the determination of the gluon helicity distribution in the nucleon from lattice quantum chromodynamics}",
    eprint = "2207.08733",
    archivePrefix = "arXiv",
    primaryClass = "hep-lat",
    reportNumber = "JLAB-THY-22-3663",
    doi = "10.1103/PhysRevD.106.094511",
    journal = "Phys. Rev. D",
    volume = "106",
    number = "9",
    pages = "094511",
    year = "2022"
}

@article{Scimemi:2019cmh,
    author = "Scimemi, Ignazio and Vladimirov, Alexey",
    title = "{Non-perturbative structure of semi-inclusive deep-inelastic and Drell-Yan scattering at small transverse momentum}",
    eprint = "1912.06532",
    archivePrefix = "arXiv",
    primaryClass = "hep-ph",
    doi = "10.1007/JHEP06(2020)137",
    journal = "JHEP",
    volume = "06",
    pages = "137",
    year = "2020"
}

@article{Shanahan:2022ifi,
    author = "Shanahan, Phiala and others",
    title = "{Snowmass 2021 Computational Frontier CompF03 Topical Group Report: Machine Learning}",
    eprint = "2209.07559",
    archivePrefix = "arXiv",
    primaryClass = "physics.comp-ph",
    reportNumber = "FERMILAB-CONF-22-719-ND-PPD-QIS-SCD",
    month = "9",
    year = "2022"
}

@article{Hagler:2009ni,
    author = "Hagler, Ph.",
    title = "{Hadron structure from lattice quantum chromodynamics}",
    eprint = "0912.5483",
    archivePrefix = "arXiv",
    primaryClass = "hep-lat",
    reportNumber = "TUM-T39-09-12",
    doi = "10.1016/j.physrep.2009.12.008",
    journal = "Phys. Rept.",
    volume = "490",
    pages = "49--175",
    year = "2010"
}

@article{Miller:2025wgr,
    author = "Miller, Joshua and Torsiello, Joseph and Anderson, Isaac and Cichy, Krzysztof and Constantinou, Martha and Delmar, Joseph and Lampreich, Sarah",
    title = "{Pion and Kaon PDFs from Lattice QCD with Complementary Approaches}",
    eprint = "2512.06121",
    archivePrefix = "arXiv",
    primaryClass = "hep-lat",
    month = "12",
    year = "2025"
}

@article{Sufian:2020vzb,
    author = "Sufian, Raza Sabbir and Egerer, Colin and Karpie, Joseph and Edwards, Robert G. and Jo{\'o}, B{\'a}lint and Ma, Yan-Qing and Orginos, Kostas and Qiu, Jian-Wei and Richards, David G.",
    title = "{Pion Valence Quark Distribution from Current-Current Correlation in Lattice QCD}",
    eprint = "2001.04960",
    archivePrefix = "arXiv",
    primaryClass = "hep-lat",
    reportNumber = "JLAB-THY-20-3131",
    doi = "10.1103/PhysRevD.102.054508",
    journal = "Phys. Rev. D",
    volume = "102",
    number = "5",
    pages = "054508",
    year = "2020"
}

@article{Ma:2014jla,
    author = "Ma, Yan-Qing and Qiu, Jian-Wei",
    title = "{Extracting Parton Distribution Functions from Lattice QCD Calculations}",
    eprint = "1404.6860",
    archivePrefix = "arXiv",
    primaryClass = "hep-ph",
    doi = "10.1103/PhysRevD.98.074021",
    journal = "Phys. Rev. D",
    volume = "98",
    number = "7",
    pages = "074021",
    year = "2018"
}

@article{Zhang:2018nsy,
    author = {Zhang, Jian-Hui and Chen, Jiunn-Wei and Jin, Luchang and Lin, Huey-Wen and Sch{\"a}fer, Andreas and Zhao, Yong},
    title = "{First direct lattice-QCD calculation of the $x$-dependence of the pion parton distribution function}",
    eprint = "1804.01483",
    archivePrefix = "arXiv",
    primaryClass = "hep-lat",
    doi = "10.1103/PhysRevD.100.034505",
    journal = "Phys. Rev. D",
    volume = "100",
    number = "3",
    pages = "034505",
    year = "2019"
}

@article{Izubuchi:2019lyk,
    author = "Izubuchi, Taku and Jin, Luchang and Kallidonis, Christos and Karthik, Nikhil and Mukherjee, Swagato and Petreczky, Peter and Shugert, Charles and Syritsyn, Sergey",
    title = "{Valence parton distribution function of pion from fine lattice}",
    eprint = "1905.06349",
    archivePrefix = "arXiv",
    primaryClass = "hep-lat",
    doi = "10.1103/PhysRevD.100.034516",
    journal = "Phys. Rev. D",
    volume = "100",
    number = "3",
    pages = "034516",
    year = "2019"
}

@article{Wilson:1974sk,
  author       = {Wilson, K. G.},
  title        = {Confinement of Quarks},
  journal      = {Phys. Rev. D},
  volume       = {10},
  year         = {1974},
  pages        = {2445--2459},
  doi          = {10.1103/PhysRevD.10.2445}
}

@article{Creutz:1980zw,
  author       = {Creutz, M.},
  title        = {Monte Carlo Study of Quantized SU(2) Gauge Theory},
  journal      = {Phys. Rev. D},
  volume       = {21},
  year         = {1980},
  pages        = {2308--2315},
  doi          = {10.1103/PhysRevD.21.2308}
}

@article{Ji:2020ect,
    author = "Ji, Xiangdong and Liu, Yu-Sheng and Liu, Yizhuang and Zhang, Jian-Hui and Zhao, Yong",
    title = "{Large-momentum effective theory}",
    eprint = "2004.03543",
    archivePrefix = "arXiv",
    primaryClass = "hep-ph",
    doi = "10.1103/RevModPhys.93.035005",
    journal = "Rev. Mod. Phys.",
    volume = "93",
    number = "3",
    pages = "035005",
    year = "2021"
}

@article{Radyushkin:2019mye,
    author = "Radyushkin, A. V.",
    title = "{Theory and applications of parton pseudodistributions}",
    eprint = "1912.04244",
    archivePrefix = "arXiv",
    primaryClass = "hep-ph",
    reportNumber = "JLAB-THY-19-3118",
    doi = "10.1142/S0217751X20300021",
    journal = "Int. J. Mod. Phys. A",
    volume = "35",
    number = "05",
    pages = "2030002",
    year = "2020"
}

@article{Accardi:2012qut,
  author       = {Accardi, A. and others},
  title        = {Electron Ion Collider: The Next QCD Frontier},
  journal      = {Eur. Phys. J. A},
  volume       = {52},
  year         = {2016},
  number       = {9},
  pages        = {268},
  doi          = {10.1140/epja/i2016-16268-9},
  eprint       = {1212.1701},
  archivePrefix= {arXiv},
  primaryClass = {nucl-ex}
}

@article{AbdulKhalek:2021gbh,
  author       = {Abdul Khalek, R. and others},
  title        = {Science Requirements and Detector Concepts for the Electron--Ion Collider},
  journal      = {Nucl. Phys. A},
  volume       = {1026},
  year         = {2022},
  pages        = {122447},
  doi          = {10.1016/j.nuclphysa.2022.122447},
  eprint       = {2103.05419},
  archivePrefix= {arXiv},
  primaryClass = {physics.ins-det}
}

@article{Aschenauer:2019kzf,
    author = "Aschenauer, Elke C. and Borsa, Ignacio and Sassot, Rodolfo and Van Hulse, Charlotte",
    title = "{Semi-inclusive Deep-Inelastic Scattering, Parton Distributions and Fragmentation Functions at a Future Electron-Ion Collider}",
    eprint = "1902.10663",
    archivePrefix = "arXiv",
    primaryClass = "hep-ph",
    doi = "10.1103/PhysRevD.99.094004",
    journal = "Phys. Rev. D",
    volume = "99",
    number = "9",
    pages = "094004",
    year = "2019"
}

@article{Ji:2015qla,
  author  = {Ji, Xiangdong and Zhang, Jian-Hui and Zhao, Yong},
  title   = {Renormalization in Large-Momentum Effective Theory of Parton Physics},
  journal = {Phys. Rev. D},
  volume  = {92},
  pages   = {014039},
  year    = {2015},
  doi     = {10.1103/PhysRevD.92.014039},
  eprint  = {arXiv:1506.00248}
}

@article{Xiong:2015nua,
  author  = {Xiong, Xiaonu and Ji, Xiangdong and Zhang, Jian-Hui and Zhao, Yong},
  title   = {One-Loop Matching for Parton Distributions: Nonsinglet Case},
  journal = {Phys. Rev. D},
  volume  = {92},
  pages   = {054037},
  year    = {2015},
  doi     = {10.1103/PhysRevD.92.054037},
  eprint  = {arXiv:1509.08016}
}

@article{Liu:2019urm,
  author  = {Liu, Yu-Sheng and Chen, Jiunn-Wei and Ji, Xiangdong and Zhang, Jian-Hui},
  title   = {Matching the Quasi Parton Distribution in a Momentum Subtraction Scheme},
  journal = {Phys. Rev. D},
  volume  = {100},
  pages   = {034006},
  year    = {2019},
  doi     = {10.1103/PhysRevD.100.034006},
  eprint  = {arXiv:1902.00407}
}

@article{Alexandrou:2020zbe,
    author = "Alexandrou, Constantia and Cichy, Krzysztof and Constantinou, Martha and Hadjiyiannakou, Kyriakos and Jansen, Karl and Scapellato, Aurora and Steffens, Fernanda",
    title = "{Unpolarized and helicity generalized parton distributions of the proton within lattice QCD}",
    eprint = "2008.10573",
    archivePrefix = "arXiv",
    primaryClass = "hep-lat",
    reportNumber = "DESY-20-150",
    doi = "10.1103/PhysRevLett.125.262001",
    journal = "Phys. Rev. Lett.",
    volume = "125",
    number = "26",
    pages = "262001",
    year = "2020"
}

@article{Alexandrou:2021bbo,
    author = "Alexandrou, Constantia and Cichy, Krzysztof and Constantinou, Martha and Hadjiyiannakou, Kyriakos and Jansen, Karl and Scapellato, Aurora and Steffens, Fernanda",
    title = "{Transversity GPDs of the proton from lattice QCD}",
    eprint = "2108.10789",
    archivePrefix = "arXiv",
    primaryClass = "hep-lat",
    doi = "10.1103/PhysRevD.105.034501",
    journal = "Phys. Rev. D",
    volume = "105",
    number = "3",
    pages = "034501",
    year = "2022"
}

@article{Chen:2019lcm,
    author = "Chen, Jiunn-Wei and Lin, Huey-Wen and Zhang, Jian-Hui",
    title = "{Pion generalized parton distribution from lattice QCD}",
    eprint = "1904.12376",
    archivePrefix = "arXiv",
    primaryClass = "hep-lat",
    doi = "10.1016/j.nuclphysb.2020.114940",
    journal = "Nucl. Phys. B",
    volume = "952",
    pages = "114940",
    year = "2020"
}

@article{Chu:2025kew,
    author = "Chu, Min-Huan and Cola{\c{c}}o, Manuel and Bhattacharya, Shohini and Cichy, Krzysztof and Constantinou, Martha and Metz, Andreas and Steffens, Fernanda",
    title = "{Generalized parton distributions from lattice QCD with asymmetric momentum transfer: Unpolarized quarks at nonzero skewness}",
    eprint = "2508.17998",
    archivePrefix = "arXiv",
    primaryClass = "hep-lat",
    doi = "10.1103/ts5s-hvb1",
    journal = "Phys. Rev. D",
    volume = "112",
    number = "9",
    pages = "094510",
    year = "2025"
}

@article{Bhattacharya:2022aob,
    author = "Bhattacharya, Shohini and Cichy, Krzysztof and Constantinou, Martha and Dodson, Jack and Gao, Xiang and Metz, Andreas and Mukherjee, Swagato and Scapellato, Aurora and Steffens, Fernanda and Zhao, Yong",
    title = "{Generalized parton distributions from lattice QCD with asymmetric momentum transfer: Unpolarized quarks}",
    eprint = "2209.05373",
    archivePrefix = "arXiv",
    primaryClass = "hep-lat",
    doi = "10.1103/PhysRevD.106.114512",
    journal = "Phys. Rev. D",
    volume = "106",
    number = "11",
    pages = "114512",
    year = "2022"
}

@article{Bhattacharya:2023jsc,
    author = "Bhattacharya, Shohini and others",
    title = "{Generalized parton distributions from lattice QCD with asymmetric momentum transfer: Axial-vector case}",
    eprint = "2310.13114",
    archivePrefix = "arXiv",
    primaryClass = "hep-lat",
    doi = "10.1103/PhysRevD.109.034508",
    journal = "Phys. Rev. D",
    volume = "109",
    number = "3",
    pages = "034508",
    year = "2024"
}

@article{Bhattacharya:2025yba,
    author = "Bhattacharya, Shohini and Cichy, Krzysztof and Constantinou, Martha and Metz, Andreas and Miller, Joshua and Petreczky, Peter and Steffens, Fernanda",
    title = "{Generalized parton distributions from lattice QCD with asymmetric momentum transfer: Tensor case}",
    eprint = "2505.11288",
    archivePrefix = "arXiv",
    primaryClass = "hep-lat",
    doi = "10.1103/tlkb-ykgp",
    journal = "Phys. Rev. D",
    volume = "112",
    number = "11",
    pages = "114504",
    year = "2025"
}

@article{Ding:2024saz,
    author = "Ding, Heng-Tong and Gao, Xiang and Mukherjee, Swagato and Petreczky, Peter and Shi, Qi and Syritsyn, Sergey and Zhao, Yong",
    title = "{Three-dimensional imaging of pion using lattice QCD: generalized parton distributions}",
    eprint = "2407.03516",
    archivePrefix = "arXiv",
    primaryClass = "hep-lat",
    doi = "10.1007/JHEP02(2025)056",
    journal = "JHEP",
    volume = "02",
    pages = "056",
    year = "2025"
}

@article{Bhattacharya:2024qpp,
    author = "Bhattacharya, Shohini and Cichy, Krzysztof and Constantinou, Martha and Metz, Andreas and Nurminen, Niilo and Steffens, Fernanda",
    title = "{Generalized parton distributions from the pseudodistribution approach on the lattice}",
    eprint = "2405.04414",
    archivePrefix = "arXiv",
    primaryClass = "hep-lat",
    reportNumber = "LA-UR-24-23903",
    doi = "10.1103/PhysRevD.110.054502",
    journal = "Phys. Rev. D",
    volume = "110",
    number = "5",
    pages = "054502",
    year = "2024"
}

@article{HadStruc:2024rix,
    author = "Dutrieux, Herv{\'e} and Edwards, Robert G. and Egerer, Colin and Karpie, Joseph and Monahan, Christopher and Orginos, Kostas and Radyushkin, Anatoly and Richards, David and Romero, Eloy and Zafeiropoulos, Savvas",
    collaboration = "HadStruc",
    title = "{Towards unpolarized GPDs from pseudo-distributions}",
    eprint = "2405.10304",
    archivePrefix = "arXiv",
    primaryClass = "hep-lat",
    reportNumber = "JLAB-THY-24-4059, JLAB-THY-24-4059",
    doi = "10.1007/JHEP08(2024)162",
    journal = "JHEP",
    volume = "08",
    pages = "162",
    year = "2024"
}

@article{Bhattacharya:2024wtg,
    author = "Bhattacharya, Shohini and Cichy, Krzysztof and Constantinou, Martha and Gao, Xiang and Metz, Andreas and Miller, Joshua and Mukherjee, Swagato and Petreczky, Peter and Steffens, Fernanda and Zhao, Yong",
    title = "{Moments of axial-vector GPD from lattice QCD: quark helicity, orbital angular momentum, and spin-orbit correlation}",
    eprint = "2410.03539",
    archivePrefix = "arXiv",
    primaryClass = "hep-lat",
    reportNumber = "LA-UR-24-29020",
    doi = "10.1007/JHEP01(2025)146",
    journal = "JHEP",
    volume = "01",
    pages = "146",
    year = "2025"
}

@article{Gao:2025inf,
    author = "Gao, Xiang and Mukherjee, Swagato and Shi, Qi and Yao, Fei and Zhao, Yong",
    title = "{Skewness-dependent moments of the pion GPD from nonlocal quark-bilinear correlators}",
    eprint = "2511.01818",
    archivePrefix = "arXiv",
    primaryClass = "hep-lat",
    doi = "10.1103/hrb2-47zm",
    journal = "Phys. Rev. D",
    volume = "113",
    number = "1",
    pages = "014505",
    year = "2026"
}

@article{Ji:2013dva,
  author       = {Ji, X.},
  title        = {Parton Physics on a Euclidean Lattice},
  journal      = {Phys. Rev. Lett.},
  volume       = {110},
  year         = {2013},
  pages        = {262002},
  doi          = {10.1103/PhysRevLett.110.262002},
  eprint       = {1305.1539},
  archivePrefix= {arXiv},
  primaryClass = {hep-ph}
}

@article{Liu:2016djw,
    author = "Liu, Keh-Fei",
    title = "{Parton Distribution Function from the Hadronic Tensor on the Lattice}",
    eprint = "1603.07352",
    archivePrefix = "arXiv",
    primaryClass = "hep-ph",
    doi = "10.22323/1.251.0115",
    journal = "PoS",
    volume = "LATTICE2015",
    pages = "115",
    year = "2016"
}

@article{Detmold:2005gg,
    author = "Detmold, William and Lin, C. J. David",
    title = "{Deep-inelastic scattering and the operator product expansion in lattice QCD}",
    eprint = "hep-lat/0507007",
    archivePrefix = "arXiv",
    reportNumber = "INT-PUB-05-16, UW-PT-05-17, NT-UW-05-05",
    doi = "10.1103/PhysRevD.73.014501",
    journal = "Phys. Rev. D",
    volume = "73",
    pages = "014501",
    year = "2006"
}

@article{Chambers:2017dov,
    author = "Chambers, A. J. and Horsley, R. and Nakamura, Y. and Perlt, H. and Rakow, P. E. L. and Schierholz, G. and Schiller, A. and Somfleth, K. and Young, R. D. and Zanotti, J. M.",
    title = "{Nucleon Structure Functions from Operator Product Expansion on the Lattice}",
    eprint = "1703.01153",
    archivePrefix = "arXiv",
    primaryClass = "hep-lat",
    reportNumber = "ADP-17-09-T1015, DESY-17-027, EDINBURGH-2017-04, LIVERPOOL-LTH-1122",
    doi = "10.1103/PhysRevLett.118.242001",
    journal = "Phys. Rev. Lett.",
    volume = "118",
    number = "24",
    pages = "242001",
    year = "2017"
}

@article{Ma:2017pxb,
    author = "Ma, Yan-Qing and Qiu, Jian-Wei",
    title = "{Exploring Partonic Structure of Hadrons Using ab initio Lattice QCD Calculations}",
    eprint = "1709.03018",
    archivePrefix = "arXiv",
    primaryClass = "hep-ph",
    reportNumber = "JLAB-THY-17-2542",
    doi = "10.1103/PhysRevLett.120.022003",
    journal = "Phys. Rev. Lett.",
    volume = "120",
    number = "2",
    pages = "022003",
    year = "2018"
}

@article{Radyushkin:2017cyf,
    author = "Radyushkin, A. V.",
    title = "{Quasi-parton distribution functions, momentum distributions, and pseudo-parton distribution functions}",
    eprint = "1705.01488",
    archivePrefix = "arXiv",
    primaryClass = "hep-ph",
    reportNumber = "JLAB-THY-17-2455",
    doi = "10.1103/PhysRevD.96.034025",
    journal = "Phys. Rev. D",
    volume = "96",
    number = "3",
    pages = "034025",
    year = "2017"
}

@article{Cichy:2018mum,
  author       = {Cichy, K. and Constantinou, M.},
  title        = {A Guide to Light-Cone PDFs from Lattice QCD},
  journal      = {Adv. High Energy Phys.},
  volume       = {2019},
  year         = {2019},
  pages        = {3036904},
  doi          = {10.1155/2019/3036904},
  eprint       = {1811.07248},
  archivePrefix= {arXiv},
  primaryClass = {hep-lat}
}

@article{Musch:2010ka,
  author       = {Musch, B. U. and others},
  title        = {Transverse Momentum Distributions of Quarks in the Nucleon from Lattice QCD},
  journal      = {Phys. Rev. D},
  volume       = {83},
  year         = {2011},
  pages        = {094507},
  doi          = {10.1103/PhysRevD.83.094507},
  eprint       = {1011.1213},
  archivePrefix= {arXiv},
  primaryClass = {hep-lat}
}

@article{Ji:2018hvs,
    author = "Ji, Xiangdong and Jin, Lu-Chang and Yuan, Feng and Zhang, Jian-Hui and Zhao, Yong",
    title = "{Transverse momentum dependent parton quasidistributions}",
    eprint = "1801.05930",
    archivePrefix = "arXiv",
    primaryClass = "hep-ph",
    doi = "10.1103/PhysRevD.99.114006",
    journal = "Phys. Rev. D",
    volume = "99",
    number = "11",
    pages = "114006",
    year = "2019"
}

@article{Prasad:2025pos,
    author = "Alexandrou, Constantia and Bacchio, Simone and Bode, Mathis and Finkenrath, Jacob and Herten, Andreas and Iona, Christos and Koutsou, Giannis and Pittler, Ferenc and Prasad, Bhavna and Spanoudes, Gregoris",
    title = "{Nucleon strange electromagnetic form factors using
                  $N_f=2+1+1$ twisted-mass fermions at the physical
                  point}",
    journal = "PoS",
    volume = "LATTICE2025",
    pages = "218",
    year = "2026"
    }

@article{Alexandrou:2026xcl,
    author = "Alexandrou, Constantia and Bacchio, Simone and Bode, Mathis and Finkenrath, Jacob and Herten, Andreas and Iona, Christos and Koutsou, Giannis and Pittler, Ferenc and Prasad, Bhavna and Spanoudes, Gregoris",
    title = "{Strangeness of nucleons from $N_f=2+1+1$ lattice QCD}",
    eprint = "2603.26600",
    archivePrefix = "arXiv",
    primaryClass = "hep-lat",
    month = "3",
    year = "2026"
}

@article{Alexandrou:2026ait,
    author = "Alexandrou, Constantia and Bacchio, Simone and Bode, Mathis and Finkenrath, Jacob and Herten, Andreas and Iona, Christos and Koutsou, Giannis and Pittler, Ferenc and Prasad, Bhavna and Spanoudes, Gregoris",
    title = "{Nucleon strange electromagnetic form factors from $N_f=2+1+1$ lattice QCD}",
    eprint = "2603.26591",
    archivePrefix = "arXiv",
    primaryClass = "hep-lat",
    month = "3",
    year = "2026"
}
\end{document}